\documentclass[11pt, a4paper]{article}
\pdfoutput=1
\usepackage[utf8]{inputenc}

\usepackage{jheppub,amsmath,amssymb,slashed,url,bm,textgreek,upgreek,hyperref}
\usepackage{jheppub}  
\usepackage{tikz,lipsum,lmodern}
\usepackage[most]{tcolorbox}
\usepackage{amssymb} 
\usepackage{amsmath}
\usepackage{mathtools}
\usepackage{amsfonts}    
\usepackage{dsfont}
\usepackage{pdfpages}
\usepackage{verbatim}
\usepackage{tensor}
\usepackage{mathrsfs}
\usepackage{textgreek} 
\usepackage[mathscr]{euscript}
\usepackage[normalem]{ulem}
\usepackage{tikz}
\usepackage{makecell}
\usepackage[verbose]{placeins}
\usepackage{subcaption}
\usetikzlibrary{3d, arrows.meta, decorations.pathreplacing, decorations.markings,calc,shapes.misc,decorations.pathmorphing,patterns.meta, math}
\usepackage{pgfplots}
\pgfplotsset{compat=1.18}
\usetikzlibrary{arrows.meta,calc}
\definecolor{ink}{HTML}{30343A}
\definecolor{quantum}{HTML}{A84F2D}

\newcommand{\beq}{\begin{equation}}
\newcommand{\eeq}{\end{equation}}
 
\newcommand{\bea}{\begin{eqnarray}}
\newcommand{\ea}{\end{eqnarray}}
\newcommand{\barr}{\begin{array}}
\newcommand{\earr}{\end{array}}

\def\be{\begin{equation}}
\def\ee{\end{equation}}
\def\ba#1\ea{\begin{align}#1\end{align}}
\def\bg#1\eg{\begin{gather}#1\end{gather}}
\def\bm#1\em{\begin{multline}#1\end{multline}}
\def\bmd#1\emd{\begin{multlined}#1\end{multlined}}

\def\vp{\chi}

\def\s{\sigma}

\def\wg{\wedge}

\def\<{\langle}
\def\>{\rangle}

\def\bea{\begin{eqnarray}}
\def\eea{\end{eqnarray}}

\newcommand{\tr}{\operatorname{tr}}

\newcommand{\zb}{{\bar z}}

\def\nn{\nonumber}

\begin{document}

\global\long\def\aad{(a\tilde{a}+a^{\dagger}\tilde{a}^{\dagger})}%

\global\long\def\ad{{\rm ad}}%

\global\long\def\bij{\langle ij\rangle}%

\global\long\def\df{\coloneqq}%

\global\long\def\bs{b_{\alpha}^{*}}%

\global\long\def\bra{\langle}%

\global\long\def\dd{{\rm d}}%

\global\long\def\dg{{\rm {\rm \dot{\gamma}}}}%

\global\long\def\ddt{\frac{{\rm d^{2}}}{{\rm d}t^{2}}}%

\global\long\def\ddg{\nabla_{\dot{\gamma}}}%

\global\long\def\del{\mathcal{\delta}}%

\global\long\def\Del{\Delta}%

\global\long\def\dtau{\frac{\dd^{2}}{\dd\tau^{2}}}%

\global\long\def\ul{U(\Lambda)}%

\global\long\def\udl{U^{\dagger}(\Lambda)}%

\global\long\def\dl{D(\Lambda)}%

\global\long\def\da{\dagger}%

\global\long\def\id{{\rm id}}%

\global\long\def\ml{\mathcal{L}}%

\global\long\def\mm{\mathcal{\mathcal{M}}}%

\global\long\def\mf{\mathcal{\mathcal{F}}}%

\global\long\def\ket{\rangle}%

\global\long\def\kpp{k^{\prime}}%

\global\long\def\lr{\leftrightarrow}%

\global\long\def\lf{\leftrightarrow}%

\global\long\def\ma{\mathcal{A}}%

\global\long\def\mb{\mathcal{B}}%

\global\long\def\md{\mathcal{D}}%

\global\long\def\mbr{\mathbb{R}}%

\global\long\def\mbz{\mathbb{Z}}%

\global\long\def\mh{\mathcal{\mathcal{H}}}%

\global\long\def\mi{\mathcal{\mathcal{I}}}%

\global\long\def\ms{\mathcal{\mathcal{\mathcal{S}}}}%

\global\long\def\mg{\mathcal{\mathcal{G}}}%

\global\long\def\mfa{\mathcal{\mathfrak{a}}}%

\global\long\def\mfb{\mathcal{\mathfrak{b}}}%

\global\long\def\mfb{\mathcal{\mathfrak{b}}}%

\global\long\def\mfg{\mathcal{\mathfrak{g}}}%

\global\long\def\mj{\mathcal{\mathcal{J}}}%

\global\long\def\mk{\mathcal{K}}%

\global\long\def\mmp{\mathcal{\mathcal{P}}}%

\global\long\def\mn{\mathcal{\mathcal{\mathcal{N}}}}%

\global\long\def\mq{\mathcal{\mathcal{Q}}}%

\global\long\def\mo{\mathcal{O}}%

\global\long\def\qq{\mathcal{\mathcal{\mathcal{\quad}}}}%

\global\long\def\ww{\wedge}%

\global\long\def\ka{\kappa}%

\global\long\def\nn{\nabla}%

\global\long\def\nb{\overline{\nabla}}%

\global\long\def\pathint{\langle x_{f},t_{f}|x_{i},t_{i}\rangle}%

\global\long\def\ppp{p^{\prime}}%

\global\long\def\qpp{q^{\prime}}%

\global\long\def\we{\wedge}%

\global\long\def\pp{\prime}%

\global\long\def\sq{\square}%

\global\long\def\vp{\varphi}%

\global\long\def\ti{\widetilde{}}%

\global\long\def\wg{\widetilde{g}}%

\global\long\def\te{\theta}%

\global\long\def\tr{{\rm Tr}}%

\global\long\def\ta{{\rm \widetilde{\alpha}}}%

\global\long\def\sh{{\rm {\rm sh}}}%

\global\long\def\ch{{\rm ch}}%

\global\long\def\Si{{\rm {\rm \Sigma}}}%

\global\long\def\sch{{\rm {\rm Sch}}}%

\global\long\def\vol{{\rm {\rm {\rm Vol}}}}%

\global\long\def\reg{{\rm {\rm reg}}}%

\global\long\def\zb{{\rm {\rm |0(\beta)\ket}}}%

\title{Quantum critical collapse in the large $D$ limit
\vspace{-0.8cm}}

\author[a]{Chih-Hung Wu}

\affiliation[a]{Department of Physics, University of Washington, Seattle, WA 98195, USA}

\emailAdd{chwu29@uw.edu}

\abstract{
In analytically tractable models of Einstein-scalar critical collapse, it has been shown that the quantum self-energy of the collapsing matter in the dominant $s$-wave sector grows with Lyapunov exponent $D-2$, eventually competing with the classical instability that controls black hole formation. This competition shifts the naked singularity threshold and produces a finite mass gap within semiclassical control, yet that control may appear threatened as the spacetime dimension $D\to\infty$. We study this quantum growing mode in both continuously (CSS) and discretely (DSS) self-similar critical spacetimes at large $D$ and establish that the universal Lyapunov exponent survives both forms of self-similarity. In CSS, the classical approach to a naked singularity is interrupted by the appearance of trapped surfaces. At large $D$, the horizon tracing reduces to a Bessel problem, yielding a quantum wall whose radius obeys $r_h\propto(D-2)^{-2}$. This radius can remain above the Planck scale at finite $D$, although maintaining this hierarchy becomes increasingly demanding as $D$ grows. In DSS, the same growth is modulated by echoing and can retain memory of previous echoes. The simplest anomaly-induced action exhibits accumulating memory but is singular at the spherical center. We construct a covariant one-loop action whose stress tensor remains smooth throughout the DSS domain and organize it into an echo-periodic structure without memory. This suggests that the smooth vacuum polarization of the genuine four-dimensional DSS Choptuik spacetime may similarly be governed by its exact echoing symmetry.}

\maketitle

%%%%%%%%%%%%%%%%%%%%%%%%%%%%%%%%%%%%%%%%%%%%%%%%%%

\section{Introduction}
\label{sec:introduction}

The \textit{weak cosmic censorship} conjecture seeks to preserve predictability by requiring that curvature singularities formed in generic gravitational collapse remain hidden behind event horizons \cite{Penrose:1969pc,Hawking:1994ss}. Critical gravitational collapse provides a particularly sharp setting to test this expectation. Consider a one-parameter family of smooth initial data labeled by $p$. Weak data disperse, while sufficiently strong data form a black hole. At the threshold $p=p_*$, the evolution approaches a universal critical spacetime with a single classical growing perturbation. For Type II collapse, the mass on the marginally supercritical side obeys
\be
    M_{\rm BH}\sim (p-p_*)^\gamma ,
    \label{eq:intro_mass_scaling}
\ee
and can be made arbitrarily small by fine-tuning. The exponent $\gamma$ is universal within a given matter model and spacetime dimension, independently of the particular family of initial data used to approach the threshold. This scaling behavior was first discovered by Choptuik in 4d Einstein gravity minimally coupled to a massless scalar field, which we refer to as the Einstein-scalar system \cite{Choptuik:1992jv}, and has since become a central example of universality in gravitational dynamics \cite{Gundlach:1995kd,Gundlach:1996eg,Gundlach:2025yje}. At the threshold itself, a \emph{self-similar} contraction reaches a zero-radius naked singularity and presents a dynamical challenge to weak cosmic censorship.

Fine-tuning $p$ closer to $p_*$ keeps the classical growing perturbation small for longer, delaying the departure from self-similarity toward dispersion or black hole formation. For supercritical data, the black hole mass is set by the scale at which this departure occurs. Self-similarity organizes the contraction in two different ways. Continuously self-similar (CSS) spacetimes possess continuous scale invariance. Discretely self-similar (DSS) spacetimes repeat their dimensionless geometry after a fixed interval $\Delta_g$ in logarithmic scale time, while the physical scale shrinks by $e^{-\Delta_g}$. The Choptuik critical solution in 4d is DSS, although critical spacetimes can exhibit either CSS or DSS depending on the matter model and spacetime dimension. CSS examples often provide greater analytic control over the approach to the endpoint~\cite{Christodoulou:1994hg,Garfinkle:2000br,Garfinkle:2002vn,Roberts:1989sk,Brady:1994aq,Brady:1994xfa,Oshiro:1994hd,Jalmuzna:2015hoa,Gundlach:2025yje}.

\paragraph{Quantum scale breaking near criticality.} As the curvature grows, this classical picture cannot remain valid through arbitrarily small scales. This motivates the question of whether the quantum self-energy of the collapsing matter changes the threshold dynamics before the evolution reaches the Planck regime where full quantum gravity becomes necessary. Quantum effects in critical spacetimes have been studied using several different approximations, reaching divergent conclusions \cite{Strominger:1993tt,Zhou:1995zj,Bose:1996pi,Peleg:1996ce,Ayal:1997ab,Brady:1998fh,Chiba,Husain:2008tc,Ziprick:2009nd,Benitez:2020szx,Benitez:2021zjs,Berczi:2020nqy,Guenther:2020kro,Berczi:2021hdh,Hoelbling:2021axl,Varnhorst:2023dew,Moitra:2022umq,Zahn:2025tnh,Xin:2026eoy,Lai:2026dvc}. A more detailed discussion of these approaches can be found in \cite{Tomasevic:2025clf}. Their differences reflect both the difficulty of quantum field theory on a time-dependent background and the physical conditions used to define the quantum state.

The choice of state is particularly important in critical collapse. Self-similarity is not an isometry and therefore does not provide a preferred quantum state. Assuming that the state inherits the classical scaling symmetry~\cite{Brady:1998fh,Zahn:2025tnh} already restricts the quantum behavior we wish to determine. The state and its renormalized stress tensor should instead be determined by physical conditions, in particular, regularity throughout the self-similar domain and boundary conditions intrinsic to the collapse problem. The quantum source determined by these conditions cannot be inferred from the background scaling alone. 

More recently, we developed an explicit one-loop treatment of the Einstein-scalar system based on these physical conditions for CSS critical spacetimes~\cite{Tomasevic:2025clf,Tomasevic:2025kqy,Wu:2026xbr}. For the 4d Choptuik solution, the classical near-threshold instability lies in the $s$-wave sector, while nonspherical perturbations decay \cite{Martin-Garcia:1998zqj}. This motivates studying the
corresponding $s$-wave quantum sector. The resulting quantum state is asymptotically Minkowskian and indeed describes the self-energy of the collapsing scalar field. In this sense, it is a Boulware-like state, without persistent incoming or outgoing Hawking fluxes \cite{Boulware:1974dm}. The relevant source is vacuum polarization produced during collapse, distinct from Hawking radiation emitted after a black hole has formed \cite{Hawking:1975vcx}.

These works cover the interior self-similar region that controls how the critical endpoint forms and the exterior naked singularity region that controls its global visibility~\cite{Tomasevic:2025clf,Tomasevic:2025kqy,Wu:2026xbr}. In the interior, the quantum stress tensor contains a universal growing contribution of the form
\be
    \langle T_{\mu\nu}^{(D)}\rangle(T,x^i)
    \sim
    \hbar \, e^{\omega_q T}\,F_{\mu\nu}(x^i),
    \qquad
    \omega_q=D-2 ,
    \label{eq:intro_previous_quantum_mode}
\ee
where $T$ is the logarithmic scale time and $x^i$ denotes the dimensionless spatial similarity coordinates. The \emph{quantum Lyapunov exponent} $\omega_q$ has a kinematical origin in the contraction of the transverse spheres \cite{Tomasevic:2025clf}. The resulting quantum growing mode is therefore a relevant deformation that competes with the classical instability, shifts the threshold, and produces a quasi-local mass gap with a Type I-like phase transition structure. The corresponding phase space picture is summarized in Figure~\ref{fig:phase-space-quantum-large-D}.

\begin{figure}[t]
\centering
\resizebox{\linewidth}{!}{%
% Schematic phase-space projections. No surface distance is a physical radius.
\begin{tikzpicture}[
  x=1cm,y=1cm,line cap=round,line join=round,
  font=\small,text=ink,
  every node/.style={inner sep=2pt},
  evolution/.style={draw=ink,line width=0.9pt,-{Latex[length=2mm,width=1.3mm]}},
  data/.style={draw=ink!70,line width=0.75pt,dash pattern=on 1pt off 2pt},
  qarrow/.style={draw=quantum,line width=1.05pt,-{Latex[length=2.1mm,width=1.4mm]}}
]
\def\sheet{(0.18,1.42)
  .. controls (0.71,1.76) and (0.68,3.24) .. (1.02,3.18)
  .. controls (2.32,2.75) and (4.70,2.76) .. (7.03,2.38)
  .. controls (6.69,1.66) and (6.60,0.10) .. (6.18,0.29)
  .. controls (4.92,0.73) and (2.08,1.61) .. (0.18,1.42) -- cycle}

% Panel headings.
\node[anchor=west,font=\normalsize] at (0.35,8.67)
  {(a)\quad Classical critical collapse (Type II)};
\node[anchor=west,font=\normalsize] at (8.55,8.67)
  {(b)\quad Quantum critical collapse (Type I-like)};

%----------------------------------------------------------
% Classical critical surface and departure.
%----------------------------------------------------------
\begin{scope}[shift={(0.35,3.20)}]
  \shade[top color=black!3,bottom color=black!12,shading angle=12]
    \sheet;
  \draw[ink!60,line width=0.65pt] \sheet;
  \node at (1.13,2.60) {$\Sigma_c$};

  % One-parameter initial data. This dotted line is not a flow.
  \draw[data] (0.60,-0.18)
    .. controls (1.02,0.53) and (1.87,1.11) .. (2.55,1.62)
    .. controls (3.04,2.13) and (3.64,3.12) .. (3.94,3.86);
  \node[anchor=east] at (0.60,0.16) {$p$};
  \fill[ink] (2.55,1.62) circle[radius=1.5pt];
  \node[anchor=south east] at (2.40,1.75) {$p_*$};

  % Tuned orbit approaches the critical solution within Sigma_c.
  \draw[evolution] (2.57,1.62)
    .. controls (3.20,1.82) and (3.93,1.84) .. (4.58,1.80);

  % Unstable departures have opposite tangent directions at Z_*.
  \draw[evolution] (4.73,1.85)
    .. controls (4.73,2.97) and (5.23,3.86) .. (6.17,4.14);
  \draw[evolution] (4.73,1.75)
    .. controls (4.73,0.70) and (4.86,-0.07) .. (5.43,-0.48);
  \fill[ink] (4.73,1.80) circle[radius=2pt];
  \node[anchor=west] at (4.90,1.83) {$Z_*$};
  \node[anchor=west,align=left,font=\footnotesize]
    at (5.68,3.35) {classical\\growing mode};
  \node[anchor=south] at (5.57,4.31) {black hole formation};
  \node[anchor=north] at (5.43,-0.59) {dispersion};
\end{scope}
\node at (4.03,1.70) {$M_{\mathrm{BH}}\propto(p-p_*)^\gamma\longrightarrow0$};

%----------------------------------------------------------
% Quantum deformation. The upper sheet is a classical reference,
% the lower sheet is a schematic local trapping threshold.
%----------------------------------------------------------
\begin{scope}[shift={(8.55,4.97)}]
  \draw[ink!45,dash pattern=on 3pt off 2pt,line width=0.65pt]
    \sheet;
  \node[text=ink!65] at (1.13,2.60) {$\Sigma_c$};
  \draw[draw=ink!55,line width=0.70pt,dashed,
        -{Latex[length=1.8mm,width=1.2mm]}]
    (3.47,1.76) .. controls (3.86,1.91) and (4.22,1.88) .. (4.62,1.80);
  \draw[draw=ink!55,fill=white,line width=0.7pt]
    (4.77,1.80) circle[radius=1.8pt];
  \node[anchor=west,text=ink!65] at (4.94,1.81) {$Z_*$};
\end{scope}
\begin{scope}[shift={(8.55,2.10)}]
  \shade[top color=quantum!3,bottom color=quantum!12,shading angle=12]
    \sheet;
  \draw[ink!60,line width=0.65pt] \sheet;
  \node at (1.13,2.60) {$\Sigma_q$};

  % The data curve crosses the deformed threshold at a new value.
  \draw[data] (0.60,-0.18)
    .. controls (1.23,0.64) and (1.81,1.12) .. (2.55,1.62)
    .. controls (3.02,2.29) and (3.03,3.64) .. (3.47,4.63)
    .. controls (3.89,5.40) and (4.40,5.90) .. (4.89,6.12);
  \node[anchor=east] at (0.60,0.16) {$p$};
  \fill[quantum] (2.55,1.62) circle[radius=1.6pt];
  \node[anchor=south east,text=quantum] at (2.40,1.75) {$p_*^{\,q}$};
  \fill[ink!65] (3.47,4.63) circle[radius=1.5pt];
  \node[anchor=south east,text=ink!65] at (3.40,4.72) {$p_*$};

  % Open ring marks a local MOTS, not a new fixed-point attractor.
  \draw[qarrow] (2.57,1.62)
    .. controls (3.30,1.82) and (3.95,1.83) .. (4.62,1.72);
  \draw[draw=quantum,fill=white,line width=1.0pt]
    (4.80,1.68) circle[radius=2.5pt];
  \node[anchor=west] at (5.03,1.72) {$r_h>0$};
\end{scope}

% Source-induced deformation arrows between the two projections.
\draw[qarrow] (10.02,6.60) -- (10.02,4.70);
\draw[qarrow] (10.81,6.52) -- (10.81,4.67);
\draw[qarrow] (14.67,6.19) -- (14.67,4.03);
\draw[qarrow] (15.30,6.11) -- (15.30,3.89);
\node[text=quantum,inner sep=3pt]
  at (13.12,5.16)
  {$\langle T^{(D)}_{\mu\nu}\rangle\propto e^{(D-2)T}$};

%----------------------------------------------------------
% A compact dimensional comparison, not another diagram.
% Radius scaling is the CSS result, not a phase-space distance.
%----------------------------------------------------------
\draw[ink!20,line width=0.45pt] (0.35,1.02) -- (15.86,1.02);
\node[anchor=west,font=\normalsize] at (0.35,0.49) {$D\to\infty$};
\node[align=center] at (4.82,0.30)
  {$\omega_q=D-2\ \uparrow$\\[4pt]\footnotesize steeper quantum growth};
\node[align=center] at (9.63,0.30)
  {$r_h/\ell_q\propto(D-2)^{-2}\ \downarrow$\\[4pt]\footnotesize smaller trapping scale};
\node[align=center] at (14.30,0.30)
  {$\ell_{P,D}\ll r_h$\\[4pt]\footnotesize requires larger $N$};
\end{tikzpicture}

}
\caption{
Phase space schematic of classical and quantum critical collapse. In (a), a one-parameter family of initial data crosses the
classical critical surface $\Sigma_c$ at $p_*$. The tuned evolution
approaches $Z_*$, while the classical growing mode leads toward dispersion
or black hole formation. Type II collapse allows the black hole mass to
vanish at threshold. Type I instead has a stationary or time-periodic
critical solution and a finite mass gap. In (b), vacuum polarization
produces $\langle T^{(D)}_{\mu\nu}\rangle\propto e^{(D-2)T}$.
The dashed sheet retains the classical critical surface, while $\Sigma_q$ represents the shifted threshold. The colored ring denotes the finite-radius MOTS onset
found in the CSS analysis. This is the Type I-like
feature studied here and extends the mechanism of
\cite{Tomasevic:2025clf,Tomasevic:2025kqy,Wu:2026xbr}.
At large $D$, the quantum growth becomes steeper while
$r_h/\ell_q\propto(D-2)^{-2}$ decreases. Maintaining
$\ell_{P,D}\ll r_h$ then requires increasingly large $N$ and does not
survive the strict large-$D$ limit at fixed $N$.
}
\label{fig:phase-space-quantum-large-D}

\end{figure}

The existence of a quantum scale makes this competition unavoidable under sufficiently precise tuning. Let $\ell_q$ denote the length scale set by the matter loop and its gravitational coupling. The local classical scale decreases as $L(T)\sim\ell e^{-T}$, so a sufficiently prolonged self-similar contraction eventually reaches $\ell_q$. This gives a direct reason to expect quantum effects to break classical self-similarity before its formal zero-radius endpoint. However, encountering $\ell_q$ does not imply a trapped surface must form. Whether and where trapped surfaces form should be determined from the geometry produced by the classical growing mode and quantum backreaction~\cite{Tomasevic:2025clf, Tomasevic:2025kqy, Wu:2026xbr}.

\paragraph{The large-$D$ limit.} The previous results raise several immediate questions. How universal is the quantum growth $\omega_q=D-2$ at general $D$? Does it survive the echoing structure of DSS? Establishing these in physical states would extend the mechanism beyond the finite-$D$ CSS examples and is essential for connecting with the genuine Choptuik spacetime. Moreover, since the exponent increases linearly with $D$, the factor $e^{(D-2)T}$ becomes arbitrarily steep as we increase $D$. Does it signal a breakdown of semiclassical control as $D\to\infty$? Rapid growth alone does not settle this question. Its physical significance depends on whether the geometric scale at which quantum effects intervene can remain well above the Planck scale.

The large-$D$ limit provides an analytic setting in which these questions can be addressed. In practice, one treats $1/D$ as a small parameter and organizes the field equations in an expansion around $D=\infty$. The expansion does not make gravity weak. Instead, gravitational dynamics becomes strongly localized in the radial direction and separates into regions with parametrically different scales, allowing the equations in each region to simplify and be matched systematically \cite{Emparan:2013moa,Emparan:2013xia,Emparan:2014aba,Emparan:2015hwa,Emparan:2020inr, Emparan:2025yfy}. Large-$D$ scalar collapse and its critical behavior were studied in \cite{Rozali:2018yrv}. CSS Einstein-scalar backgrounds and their perturbations have been constructed at finite $D$ and studied in a large-$D$ expansion \cite{Clark:2025tqi}. Analytic DSS backgrounds have also been found at large $D$ \cite{Ecker:2026akf},
while finite-$D$ DSS critical solutions and their unstable modes have been
constructed numerically \cite{Ecker:2026joz}. These developments allow CSS and DSS to be studied within a common semiclassical framework.

We focus throughout on the self-similar interior. We will see that self-similarity amplifies the quantum effects in the same kinematical manner in both backgrounds. In CSS, radial localization isolates the region where the classical and quantum contributions compete, allowing the trapping scale to be determined from their limiting radial profiles. The resulting semiclassical spacetime depends on how vacuum polarization is distributed across the localized radial layers. In DSS, the simplified radial geometry allows us to retain the full evolution within and between echoes. This lets us determine how the quantum source retains memory of previous echoes.

\paragraph{Overview of main results.} For CSS, we show that physical boundary conditions select a Boulware-like state whose stress tensor grows with the exact exponent $D-2$ at finite $D$. This establishes the universal quantum growth without imposing self-similarity on the state.

To determine when the quantum mode becomes dynamically important, we compare it with the classical departure from criticality. For the classical perturbations studied in \cite{Clark:2025tqi}, the continuous spectrum extends up to the limiting growth rate $\omega_c=(D-2)/2$. We use this fastest classical sector to study its competition with the quantum growing mode. Fine-tuning $p$ toward $p_*$ suppresses the initial classical amplitude proportional to $(p-p_*)$. The quantum contribution grows twice as rapidly and, at sufficiently precise tuning, catches up before the classical departure ends the self-similar evolution. 

We solve the semiclassical Einstein equation and trace the null expansions of the resulting geometry. The classical approach to zero radius is interrupted by the local appearance of a future marginally outer trapped surface (MOTS) at finite areal radius. At large $D$, the competition reduces to a Bessel problem that determines the local onset of trapping. Denoting the onset radius by $r_h$, we find
\be
    r_h
    \propto
    \frac{\ell_q}{(D-2)^2}.
    \label{eq:intro_css_wall_scale}
\ee
The inverse-square scaling separates two physical scales. Vacuum polarization first becomes appreciable near $\ell_q$, while the outgoing expansion reaches zero at the much smaller radius $r_h$. The narrow radial region in which this occurs defines a \emph{quantum wall}. Thus the steep quantum growth has a concrete geometric consequence, concentrating the departure from the classical zero-radius behavior into a sharply localized trapping region. The finite-$D$ construction begins at $D=12$, which is the first integer on the branch connected to the large-$D$ solution for which the marginal surface has the required future orientation. We follow this branch through every integer dimension up to $D=129$, where it has entered the large-$D$ scaling regime and provides a bridge to the independent asymptotic analysis. The finite-$D$ calculation and the independent large-$D$
analysis support the same local picture.

This localization makes the question of semiclassical control concrete. At each fixed finite $D$, a sufficiently large number $N$ of matter fields can keep $r_h$ above the Planck scale and suppress graviton loops relative to the matter loop. As $D$ increases, however, \eqref{eq:intro_css_wall_scale} moves the onset radius deeper below $\ell_q$, and maintaining its separation from the Planck scale requires increasingly demanding large $N$. Rapid quantum growth can therefore coexist with trapping above the Planck scale, although this separation cannot be maintained in the strict large-$D$ limit at fixed $N$.

DSS carries the same quantum growth through a sequence of echoes, but the quantum state can retain information about the earlier evolution. The state choice is part of a nonlocal boundary-value problem. The inverse operators in the one-loop action are defined through retarded Green's functions, which integrate the source over its past history. The resulting quantum fields can therefore retain memory of previous echoes even when the local geometry returns to the same echo phase. This allows the quantum stress tensor to distinguish successive echoes beyond the common factor $e^{(D-2)T}$. The resulting renormalized stress tensor can in general be written as
\be
    \langle T_{\mu\nu}^{(D)}\rangle
    =
    e^{(D-2)T}
    \sum_{j=0}^{J}(T-T_i)^j P_{\mu\nu}^{(j)}(T,x),
    \qquad
    P_{\mu\nu}^{(j)}(T+\Delta_s,x)
    =
    P_{\mu\nu}^{(j)}(T,x),
    \label{eq:intro_dss_general_source}
\ee
where $\Delta_s$ is the period respected by the source and $T_i$ marks the initial echo. The periodic profiles describe the evolution within each echo, while the powers of $T-T_i$ record accumulated memory. Every term in~\eqref{eq:intro_dss_general_source} still has the same quantum growth $D-2$.

The simplest effective action induced by the trace anomaly realizes $J=1$, with memory growing linearly with the number of elapsed echoes. However, its stress tensor is singular at the spherical center, where the classical DSS geometry is smooth at every finite $T$. This is unphysical and should be ruled out as an admissible quantum source. It cannot represent the vacuum polarization of a collapse geometry that is regular at the center.\footnote{While this work was being completed,~\cite{Lai:2026dvc} studied the same large-$D$ DSS background using the intrinsic 2d Polyakov action, with explicit calculations based on a particular
periodic representative. This is not the $s$-wave sector obtained by spherical reduction of the Einstein-scalar system, which is dilaton-coupled. The corresponding $D$-dimensional quantum stress tensor is also singular at the regular center. We will demonstrate how to obtain a smooth one from the full dilaton-coupled matter theory in Section~\ref{sec:dss}.} We construct a broader family of covariant one-loop effective actions whose stress tensors remain smooth throughout the DSS interior. In this family, the accumulated term is absent and $J=0$. Modulo the universal factor $e^{(D-2)T}$, the stress tensor is therefore bounded and echo-periodic.

The next geometric question is whether these smooth sources drive the outgoing null expansion to zero and produce a DSS counterpart of the CSS quantum wall. Our preliminary analysis shows that the direct quantum contribution can favor or oppose trapping. Answering the full question also requires the coupled metric and scalar backreaction, together with the classical growing perturbation, which has not yet been determined for the large-$D$ background used here. The present analysis remains in the untrapped regime, so whether a DSS quantum wall forms remains open.

Physical boundary conditions and smoothness across the self-similar domain emerge as guiding principles for constructing the quantum source. To our knowledge, this is the first explicit construction of higher-dimensional one-loop stress tensors that remain smooth throughout a DSS critical collapse domain. The resulting stress tensor has echo-periodic
profiles, showing that DSS does not require secular
memory between echoes. This leads us to conjecture that the smooth quantum stress tensor of the genuine 4d DSS Choptuik spacetime is similarly governed by its exact echoing symmetry and carries the same universal growing mode. These principles may guide the construction of a smooth quantum state directly on that spacetime.

The paper is organized as follows. In Section~\ref{sec:framework}, we review self-similar critical collapse and the one-loop effective theory for the Einstein-scalar system. Section~\ref{sec:css} studies the semiclassical CSS geometry and derives the finite-$D$ and large-$D$ results. We turn to the DSS problem in Section~\ref{sec:dss}, where we analyze the imprint of echoing on the quantum source and construct the smooth stress tensor. We then discuss their connection to the backreaction and horizon tracing problem. We conclude in Section~\ref{sec:discussion} with a discussion of the physical implications and future directions. Technical and numerical details are collected in the appendices.

\section{Self-similar critical collapse and one-loop vacuum polarization}
\label{sec:framework}

This section develops the framework common to the CSS and DSS calculations. We first describe how self-similarity organizes classical perturbations and Type II scaling. The one-loop theory is then formulated, including the ingredients needed to derive the renormalized stress tensor. This makes the kinematical origin of the quantum Lyapunov exponent manifest, while the explicit backgrounds and their semiclassical dynamics are treated separately in Sections~\ref{sec:css} and \ref{sec:dss}.

\subsection{Critical scaling and self-similarity}
\label{subsec:self-similarity}

Critical collapse is naturally described as a dynamical system on the phase space of initial data. The boundary between the basins of dispersion and black hole formation is a codimension-one critical surface. A generic one-parameter family labeled by $p$ crosses this surface at $p=p_*$. The critical solution acts as an attractor within the surface. Near the threshold, the evolution first approaches the universal critical spacetime and later departs from it along the growing classical perturbation. Physical perturbations tangent to the critical surface decay. The single growing perturbation is transverse to it, and its sign determines dispersion or black hole formation. The intermediate critical geometry and its growing exponent remain universal within a given matter model and spacetime dimension \cite{Choptuik:1992jv,Gundlach:1995kd,Gundlach:1996eg,Gundlach:2025yje}. Further discussion of this phase space picture, including the separation between the interior self-similar problem and the exterior problem controlling global visibility, can be found in~\cite{Tomasevic:2025clf,Wu:2026xbr}. Here we establish the common language for CSS and DSS, their perturbations, and the scaling quantities appropriate to general $D$.

\paragraph{CSS spacetimes.} The approach to the critical spacetime is most conveniently expressed in terms of logarithmic scale time $T$. We introduce an arbitrary reference length $\ell$ and choose the orientation of $T$ so that increasing $T$ corresponds to observing the same dimensionless geometry at a smaller physical scale, 
\be
    L(T)\sim \ell e^{-T}.
    \label{eq:scale_time}
\ee
The self-similar endpoint is approached as $T\to\infty$. Let $x^i$ denote dimensionless similarity coordinates held fixed in this limit. A CSS spacetime admits a homothetic Killing vector $\xi$ satisfying
\be
    \mathcal{L}_{\xi}g_{\mu\nu}=2g_{\mu\nu}.
    \label{eq:homothetic_killing_vector}
\ee
Choosing coordinates adapted to the homothety so that $\xi=-\partial_T$, the metric can be written as
\be
    g_{\mu\nu}(T,x^i)
    =
    \ell^2e^{-2T}\widetilde g_{\mu\nu}(x^i).
    \label{eq:general_css_metric}
\ee
Without the overall scale, $\widetilde g_{\mu\nu}(x^i)$ is independent of $T$.

Let $Z$ denote any corresponding scale-invariant variable built from the metric and matter fields. A linear perturbation can then be decomposed into modes of the form
\be
    \delta Z(T,x^i)=e^{kT}Z_k(x^i).
    \label{eq:general_css_perturbation}
\ee
The exponent $k$ can be complex. Its real part determines whether the perturbation grows or decays toward the self-similar endpoint, while its imaginary part describes oscillations in logarithmic scale time. Here we reserve \emph{Lyapunov exponent} for the real growth rate and denote the classical and quantum rates by $\omega_c$ and $\omega_q$. For the standard codimension-one critical solution, the single growing direction has a real $k$. The large-$D$ CSS spacetime studied in Section~\ref{sec:css} instead has a continuous perturbation spectrum that also allows complex $k$.

\paragraph{DSS spacetimes.} On the other hand, a DSS spacetime retains a periodic dependence on logarithmic scale time,
\be
    g_{\mu\nu}(T,x^i)
    =
    \ell^2e^{-2T}\widetilde g_{\mu\nu}(T,x^i),
    \qquad
    \widetilde g_{\mu\nu}(T+\Delta_g,x^i)
    =
    \widetilde g_{\mu\nu}(T,x^i).
    \label{eq:general_dss_metric}
\ee
Here $\Delta_g$ is the geometric echoing period. We denote the full period of the collapsing matter scalar $f$ by $\Delta$. The Einstein-scalar equations are invariant under the $\mathbb{Z}_2$ reflection $f\to-f$ and under constant shifts of $f$. After fixing the additive shift, the Choptuik solution realizes this symmetry as a half-echo reflection, and the large-$D$ DSS family we will consider admits the same structure \cite{Choptuik:1992jv,Martin-Garcia:2003xgm,Ecker:2026akf,Ecker:2026joz}. For classical background fields, we denote the critical solution by a subscript $*$. The half-echo relation is
\be
    f_*(T+\Delta/2,x^i)=-f_*(T,x^i),
    \qquad
    \Delta_g=\frac{\Delta}{2}.
    \label{eq:scalar_half_echo}
\ee
Local quantities built from the classical background and even under scalar reflection inherit the shorter period $\Delta_g$. Without the half-echo symmetry, the geometry generally has only the full period $\Delta$. The echo structure of the quantum source also depends on the state and the one-loop effective action.

Linearization about a DSS background produces an operator that is periodic in logarithmic scale time. Linear perturbations can therefore be written in Floquet form \cite{Gundlach:1996eg,Gundlach:2025yje},
\be
    \delta Z(T,x^i)
    =
    e^{kT}P_k(T,x^i),
    \qquad
    P_k(T+\Delta_s,x^i)=P_k(T,x^i).
    \label{eq:general_dss_perturbation}
\ee
The period is $\Delta_s=\Delta_g$ in a reflection-even sector and $\Delta_s=\Delta$ when only the full-echo symmetry is preserved. The Floquet multiplier from one echo to the next is $e^{k\Delta_s}$. The exponent $k$ is therefore defined up to shifts by $2\pi i m/\Delta_s$, with $m\in\mathbb{Z}$, while its real part gives the common growth rate. The periodic profile resolves the dynamics within an echo.

Nonlocal inverse operators in the one-loop action can also carry information from one echo to the next, as we will show in Section~\ref{sec:dss}. Let the evolution through one echo define a Floquet map. If the relevant multiplier belongs to a finite Jordan chain, repeated echoes generate polynomial factors in the number of elapsed echoes. The corresponding quantum source or backreaction can take the generalized Floquet form
\be
    \mathcal{Q}(T,x^i)
    =
    e^{kT}
    \sum_{j=0}^{J}(T-T_i)^j P_j(T,x^i),
    \qquad
    P_j(T+\Delta_s,x^i)=P_j(T,x^i),
    \label{eq:generalized_echo_growth}
\ee
with quantum Lyapunov exponent $\omega_q=\operatorname{Re}k$. Here $T_i$ specifies the initial echo, while $J$ is the highest nonvanishing power of $T-T_i$. A finite Jordan chain of length $J+1$ generates powers up to $(T-T_i)^J$, and we refer to $J$ as the Jordan degree. The case $J=0$ is an ordinary Floquet mode with bounded periodic modulation. A nonzero $J$ records accumulated memory from earlier echoes. In particular, $J=1$ gives memory that grows linearly with the number of elapsed echoes. These polynomial factors modify the echo modulation without changing the real exponential growth rate $\omega_q$. In Section~\ref{sec:dss}, the minimal anomaly-induced action will realize $J=1$, while the smooth quantum sources realize $J=0$.

\paragraph{Critical scaling.} The classical growing perturbation relates this self-similar evolution to Type II scaling. Suppose the critical solution has one isolated growing mode with Lyapunov exponent $\omega_c>0$. For a generic one-parameter family, its amplitude is proportional to $p-p_*$ near the threshold. We absorb the proportionality constant into the normalization of the mode. The solution leaves the critical regime at a logarithmic time $T_*$ determined by
\be
    |p-p_*|e^{\omega_cT_*}\sim1,
    \qquad
    L_*\sim\ell e^{-T_*}
    \sim\ell|p-p_*|^{1/\omega_c}.
    \label{eq:critical_exit_scale}
\ee
It is useful to define the length exponent
\be
    \gamma_L=\frac{1}{\omega_c}.
    \label{eq:length_exponent}
\ee
Keeping the $D$-dimensional Newton constant fixed, the black hole mass scales as
\be
    G_D M_{\rm BH}\sim L_*^{D-3},
    \qquad
M_{\rm BH}\propto(p-p_*)^\gamma,
    \qquad
    \gamma=(D-3)\gamma_L
    =\frac{D-3}{\omega_c},\qquad p>p_*.
    \label{eq:mass_length_exponents}
\ee
This gives the mass scaling of Type II critical collapse. Type I collapse instead has a finite mass gap at the threshold. The mass exponent $\gamma$ equals the length exponent only in 4d. In DSS collapse, the echo phase produces a periodic modulation of
$\log L_*$ and $\log M_{\rm BH}$ as functions of $\log|p-p_*|$,
without changing the leading exponents $\gamma_L$ or $\gamma$.

The classical growth comes from a \emph{homogeneous} perturbation. When a critical solution has one isolated unstable mode, $\omega_c$ denotes its real growth rate and determines the exit scale in~\eqref{eq:critical_exit_scale}. The CSS background studied in Section~\ref{sec:css} is different. Its CSS-breaking perturbations form the
continuous spectrum found in~\cite{Clark:2025tqi}, and we will use
$\omega_c\equiv\sup\operatorname{Re}k$ to denote its upper spectral edge.

The one-loop stress tensor provides a different kind of growth. It acts as a source in the semiclassical Einstein equation and generates an \emph{inhomogeneous} deformation with Lyapunov exponent $\omega_q$. For the $s$-wave matter sector studied here, $\omega_q=D-2$, whose origin will be discussed in the next subsection. Once the one-loop theory and its boundary data are specified, the corresponding particular deformation is fixed by the source and is distinct from the freely adjustable homogeneous classical perturbations. The comparison between $\omega_c$ and $\omega_q$ therefore measures the competition between the fastest classical envelope and the semiclassical backreaction. Large $D$ makes this comparison tractable by separating radial scales and localizing the self-similar dynamics \cite{Emparan:2013moa}.

\subsection{The Einstein-scalar system and the one-loop effective theory}
\label{subsec:one-loop}

We consider Einstein gravity in $D$ spacetime dimensions minimally coupled to $N$ free massless scalar fields, which we refer to as the Einstein-scalar system. In a normalization convenient for the gravitational equations, the classical action is
\be
    S_D
    =
    \frac{1}{16\pi G_D}
    \int d^Dx\sqrt{-g^{(D)}}
    \left[
        R^{(D)}
        -
        \sum_{I=1}^{N}(\nabla f_I)^2
    \right].
    \label{eq:einstein_scalar_action}
\ee
We choose the common classical profile as $f_I=f/\sqrt{N}$, so that
$\sum_I(\nabla f_I)^2=(\nabla f)^2$ and the classical geometry and the
equations for $f$ are independent of $N$. By contrast, in the matter-loop approximation used here, integrating out the $N$ identical scalar fluctuations with the same state and subtraction scheme produces $N$ identical matter determinants, so the matter one-loop effective action is proportional to $N$. 

The background equations reduce to
\be
    R^{(D)}_{\mu\nu}
    =
    \nabla_\mu f\nabla_\nu f,
    \qquad
    \Box_D f=0.
    \label{eq:einstein_scalar_equations}
\ee
Thus $f$ denotes the classical scalar supporting the self-similar geometry, whereas $N$ controls the strength of the spherical matter loop. To isolate the $s$-wave sector, we impose spherical symmetry and decompose the metric as~\cite{Kummer:1998dc, Kummer:1999zy, Grumiller:2002nm}
\be
    ds_D^2
    =
    g_{ab}(x)dx^a dx^b
    +
    r^2(x)d\Omega_n^2,
    \qquad
    n\equiv D-2.
    \label{eq:spherical_metric_decomposition}
\ee
Greek indices refer to the full spacetime, $a,b=0,1$ label the 2d Lorentzian sector, and $i,j$ label the transverse sphere. Here $r$ is the physical areal radius and $\Omega_n$ denotes the area of the unit $n$-sphere. We encode the transverse volume in the dilaton $\phi$ through
\be
    r=\ell e^{-2\phi/n},
    \qquad
    e^{-2\phi}=\frac{r^n}{\ell^n}.
    \label{eq:dilaton_definition}
\ee
Integrating over the sphere gives the 2d action
\be
    S_{\rm 2d}
    ={}
    \frac{\Omega_n\ell^n}{16\pi G_D}
    \int d^2x\sqrt{-g}\,e^{-2\phi}
    \Bigg[
        R^{(2)}
        +
        \frac{4(n-1)}{n}(\nabla\phi)^2
        +
        \frac{n(n-1)}{\ell^2}e^{4\phi/n}
        -
        (\nabla f)^2
    \Bigg].
    \label{eq:spherical_reduced_action}
\ee
Boundary terms from the spherical reduction are omitted,
since they do not affect the local equations below. The prescription adopted here agrees with the discussion in~\cite{Tomasevic:2025clf, Tomasevic:2025kqy,Wu:2026xbr}.

\paragraph{The one-loop effective theory.} Due to the overall $e^{-2 \phi}$, the spherical matter field is therefore a dilaton-coupled 2d scalar, distinct from an ordinary conformal scalar on $g_{ab}$. Its classical kinetic term remains Weyl invariant in 2d, while the functional measure retains the dependence on $\phi$ inherited from the $D$-dimensional geometry. We quantize the $N$ matter fields in this $s$-wave sector on the classical self-similar background. The resulting one-loop effective action has the general decomposition
\be
    \Gamma_{\rm 1-loop}
    =
    \Gamma_{\rm anom}
    +
    \Gamma_W
    +
    \Gamma_{\rm ct}.
    \label{eq:one_loop_decomposition}
\ee
The anomaly-induced term $\Gamma_{\rm anom}$ is a particular action whose Weyl variation reproduces the local trace anomaly. In the conformal frame and functional measure adopted here, the heat-kernel calculation gives~\cite{Mukhanov:1994ax, Bousso:1997cg, Grumiller:2002nm, Kummer:1998dc, Kummer:1999zy, Balbinot:2000iy, Fabbri:2003vy, Hofmann:2004kk}
\be
    \langle T^{(2)a}{}_{a}\rangle
    =
    \frac{N\hbar}{24\pi}
    \left[
        R^{(2)}
        -
        6(\nabla\phi)^2
        +
        6\Box\phi
    \right].
    \label{eq:original_dilaton_trace_anomaly}
\ee
The anomaly-induced part is therefore
\be
    \Gamma_{\rm anom}
    =
    -\frac{N\hbar}{96\pi}
    \int d^2x\sqrt{-g}
    \left[
        R^{(2)}\frac{1}{\Box}R^{(2)}
        -
        12(\nabla\phi)^2\frac{1}{\Box}R^{(2)}
        +
        12\phi R^{(2)}
    \right],
    \label{eq:anomaly_induced_action}
\ee
containing the nonlocal operator $\Box^{-1}$. Although the final term in~\eqref{eq:anomaly_induced_action} is local, its coefficient is part of this anomaly-induced sector and should not be confused with independently chosen finite counterterms~\cite{Kummer:1999zy,Shi:2026wnk}. The functional $\Gamma_W$ is Weyl invariant and contains information not fixed by the anomaly. It can contribute to the state-dependent, normal-ordered part of the stress tensor, whose definition depends on the choice of reference vacuum \cite{Wu:2023uyb,Tomasevic:2025clf, Shi:2026wnk,Wu:2026xbr}. The term $\Gamma_{\rm ct}$ contains finite local covariant terms and fixes the subtraction scheme. The finite completion and the associated renormalized couplings must therefore be specified before the one-loop source is determined.\footnote{A renormalization scale $\mu$ can appear in logarithmic terms
of the one-loop effective action. Its arbitrary dependence is canceled
by the running of the corresponding renormalized couplings, so the
semiclassical equations are independent of $\mu$ to the order considered.}

The inverse operator in~\eqref{eq:anomaly_induced_action} can be represented by auxiliary fields, with useful prescriptions discussed in~\cite{Wu:2023uyb,Tomasevic:2025clf,Shi:2026wnk}. For convenience in the large-$D$ calculations, we introduce potentials satisfying
\be
    \Box U=R^{(2)},
    \qquad
    \Box V=(\nabla\phi)^2.
    \label{eq:auxiliary_potentials}
\ee
The choice of Green function and the homogeneous parts of $U$ and $V$ specify the quantum-state data. Together with the subtraction convention, they determine the reference vacuum used in the normal-ordering prescription.

We use the same Minkowski normalization in the CSS and DSS calculations. The complete prescription, including its 2d effective action and Minkowski-vacuum auxiliary data, must give a vanishing renormalized stress tensor on the spherical reduction of Minkowski space. In the convention of~\eqref{eq:dilaton_definition}, we include the finite subtraction 
\be
    \Gamma_{\rm ct}^{(\phi R)}
    =
    \frac{N\hbar(1-n/2)}{8\pi}
    \int d^2x\sqrt{-g}\,\phi R^{(2)}.
    \label{eq:minkowski_counterterm}
\ee
Its coefficient vanishes in 4d, where $n=2$. This finite normalization is related to the matching issue associated with the dimensional reduction anomaly, since dimensional reduction and renormalization need not commute \cite{Balbinot:2000iy,Frolov:1999an,Cognola:2000xp}. The subtraction is a consistent covariant choice within the 2d theory.

The minimal one-loop theory is therefore
\be
    \Gamma_{\rm min}
    \equiv
    \Gamma_{\rm anom}
    +
    \Gamma_{\rm ct}^{(\phi R)},
    \qquad
    \Gamma_W=0,
    \label{eq:minimal_anomaly_theory}
\ee
which is the simplest with no further Weyl-invariant or finite local functionals. Here ``minimal" specifies the functional content of the effective theory, as the Green's function and homogeneous auxiliary data will still encode the choice of quantum state. The trace of this theory is
\be
    \langle T^{(2)a}{}_{a}\rangle
    =
    \frac{N\hbar}{24\pi}
    \left[
        R^{(2)}
        -
        6(\nabla\phi)^2
        +
        3n\Box\phi
    \right].
    \label{eq:minkowski_normalized_trace}
\ee
Any additional terms in $\Gamma_W$ or $\Gamma_{\rm ct}$ must preserve the same Minkowski normalization. In particular, the DSS construction in Section~\ref{sec:dss} will allow a broader class of finite functionals in $\Gamma_W+\Gamma_{\rm ct}$.

\paragraph{Stress tensor and conservation laws.} The complete 2d one-point functions and field sources are defined by
\be
    \langle T^{(2)}_{ab}\rangle
    =-\frac{2}{\sqrt{-g}}
    \frac{\delta\Gamma_{\rm 1-loop}}{\delta g^{ab}}, \qquad
    E_\phi
    =\frac{1}{\sqrt{-g}}
    \frac{\delta\Gamma_{\rm 1-loop}}{\delta\phi},
    \qquad
    E_f
    =\frac{1}{\sqrt{-g}}
    \frac{\delta\Gamma_{\rm 1-loop}}{\delta f}.
\label{eq:reduced_one_point_functions}
\ee
Since $\Gamma_{\rm min}[g,\phi]$ has no explicit dependence on the background scalar $f$, the minimal theory has $E_f=0$. More general finite functionals can instead produce a nonzero scalar source. Diffeomorphism invariance gives the 2d conservation law
\be
    \nabla_a\langle T^{(2)a}{}_{b}\rangle
    -
    E_\phi\nabla_b\phi
    -
    E_f\nabla_b f
    =0.
    \label{eq:reduced_ward_identity}
\ee
This conservation law keeps the radial stress, angular pressure, and scalar equation tied to a single variational principle. 

To determine the semiclassical backreaction, we need the corresponding
$D$-dimensional quantum source. Varying the effective action with respect
to the 2d metric $g_{ab}$ at fixed $r$ gives its radial components, while
varying the transverse sphere, or equivalently the dilaton, gives the
angular components,
\be
    \langle T^{(D)}_{ab}\rangle
    =
    \frac{\langle T^{(2)}_{ab}\rangle}
         {\Omega_n r^n},
    \qquad
    \langle T^{(D)}_{ij}\rangle
    =
    -\frac{E_\phi}{2\Omega_n r^n}
    g^{(D)}_{ij}.
    \label{eq:spherical_uplift}
\ee
If the effective action depends explicitly on $f$, the corresponding scalar source in $D$ dimensions is $E_f/(\Omega_n r^n)$. The conservation law in~\eqref{eq:reduced_ward_identity} then becomes
\be
    \nabla^{(D)}_\mu
    \langle T^{(D)\mu}{}_{\nu}\rangle
    -
    \frac{E_f}{\Omega_n r^n}\nabla_\nu f
    =0.
    \label{eq:full_dimensional_conservation}
\ee
For $E_f=0$, the one-loop stress tensor is conserved by itself. When $E_f\neq0$, it exchanges energy and momentum with the classical scalar. With the normalization in~\eqref{eq:einstein_scalar_action}, varying the complete action with respect to $f$ gives
\be
    \Box_D f
    +
    \frac{8\pi G_D E_f}{\Omega_n r^n}
    =0.
    \label{eq:corrected_scalar_equation}
\ee
The quantum correction therefore supplies a source for the classical scalar and changes its profile. By~\eqref{eq:corrected_scalar_equation}, the covariant divergence of the classical scalar stress tensor is
$-E_f\nabla_\nu f/(\Omega_n r^n)$, which exactly cancels that of the one-loop stress tensor in~\eqref{eq:full_dimensional_conservation}. The total matter source entering the semiclassical Einstein equation is therefore conserved.

\paragraph{Quantum growth and semiclassical control.} Now we are ready to discuss the origin of the quantum growing mode. We first need the time dependence of the 2d stress tensor. In our earlier finite-$D$ CSS analyses, physical boundary conditions selected a Boulware-like vacuum polarization state incorporating the self-energy of the collapsing matter \cite{Tomasevic:2025clf,Tomasevic:2025kqy}. Its 2d stress tensor was homothetically invariant,
\be
    \partial_T\langle T_{ab}^{(2)}\rangle=0
    \qquad
    \text{at fixed similarity coordinates}.
    \label{eq:css_reduced_stress_scaling}
\ee
This property followed from regularity requirements, without imposing self-similarity on the quantum state. It is this result that allows the quantum growth to be read from the shrinking transverse area according to~\eqref{eq:spherical_uplift}.

The relation between the 2d and $D$-dimensional stress tensors introduces the factor $1/(\Omega_n r^n)$. Together with the matter-loop normalization $N\hbar$ and the gravitational coupling $8\pi G_D$, this gives the coefficient $8\pi G_DN\hbar/(\Omega_n r^n)$ in the semiclassical Einstein equation. We therefore define
\be
    \ell_q^n
    \equiv
    \frac{8\pi G_DN\hbar}{\Omega_n},
    \qquad
    g_q
    \equiv
    \left(\frac{\ell_q}{\ell}\right)^n.
    \label{eq:quantum_length}
\ee
The factor $(\ell_q/r)^n$ measures the gravitational strength of the matter loop multiplying the 2d source profile, and $g_q$ gives this coupling at the reference scale $\ell$. For a fixed source profile, $g_q\ll1$ makes the quantum correction small at that scale. The length $\ell_q$ is neither a renormalization scale nor, by itself, a quantum gravity cutoff or a trapping radius.

Self-similarity then amplifies the source. In CSS, the areal radius takes the form
\be
    r=\ell e^{-T}\mathcal R(x),
    \qquad
    r^{-n}=\ell^{-n}e^{nT}\mathcal R^{-n}(x).
    \label{eq:inverse_area_scaling}
\ee
Combining this geometric factor with \eqref{eq:css_reduced_stress_scaling} gives a stress tensor proportional to $e^{nT}$ at fixed similarity radius, with quantum Lyapunov exponent $\omega_q=n=D-2$.\footnote{This exponent refers to covariant components in similarity coordinates. Raising one index or using an orthonormal frame introduces an additional factor $e^{2T}$ from the metric.} The exponent has a kinematical origin once the time dependence of the 2d quantum source has been established.

In DSS, the areal radius instead takes the form
\be
    r=\ell e^{-T}\mathcal R(T,x),
    \qquad
    r^{-n}=\ell^{-n}e^{nT}\mathcal R^{-n}(T,x),
    \label{eq:dss_inverse_area_scaling}
\ee
where $\mathcal R(T+\Delta_g,x)=\mathcal R(T,x)$. The 2d stress tensor must again be determined from the state and the one-loop effective action. For sources with periodic time dependence multiplied by a finite polynomial in $T-T_i$, as in \eqref{eq:generalized_echo_growth}, the same inverse-area factor gives $\omega_q=D-2$. Such polynomial factors can encode memory of
previous echoes. They are
subexponential and therefore modify the echo-to-echo amplitude without
changing the Lyapunov exponent.

Now we discuss the validity of the semiclassical approximation. The coupling $g_q$ measures the strength of the matter loop at the reference scale $\ell$. Its separation from the Planck scale follows by defining
\be
    \ell_{P,D}^n=G_D\hbar,
    \qquad
    \frac{\ell_q}{\ell_{P,D}}
    =
    \left(\frac{8\pi N}{\Omega_n}\right)^{1/n}.
    \label{eq:quantum_planck_separation}
\ee
At fixed $D$, a large number of matter fields can separate $\ell_q$ from the Planck length and suppress graviton loops relative to the matter loop. Curvature in Planck units must nevertheless be checked where the backreaction becomes important, since that scale need not coincide with $\ell_q$. In particular, a quantum contribution comparable to the classical one can occur while the curvature remains weak.

\section{Quantum critical collapse in the large-$D$ CSS spacetime}
\label{sec:css}

The exact CSS background allows us to determine the vacuum polarization state from the collapse boundary conditions and compute its semiclassical backreaction at finite $D$. We first obtain the quantum growing mode with exponent $\omega_q=D-2$. The resulting backreaction is then combined with the classical perturbation to locate, through horizon tracing, the appearance of a future MOTS at finite areal radius. The large-$D$ limit isolates the radial region where this competition takes place and reveals the quantum wall.

\subsection{The critical CSS background and the quantum growing mode}
\label{subsec:css-background}

\begin{figure}[t]
\centering
\resizebox{0.45\linewidth}{!}{%
\begin{tikzpicture}[
  x=0.8cm, y=0.8cm,
  line cap=round,
  line join=round,
  font=\small,
  every node/.style={inner sep=1.5pt},
  boundary/.style={draw=black, line width=1.0pt},
  matching/.style={draw=black, line width=0.85pt},
  singularity/.style={
    draw=black,
    line width=1.05pt,
    decorate,
    decoration={zigzag, segment length=4.5pt, amplitude=1.5pt}
  }
]

%------------------------------------------------------------
% Classical background.
% The right diamond is u<0, v>0.
% The lower-left triangle is the Minkowski region u<=v<0.
%------------------------------------------------------------
\coordinate (L) at (0,0);
\coordinate (U) at (4,4);
\coordinate (R) at (8,0);
\coordinate (B) at (4,-4);
\coordinate (P) at (0,-8);

% Region fills.
\fill[gray!10] (L) -- (U) -- (R) -- (B) -- cycle;
\fill[gray!3]  (L) -- (B) -- (P) -- cycle;

% Boundaries and the internal matching surface.
\draw[boundary]    (P) -- (R);  % u -> -infinity
\draw[boundary]    (U) -- (R);  % v -> +infinity
\draw[boundary]    (P) -- (L);  % regular center
\draw[matching]    (B) -- (L);  % v=0
\draw[singularity] (L) -- (U);  % outgoing null singularity

% Uniform offsets from the null boundaries.
\def\din{0.24}
\def\dout{0.27}

%------------------------------------------------------------
% Boundary labels.
%------------------------------------------------------------

% Upper-left boundary.
\node[rotate=45]
  at ($(L)!0.5!(U)+(\din,-\din)$)
  {$u=0,\quad r_*=0$};

% Upper-right boundary.
\node[rotate=-45]
  at ($(U)!0.5!(R)+(-\din,-\din)$)
  {$v\to+\infty$};

\node[rotate=-45]
  at ($(U)!0.5!(R)+(\dout,\dout)$)
  {$x\to+\infty$};

% Lower-right boundary of the CSS diamond.
\node[rotate=45]
  at ($(B)!0.5!(R)+(-\din,\din)$)
  {$u\to-\infty$};

\node[rotate=45]
  at ($(B)!0.5!(R)+(\dout,-\dout)$)
  {$T\to-\infty$};

% Matching surface.
% Since x=log(-v/u), v=0 corresponds to x -> -infinity.
\node[rotate=-45]
  at ($(L)!0.5!(B)+(\din,\din)$)
  {$v=0$};

\node[rotate=-45]
  at ($(L)!0.5!(B)+(-\dout,-\dout)$)
  {$x\to-\infty$};

% Regular center and past timelike infinity.
\node[rotate=90] at (-0.40,-4.0)
  {regular center $(r_*=0)$};

\node[anchor=north] at (P)
  {$i^-$};

%------------------------------------------------------------
% Interior labels.
%------------------------------------------------------------
\node at (4,0.35)
  {$u<0,\quad v>0$};

\node at (4,-0.40)
  {$\theta_+>0,\quad\theta_-<0$};

\node[align=center] at (1.75,-4.05)
  {Minkowski\\[3pt]
   $v<0$};

\end{tikzpicture}%
}
\caption{
Penrose diagram of the critical Roberts-type CSS background.
The shaded diamond is the self-similar region $u<0$, $v>0$, matched to
Minkowski spacetime across the ingoing null surface $v=0$
($x\to-\infty$). The regular center in the Minkowski region ends at
$u=v=0$. The jagged outgoing null boundary $u=0$, $v>0$ is a curvature
singularity with $r_*=0$. The classical CSS region is untrapped,
with positive outgoing expansion $\theta_+>0$ and negative ingoing
expansion $\theta_-<0$.
}
\label{fig:CSSRobertsCausal}
\end{figure}
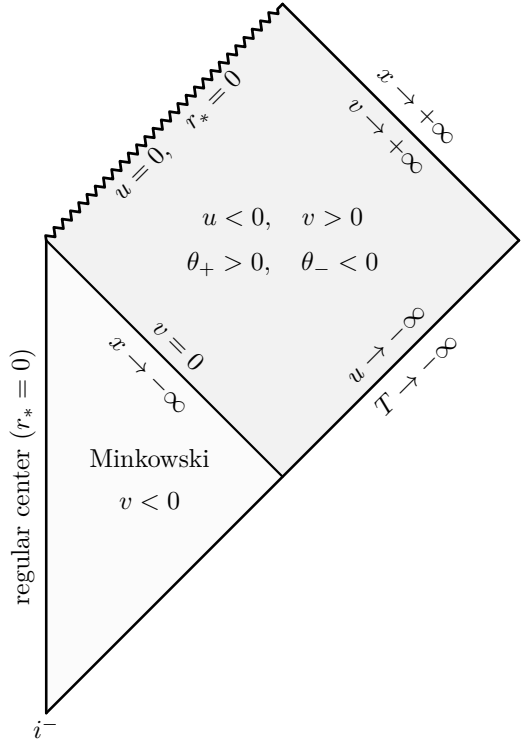

We follow the exact CSS Einstein-scalar solutions of \cite{Clark:2025tqi}, with notation adapted to \cite{Tomasevic:2025clf, Tomasevic:2025kqy,Wu:2026xbr}. These solutions generalize the 4d Roberts geometry \cite{Roberts:1989sk,Brady:1994aq, Brady:1994xfa,Oshiro:1994hd}. The critical member joins a Minkowski past across an ingoing null surface and ends at an outgoing null singularity, with no apparent horizon in the self-similar region. Its causal structure is shown in Figure~\ref{fig:CSSRobertsCausal}. It provides an exact analytic model for studying quantum effects near the classical singularity threshold. We refer to~\cite{Tomasevic:2025clf,Wu:2026xbr} for a more detailed discussion of the relation between Roberts-type solutions and true critical collapse.

\paragraph{The critical solution.} In double-null coordinates, the metric takes the form
\be
    ds_D^2
    =
    -2e^{2\rho(u,v)}du\,dv
    +r^2(u,v)d\Omega_n^2,
    \qquad
    n=D-2\geq2.
    \label{eq:css_double_null_metric}
\ee
We study the self-similar region $u<0$ and $v>0$. The scalar influx begins at $v=0$, across which the solution joins the Minkowski past with
\be
    \rho(u,0)=0,
    \qquad
    r(u,0)=-\frac{u}{\sqrt{2}},
    \qquad
    f(u,0)=0.
    \label{eq:css_flat_past_data}
\ee
We introduce the logarithmic scale time $T$ and the similarity coordinate $x$ through
\be
    u=-\ell e^{-T},
    \qquad
    v=\ell e^{-T+x},
    \qquad
    x=\log\!\left(-\frac{v}{u}\right).
    \label{eq:css_adapted_coordinates}
\ee
Increasing $T$ moves toward smaller physical scales at fixed $x$. The homothetic Killing vector is
\be
    \xi
    =
    u\partial_u+v\partial_v
    =
    -\partial_T.
    \label{eq:css_homothetic_vector}
\ee
The classical scalar is invariant under this homothety. Again the subscript $*$ denotes the critical background,
\be
    \rho_*=0, \quad
    r_*(T,x)=\ell e^{-T}R(x), \quad
    R(x)=e^{x/2}\sqrt{y(x)}, \quad
    f_*(T,x)=f_*(x).
    \label{eq:css_critical_background}
\ee
The radial function $y$ is the one introduced in \cite{Clark:2025tqi}. With the classical scalar invariant under the homothety, the $uu$ and $vv$ Einstein equations imply that $\rho$ is constant, and the flat-past normalization fixes $\rho=0$. The remaining equation for $y$ decouples from the scalar profile and has the first integral
\be
    (y')^2
    =
    y^2-2y+(2q)^ny^{2-n},
    \qquad
    y'<0,
    \label{eq:css_exact_first_integral}
\ee
where $q$ is an integration constant and a prime denotes $d/dx$. The negative branch is selected by the flat-past behavior $y\sim\frac{1}{2}e^{-x}$ as $x\to-\infty$. Once $y$ is known, the remaining Einstein equations determine the scalar gradient, so only this one equation needs to be solved.

This simplification is exact at finite $D$. The large-$D$ limit further separates the nearly flat region from a narrow transition region where the scalar backreaction becomes important. This separation allows the background to be studied through approximations on the different radial scales and their matching \cite{Clark:2025tqi}. We will use the same separation of scales to analyze the quantum backreaction.

As in the Roberts family, $q$ labels different exact CSS backgrounds. Changing $q$ preserves CSS and therefore does not describe the growing departure controlled by the initial-data parameter $p-p_*$. Hence we use the critical member $q=q_*$ and introduce the classical perturbation separately in Section~\ref{subsec:css-horizon}. On this solution, $y$ decreases toward a positive limiting value $y_s$ as $x\to+\infty$. The RHS of \eqref{eq:css_exact_first_integral} must have a double root at $y=y_s$, which gives
\be
    \alpha\equiv1-\frac{1}{n},
    \qquad
    y_s=2\alpha,
    \qquad
    q_*
    =
    \frac{(n-1)^{(n-1)/n}}{n}.
    \label{eq:css_alpha_definition}
\ee
The critical solution approaches this root with $y'<0$ throughout the self-similar interior.

From now on, $q=q_*$. The remaining translation of $x$ and the additive constant in $f_*$ are fixed by the matching at $v=0$,
\be
    y(x)
    =
    \frac{1}{2}e^{-x}+1+O(e^x),
    \qquad
    f_*(x)\longrightarrow0,
    \qquad
    x\longrightarrow-\infty.
    \label{eq:css_flat_past_normalization}
\ee
The $uu$ and $vv$ Einstein equations then give the scalar gradient,
\be
    f_*'
    =
    \frac{n}{2}\sqrt{\alpha}
    \left(\frac{2q_*}{y}\right)^{n/2}.
    \label{eq:css_background_scalar_derivative}
\ee
Here we choose $f_*'>0$. The opposite sign is obtained by $f_*\to-f_*$ and gives the same geometry.

\paragraph{Radial geometry.}

To follow the backreaction, we need the radial profiles throughout the self-similar region and their behavior at both ends. It is convenient to map the unbounded range of $x$ to a finite interval, where we define
\be
    Q=\frac{2\alpha}{y},
    \qquad
    \chi(Q)
    =
    \left(
        1-\frac{nQ}{n-1}
        +\frac{Q^n}{n-1}
    \right)^{1/2}.
    \label{eq:css_compact_coordinate}
\ee
The coordinate $Q$ increases monotonically from the flat past toward the null singularity,
\be
    x\to-\infty
    \quad\longleftrightarrow\quad
    Q\to0,
    \qquad
    x\to+\infty
    \quad\longleftrightarrow\quad
    Q\to1.
    \label{eq:css_compact_endpoints}
\ee
The exact relations needed below are
\be
    \frac{y'}{y}=-\chi, \qquad
    Q'=Q\chi, \qquad
    \frac{R'}{R}=\frac{1-\chi}{2},
\ee
and
\be
    h
    \equiv
    \left(\frac{2q_*}{y}\right)^{n/2}
    =
    \frac{Q^{n/2}}{\sqrt{n-1}},\qquad 
    f_*'=\frac{\sqrt n}{2}Q^{n/2}.
    \label{eq:css_compact_background_identities}
\ee
The function $h$ describes the radial falloff of the scalar gradient. Its square will also enter the quantum source and the normalization of the backreaction. The factor $Q^{n/2}$ already displays the separation of radial scales at large $D$, suppressing the scalar gradient at fixed $Q<1$ and concentrating it toward $Q=1$. The scalar profile itself is
\be
    f_*(Q)
    =
    \frac{\sqrt n}{2}
    \int_0^Q
    \frac{s^{n/2-1}}{\chi(s)}\,ds.
    \label{eq:css_exact_scalar_profile}
\ee

Now we discuss horizon tracing for the classical background. The two radial null derivatives have definite signs. Since $\rho_*=0$,
we choose the future-directed null normals $k_+=\partial_v$ and
$k_-=\partial_u$, with $k_+\cdot k_-=-1$. The corresponding outgoing
and ingoing null expansions are
\be
\begin{aligned}
    \partial_v r_*
    &=
    \frac{1}{2}e^{-x}R(1-\chi)>0,
    &\qquad
    \theta_+
    &\equiv
    \frac{n}{r_*}\partial_v r_*>0,
    \\
    \partial_u r_*
    &=
    -\frac{1}{2}R(1+\chi)<0,
    &
    \theta_-
    &\equiv
    \frac{n}{r_*}\partial_u r_*<0.
\end{aligned}
\label{eq:css_background_null_derivatives}
\ee
It is also useful to define the horizon tracing quantity
\be
    H
    \equiv
    (\nabla r)^2
    =
    g^{\mu\nu}\partial_\mu r\,\partial_\nu r.
    \label{eq:horizon_scalar_definition}
\ee
On the critical background,
\be
    H_*
    =
    -\frac{2r_*^2}{n^2}\theta_+\theta_-
    =
    1-\frac{Q^{n-1}}{n}>0.
    \label{eq:css_background_horizon_scalar}
\ee
Thus every sphere in the critical self-similar region is untrapped, and the background contains no apparent horizon. At fixed $v>0$,
approaching $u=0$ sends $Q\to1$ and
\be
    r_*^2\sim-2\alpha uv,
    \qquad
    u\to0^-.
    \label{eq:css_null_endpoint_radius}
\ee
The areal radius vanishes and the curvature diverges, giving the null
singularity described above. The quantum calculation will determine how
the approach to this endpoint changes under vacuum polarization.

At $n=2$, the general solution reduces to the 4d Roberts geometry,
\be
    q_*=\frac{1}{2},
    \qquad
    y(x)=1+\frac{1}{2}e^{-x},
    \qquad
    r_*^2=\frac{u^2}{2}-uv,
    \label{eq:css_4d_geometry}
\ee
with
\be
    f_*(x)
    =
    \frac{1}{\sqrt{2}}
    \log\!\left(1+2e^x\right).
    \label{eq:css_4d_scalar}
\ee
This recovers the critical Roberts spacetime used in \cite{Tomasevic:2025clf, Tomasevic:2025kqy,Wu:2026xbr}.\footnote{There is a difference in the normalization of the null
coordinates. It is removed by
$\bar u=u/\sqrt{2}$ and $\bar v=\sqrt{2}v$, which preserve $du\,dv=d\bar u\,d\bar v$ and give
$r_*^2=\bar u^2-\bar u\bar v$ and
$f_*=2^{-1/2}\log(1-\bar v/\bar u)$.}

\paragraph{The quantum growing mode.}

We now evaluate the one-loop theory on this critical background, with the minimal action~\eqref{eq:minimal_anomaly_theory} introduced in Section~\ref{subsec:one-loop}. Following \cite{Tomasevic:2025clf, Tomasevic:2025kqy,Wu:2026xbr}, we require a state that is regular in the physical self-similar region and agrees with the Minkowski vacuum in the flat past, with no independently prescribed incoming or outgoing quantum flux. These conditions describe the Boulware-like vacuum polarization of the collapsing matter. We impose them on the auxiliary solutions below, without requiring the quantum state to inherit the classical self-similarity.

Since $\rho_*=0$, the 2d metric is flat and the Minkowski auxiliary data give $U=0$. Quantum effects remain nontrivial because the matter field couples to the varying dilaton,
$\phi=\frac{n}{2}[T-\log R(x)]$.
Using the potential $V$ introduced in Section~\ref{subsec:one-loop}, we define
\be
    U=0,
    \qquad
    \Box V=(\nabla\phi)^2,
    \qquad
    W \equiv V-\frac{n}{2}\phi.
    \label{eq:css_one_potential_definition}
\ee
In terms of $W$, the 2d stress tensor is
\be
    \langle T^{(2)}_{ab}\rangle
    =
    \frac{N\hbar}{4\pi}
    \left(
        \nabla_a\nabla_bW-g_{ab}\Box W
    \right).
    \label{eq:css_2d_stress_tensor}
\ee
The equation for $W$ is
$\Box W=(\nabla\phi)^2-\frac{n}{2}\Box\phi$.
On the CSS background, it reduces to
$uv\,\partial_u\partial_vW=n^2Q^n(x)/16$.
It therefore admits a particular solution $W_0$ depending only on $x$, and the general solution is
\be
    W=W_0(x)+F(u)+G(v).
    \label{eq:css_chiral_auxiliary_data}
\ee
The functions $F$ and $G$ carry the remaining state dependence. Their contributions can be displayed directly in the null components,
\be
\begin{aligned}
    \langle T^{(2)}_{uu}\rangle
    &=
    \frac{N\hbar}{4\pi}
    \left[
        \frac{W_0''+W_0'}{u^2}
        +\frac{d^2F}{du^2}
    \right],
    \\
    \langle T^{(2)}_{vv}\rangle
    &=
    \frac{N\hbar}{4\pi}
    \left[
        \frac{W_0''-W_0'}{v^2}
        +\frac{d^2G}{dv^2}
    \right],
    \\
    \langle T^{(2)}_{uv}\rangle
    &=
    \frac{N\hbar}{4\pi uv}W_0''.
\end{aligned}
    \label{eq:css_2d_null_stress}
\ee
Matching to the flat past and excluding independently added quantum flux give
$d^2F/du^2=d^2G/dv^2=0$.
The remaining constant and linear terms in $F$ and $G$ make no contribution to these components, so we may take $F=G=0$.

The collapsing scalar determines the particular solution through
\be
    W_0''
    =
    -\frac{n}{4}(f_*')^2
    =
    -\frac{n^2(n-1)}{16}h^2
    =
    -\frac{n^2}{16}Q^n,
    \qquad
    W_0'(Q=0)=0.
    \label{eq:css_auxiliary_equation}
\ee
Integrating from the flat past gives
\be
    W_0'(Q)
    =
    -\frac{n^2}{16}
    \int_0^Q
    \frac{s^{n-1}}{\chi(s)}\,ds.
    \label{eq:css_auxiliary_integral}
\ee
Only $W_0'$ and $W_0''$ enter the stress tensor, so this integral determines the source completely. Near $Q=0$,
\be
    W_0'
    =
    -\frac{n}{16}Q^n+O(Q^{n+1}),
    \qquad
    \lim_{Q\to0}
    \frac{W_0'}{n^2h^2}
    =
    -\frac{\alpha}{16}.
    \label{eq:css_auxiliary_past_limit}
\ee
Thus the source profiles vanish as $Q^n$, while the coefficient remaining after this falloff is removed approaches a finite, nonzero constant. This limit will enter the past boundary conditions for the normalized backreaction equations.

At finite $D$, $\chi(Q)>0$ for $0\leq Q<1$, so \eqref{eq:css_auxiliary_integral} defines analytic profiles throughout $0<Q<1$. Together with the analytic background, this gives a stress tensor that is real analytic in the open self-similar interior $u<0$, $v>0$. The null components also have finite limits as $v\to0^+$ at fixed $u<0$, since their numerators vanish as $Q^n=O(v^n)$ with $n\geq2$. The null singularity at $u=0$ of course lies outside this regular domain.

At $n=2$, we have $\chi=1-Q$, and the integral becomes elementary,
\be
    Q=\frac{2e^x}{1+2e^x},
    \qquad
    W_0'
    =
    \frac{1}{4}\left[Q+\log(1-Q)\right],
    \qquad
    W_0''=-\frac{Q^2}{4}.
    \label{eq:css_4d_auxiliary_profile}
\ee
Substituting these expressions into \eqref{eq:css_2d_null_stress} with $F=G=0$ and transforming to $(\bar u,\bar v)$ reproduces the Boulware-like Roberts stress tensor of \cite{Tomasevic:2025clf}.\footnote{In that paper, $\gamma$ parametrizes a residual homogeneous contribution to the auxiliary fields. Regularity of the semiclassical geometry fixes $\gamma=-3/4$, corresponding to the state used here.}
The finite-$D$ construction therefore extends the same vacuum polarization state used in the earlier 4d analysis.

In the similarity coordinates $(T,x)$, the stress tensor becomes
\be
\begin{aligned}
    \langle T^{(2)}_{TT}\rangle
    &=
    \frac{N\hbar}{\pi}W_0'',
    \\
    \langle T^{(2)}_{Tx}\rangle
    &=
    \frac{N\hbar}{4\pi}
    \left(W_0'-2W_0''\right),
    \\
    \langle T^{(2)}_{xx}\rangle
    &=
    \frac{N\hbar}{4\pi}
    \left(W_0''-W_0'\right).
\end{aligned}
    \label{eq:css_2d_adapted_stress}
\ee
Every component depends only on $x$, so the 2d stress tensor is invariant under the CSS homothety, as found in \cite{Tomasevic:2025clf, Tomasevic:2025kqy,Wu:2026xbr}. Here this property follows explicitly from the Boulware-like state fixed by the collapse conditions.

The tensor in \eqref{eq:css_2d_stress_tensor} is separately conserved with $\nabla^a \langle T^{(2)}_{ab}\rangle=0$ because the 2d metric is flat. Since the minimal action has $E_f=0$ and the background dilaton is nonconstant, \eqref{eq:reduced_ward_identity} gives $E_\phi=0$. The angular pressure consequently vanishes. Using
$\langle T^{(D)}_{ab}\rangle
=\langle T^{(2)}_{ab}\rangle/(\Omega_n r_*^n)$,
the full $D$-dimensional stress tensor has components
\be
\begin{aligned}
    \langle T^{(D)}_{TT}\rangle
    &=
    \frac{N\hbar e^{nT}}
         {\pi\Omega_n\ell^nR^n(x)}W_0'',
    \\
    \langle T^{(D)}_{Tx}\rangle
    &=
    \frac{N\hbar e^{nT}}
         {4\pi\Omega_n\ell^nR^n(x)}
    \left(W_0'-2W_0''\right),
    \\
    \langle T^{(D)}_{xx}\rangle
    &=
    \frac{N\hbar e^{nT}}
         {4\pi\Omega_n\ell^nR^n(x)}
    \left(W_0''-W_0'\right),
    \\
    \langle T^{(D)}_{Ti}\rangle
    &=
    \langle T^{(D)}_{xi}\rangle
    =
    \langle T^{(D)}_{ij}\rangle
    =
    0,
\end{aligned}
    \label{eq:css_quantum_stress_uplift}
\ee
where $i,j=1,\ldots,n$ label the angular directions. Together, \eqref{eq:css_auxiliary_equation} and
\eqref{eq:css_auxiliary_integral} determine the quantum stress tensor
exactly at finite $D$. To evaluate it for general finite $D$, the only integral to do is~\eqref{eq:css_auxiliary_integral}.
Its integrand is nonsingular for every $0\leq Q<1$, so the quantum source can be evaluated directly throughout the self-similar interior.

With the state fixed, the entire time dependence appears in the common factor $e^{nT}$. The collapse conditions first determine a homothetically invariant 2d source, and self-similar contraction then amplifies it through the inverse transverse area. The quantum Lyapunov exponent is therefore
\be
    \omega_q=n=D-2.
    \label{eq:css_quantum_growth_rate}
\ee
This is exact at finite $D$. The large-$D$ limit will determine how this growth is distributed radially and where its competition with the classical perturbation changes the trapping geometry.

\subsection{The quantum wall from large-$D$ horizon tracing}
\label{subsec:css-horizon}

We now determine how the quantum growing mode changes the local geometry. Combining its backreaction with the classical departure from criticality allows us to trace the appearance of a future MOTS. The large-$D$ limit shows that the competition is concentrated in a parametrically thin radial layer, where a future MOTS first appears at finite areal radius. We refer to this layer as the quantum wall.

\paragraph{Classical and quantum perturbations.}

The classical perturbations studied in \cite{Clark:2025tqi} fall into two classes. Changing $q$ moves within the exact CSS family, while CSS-breaking perturbations describe departures from it. In critical collapse, the solution approaches the critical geometry and then leaves it along a growing perturbation. The time spent near criticality depends on the amplitude of this perturbation and sets the classical departure scale. Changing $q$ preserves CSS throughout the evolution and cannot describe this mechanism.\footnote{The supercritical members of the exact CSS family also have the familiar global limitations of Roberts-type solutions, discussed in detail in \cite{Tomasevic:2025clf}.} We therefore hold $q=q_*$ and use the CSS-breaking sector as the classical competitor, following the approach of \cite{Tomasevic:2025clf,Tomasevic:2025kqy,Wu:2026xbr}.

For a one-parameter family of initial data, we absorb the normalization of the classical perturbation into the local definition of $p-p_*$. With the classical and quantum Lyapunov exponents $\omega_c$ and $\omega_q$, the geometry and scalar field take the form
\be
\begin{aligned}
    \rho(T,x)
    &=(p-p_*)e^{\omega_cT}\rho_c(x)
      +g_qe^{\omega_qT}\rho_q(x),
    \\
    r(T,x)
    &=\ell e^{-T}R(x)
      \left[1+(p-p_*)e^{\omega_cT}r_c(x)
      +g_qe^{\omega_qT}r_q(x)\right],
    \\
    f(T,x)
    &=f_*(x)+(p-p_*)e^{\omega_cT}f_c(x)
      +g_qe^{\omega_qT}f_q(x).
\end{aligned}
\label{eq:css_semiclassical_ansatz}
\ee
The subscripts $c$ and $q$ denote the classical and quantum perturbations. The functions $\rho_c,\rho_q$ perturb the conformal factor, $r_c,r_q$ are fractional changes of the areal radius, and $f_c,f_q$ perturb the scalar field.

The classical and quantum perturbations are governed by the same linearized Einstein-scalar operator. The classical perturbation satisfies the homogeneous equations, while the quantum one satisfies the inhomogeneous equations sourced by the one-loop stress tensor. Writing $Z_k=(r_k,\rho_k,f_k)^{\rm T}$, we have
\be
    \mathcal L(\omega_c)Z_c=0,
    \qquad
    \mathcal L(\omega_q)Z_q=S_q.
\label{eq:css_sourced_backreaction}
\ee
Here $\mathcal L(k)$ is the radial linearized operator obtained after substituting a perturbation proportional to $e^{kT}$ and factoring out this common time dependence, while $S_q$ is determined by \eqref{eq:css_2d_adapted_stress}. Matching to the flat past selects the sourced solution and excludes an independent homogeneous mode at $\omega_q$.

For a classical perturbation proportional to $e^{kT}$, the radial profiles compatible with the flat-past boundary behave as
\be
    r_k,\rho_k=O\!\left(e^{(n-k)x}\right),
    \qquad
    f_k=O\!\left(e^{(n/2-k)x}\right),
    \qquad x\to-\infty.
\label{eq:css_past_mode_orders}
\ee
The scalar boundary condition gives $\operatorname{Re}k<n/2$. At the opposite end  $x\to\infty$, boundedness near the CSS endpoint further restricts the allowed region in the complex $k$ plane. The resulting continuous spectrum has the upper edge \cite{Clark:2025tqi}
\be
    \omega_c\equiv\sup\operatorname{Re}k=\frac{n}{2}.
\label{eq:css_classical_spectral_edge}
\ee
This asymptotic analysis assumes that matching through finite $x$ adds no further spectral restriction, and we make the same assumption.

In the dimensions considered below, the spectral edge can be approached along real $k$. For every $k<n/2$, the scalar perturbation vanishes toward the flat past, while the limiting profile at $k=n/2$ has a constant leading scalar term. We therefore use the bounded, logarithm-free profile obtained as $k\to n/2^-$, normalized so that this constant is unity. This choice is the smooth continuation of the admissible real modes to the spectral edge. Even though the endpoint $k=n/2$ itself is excluded by the strict scalar boundary condition, here it is used as the limiting profile of the fastest real growth. Its construction and the corresponding past data are discussed in Appendix~\ref{app:css-response}, with the normalization given in \eqref{eq:css_classical_past_data}.\footnote{Since $k=n/2$ is a spectral edge, a more general classical prescription is described by a finite-width packet of nearby allowed exponents. Such a packet retains the leading exponential rate $e^{nT/2}$ but can acquire a power-law prefactor in $T$, producing logarithmic corrections to the threshold relation. See the discussion in Appendix~\ref{app:css-response}. A complete packet calculation must also retain the $k$ dependence of the radial profiles and spectral weights. In particular, the dominant 4d Roberts perturbation has a complex exponent~\cite{Frolov:1997uu, Frolov:1999fv}, so a real perturbation depends on a phase and oscillates in $T$~\cite{Tomasevic:2025clf}. The prescription discussed here cannot be applied directly to the low-dimensional problems.}

Now, the classical term in \eqref{eq:css_semiclassical_ansatz} grows as $(p-p_*)e^{nT/2}$, while the quantum backreaction grows as $g_qe^{nT}$. Thus
\be
    \omega_c=\frac{n}{2},
    \qquad
    \omega_q=n,
    \qquad
    \frac{\omega_q}{\omega_c}=2.
\label{eq:css_growth_rate_comparison}
\ee
In the RG language, both are relevant deformations of the CSS solution, but the quantum term grows twice as fast toward smaller scales. Fine-tuning $p$ toward $p_*$ delays the classical departure and allows the evolution to remain close to the critical geometry for longer. The quantum amplitude $g_q$ is fixed, so sufficiently precise tuning allows the faster quantum backreaction to catch up with the classical perturbation. Although both growth rates increase linearly with $n$, their ratio is fixed at large $D$.\footnote{In principle, a resonance of the inhomogeneous equations at $k=n$ could generate additional powers of $T$ multiplying $e^{nT}$. The bounded solution selected here contains no such enhancement, as shown explicitly in the large-$D$ analysis of Appendix~\ref{app:css-response}.}

\paragraph{Finite-radius MOTS formation.}

The growth rates determine when the classical and quantum perturbations become comparable, but trapping also depends on their radial profiles. We construct these profiles from the flat-past boundary at $Q=0$ and integrate them across $0<Q<1$. Here we summarize the physical results needed for horizon tracing, while the details are collected in Appendix~\ref{app:css-response}. We then use the resulting first-order corrected geometry to trace the formation of a future MOTS.

In the double-null metric \eqref{eq:css_double_null_metric},
the horizon quantity \eqref{eq:horizon_scalar_definition} is
\be
    H\equiv(\nabla r)^2
    =-2e^{-2\rho}\partial_u r\,\partial_v r .
\label{eq:css_H_null}
\ee
The outgoing and ingoing null expansions have the signs of $\partial_v r$ and $\partial_u r$, respectively. A future MOTS therefore lies on the branch
\be
    \partial_v r=0,
    \qquad
    \partial_u r<0.
\label{eq:css_future_mots_conditions}
\ee
To make these conditions explicit, write
\be
    r(T,x)=\ell e^{-T}R(x)\mathcal F(T,x),
    \qquad
    \mathcal F
    =1+(p-p_*)e^{nT/2}r_c+g_qe^{nT}r_q.
\label{eq:css_corrected_areal_radius}
\ee
It is useful to isolate the factor controlling the outgoing expansion
\be
\begin{aligned}
    \mathcal V(T,x)
    &\equiv\frac{R'}{R}\mathcal F+\mathcal F'
    \\
    &=\mathcal V_0(x)
      +(p-p_*)e^{nT/2}\mathcal V_c(x)
      +g_qe^{nT}\mathcal V_q(x),
\end{aligned}
\label{eq:css_outgoing_factor}
\ee
where
\be
    \mathcal V_0=\frac{R'}{R}=\frac{1-\chi}{2},
    \qquad
    \mathcal V_c=\mathcal V_0r_c+r_c',
    \qquad
    \mathcal V_q=\mathcal V_0r_q+r_q'.
\label{eq:css_outgoing_profiles}
\ee
Using the coordinate relations \eqref{eq:css_adapted_coordinates}, the null derivatives become
\be
\begin{aligned}
    \partial_vr&=e^{-x}R\mathcal V,
    \\
    \partial_ur&=-R\left(\mathcal F-\partial_T\mathcal F-\mathcal V\right),
\end{aligned}
\label{eq:css_null_derivatives}
\ee
and hence
\be
    H
    =2e^{-2\rho-x}R^2
      \left(\mathcal F-\partial_T\mathcal F-\mathcal V\right)\mathcal V.
\label{eq:css_exact_null_identity}
\ee
Since $\ell e^{-T}R>0$, $\mathcal F>0$ ensures the areal radius $r$ is always positive. A zero of $\mathcal V$ is precisely a zero of the outgoing expansion. At such a point, $\mathcal F-\partial_T\mathcal F>0$ gives $\partial_u r<0$ and therefore the future orientation. If this sign remains positive nearby, $\mathcal V<0$ corresponds to a future-trapped region.

We now insert the first-order classical and quantum profiles into the full horizon condition $\mathcal V=0$. The critical background is everywhere untrapped, so there is no pre-existing marginal surface whose position is merely shifted by the perturbations. Instead, a new marginal surface can appear where the positive background outgoing factor $\mathcal V_0$ becomes sufficiently small for the classical and quantum contributions to cancel it. Although the profiles themselves are obtained from linearized equations, we impose the horizon condition on the complete first-order expression in \eqref{eq:css_outgoing_factor}.

At fixed $g_q$, the quantum backreaction is already present even when the classical tuning reaches $p=p_*$. The value of $p$ at which a marginal surface first appears is therefore shifted from the classical threshold, and we denote this local quantum threshold by $p_*^{\,q}(g_q)$. This picture has already been discussed in~\cite{Tomasevic:2025clf, Tomasevic:2025kqy, Wu:2026xbr}.

For $p<p_*^{\,q}$, the outgoing factor $\mathcal V(T,x)$ remains positive in a neighborhood of this component. As $p$ is increased, the classical focusing term lowers $\mathcal V$ until a local minimum first reaches zero. At the threshold point $(T_h,x_h)$, the zero is therefore stationary in both the time and radial directions,
\be
    \mathcal V_h=0,
    \qquad
    (\partial_T\mathcal V)_h=0,
    \qquad
    \mathcal V_h'=0.
\label{eq:css_stationary_conditions}
\ee
The first condition is the MOTS condition. The second and third express the fact that the marginal sphere is created when a local extremum of $\mathcal V$ first reaches zero, since the classical background has no pre-existing zero to shift. The first two conditions determine the classical and quantum amplitudes at the marginal sphere,
\be
    (p_*^{\,q}-p_*)e^{nT_h/2}
    =-\frac{2\mathcal V_{0,h}}{\mathcal V_{c,h}},
    \qquad
    g_qe^{nT_h}
    =\frac{\mathcal V_{0,h}}{\mathcal V_{q,h}}.
\label{eq:css_stationary_amplitudes}
\ee
At the onset of trapping, the background and quantum terms each contribute $+\mathcal V_{0,h}$, while the classical focusing term contributes $-2\mathcal V_{0,h}$, so the three terms cancel exactly. The third condition then determines the radial position,
\be
    \left.
    \frac{d}{dx}\log\!\left(
        \frac{\mathcal V_0\mathcal V_q}{\mathcal V_c^2}
    \right)
    \right|_{x_h}=0.
\label{eq:css_radial_fold_equation}
\ee
The horizon problem has therefore been reduced to a single radial equation. Once $x_h$ is known, \eqref{eq:css_stationary_amplitudes} fixes both the threshold time $T_h$ and the threshold shift $p_*^{\,q}-p_*$.

We choose the orientation of $p$ so that increasing $p$ corresponds to positive classical focusing. The physical branch then satisfies
\be
    \mathcal V_{c,h}<0<\mathcal V_{q,h},
    \qquad
    \mathcal F_h>0,
    \qquad
    \left(\mathcal F-\partial_T\mathcal F\right)_h>0.
\label{eq:css_positive_amplitude_condition}
\ee
The first inequality ensures positive classical and quantum amplitudes in \eqref{eq:css_stationary_amplitudes}. The second condition gives a positive areal radius, while the last condition gives $\partial_u r<0$ and therefore the future orientation. The root must also lie inside the self-similar domain $0<Q<1$ where the radial solutions are constructed.

To determine whether this stationary zero actually marks the onset of trapping, we examine the local shape of $\mathcal V$ in the $(T,x)$ plane. We require the Hessian at $(T_h,x_h)$ to be positive definite, so that the stationary zero is a local minimum. Since $\mathcal V_{c,h}<0$, increasing the positive classical focusing amplitude lowers this minimum. For $p<p_*^{\,q}$ it remains above zero and there is no nearby marginal surface. At $p=p_*^{\,q}$ the minimum just reaches zero, producing a degenerate future MOTS at $(T_h,x_h)$. Increasing $p$ further pushes the minimum below zero. Two nearby zeros of $\mathcal V$ then bound a region with $\mathcal V<0$, where both future null expansions are negative as long as the ingoing sign remains unchanged. Near the threshold, the two marginal branches separate from $(T_h,x_h)$ with the characteristic square-root behavior $(p-p_*^{\,q})^{1/2}$, as follows from the quadratic expansion given in Appendix~\ref{app:css-response}. This is the local saddle-node structure that underlies the whole mechanism.

This local bifurcation is the sense in which we use the term Type I-like. In classical Type II collapse, tuning toward the threshold sends the characteristic scale of departure to zero. Here the marginal surface instead appears at a nonzero radius at the quantum-shifted threshold. It is only Type I-like as we have not established a metastable Type I critical solution or the associated lifetime scaling~\cite{Gundlach:2025yje}.

The stationary amplitudes also determine when the quantum backreaction becomes important and how far it shifts the classical threshold. From \eqref{eq:css_stationary_amplitudes},
\be
    T_h=\frac{1}{n}
        \log\!\left(\frac{\mathcal V_{0,h}}
        {g_q\mathcal V_{q,h}}\right),
    \qquad
    p_*^{\,q}-p_*
    =-\frac{2\sqrt{\mathcal V_{0,h}\mathcal V_{q,h}}}
    {\mathcal V_{c,h}}\,g_q^{1/2}.
\label{eq:css_shifted_fold}
\ee
For weak quantum coupling, the onset is delayed to
$T_h\sim n^{-1}\log(1/g_q)$, so the evolution must remain close to the critical solution down to a smaller physical scale before the quantum term becomes important. The square-root shift in $p_*^{\,q}-p_*$ follows directly from the ratio
$\omega_c/\omega_q=1/2$. The growth rates fix this power of $g_q$, while the coefficient depends on the radial profiles evaluated at the onset. 

The same balance determines the physical radius of the marginal sphere. Using
$g_q=(\ell_q/\ell)^n$, we find
\be
    r_h=\ell e^{-T_h}R_h\mathcal F_h
       =C_n\ell_q,
    \qquad
    C_n=R_h\mathcal F_h
        \left(\frac{\mathcal V_{q,h}}
        {\mathcal V_{0,h}}\right)^{1/n}.
\label{eq:css_marginal_radius}
\ee
The arbitrary reference scale $\ell$ has therefore dropped out. The onset radius is set by the quantum scale $\ell_q$, multiplied by the dimensionless factor $C_n$ determined by the radial horizon problem. This factor depends on $n$, the quantum prescription, and the shape of the classical radial component, but not on its overall normalization. Rescaling the classical profile changes the numerical coefficient in the threshold shift, with the compensating change absorbed into $p-p_*$, while leaving $r_h$ unchanged. This is exactly the same physical picture discovered in~\cite{Tomasevic:2025clf, Tomasevic:2025kqy, Wu:2026xbr}. The radius $r_h$ also defines a finite quasi-local mass scale for the marginal sphere. Its relation to the final
black hole mass requires the subsequent evolution.

\paragraph{The large-$D$ trapping scale.}

The conditions above determine how a marginal surface can appear at finite $D$, but they do not yet reveal where in the radial profile the relevant balance occurs. The large-$D$ limit makes this structure analytic. The flat-past expansion of the classical scalar perturbation begins as
$f_c=1-n^2e^x/4+\cdots$, so its radial dependence first becomes appreciable when $n^2e^x=O(1)$. The corresponding region moves toward the flat past as $n$ increases. We therefore introduce the stretched coordinate
\be
    z=ne^{x/2},
\label{eq:css_outer_coordinates}
\ee
which keeps this region at finite position as $n\to\infty$.

At fixed $z$, one has $x=-2\log n+O(1)$ and hence $Q=O(n^{-2})$. Using $Q'=Q\chi$ together with the flat-past normalization in \eqref{eq:css_flat_past_normalization},
\be
    Q=\frac{4z^2}{n^2}
      \left[1-\frac{1}{n}+O(n^{-2})\right],
    \qquad
    n^2\mathcal{V}_0=z^2+O(n^{-2}).
\label{eq:css_outer_background_expansion}
\ee
Thus the relevant radial region is pushed into the $Q\ll1$ tail of the CSS geometry, where the background outgoing factor is only $O(n^{-2})$. This small background expansion is what allows the classical and quantum corrections to compete with it. At the same time,
$h^2=Q^n/(n-1)$ is suppressed faster than any power of $1/n$, so the local classical scalar stress is negligible there. The relevant balance therefore occurs in the near-vacuum region with $Q\ll1$, close to the flat-past matching surface, well separated from the narrow matter region near $Q=1$. As $n$ increases, this region moves to $x\simeq-2\log n$, so taking $n\to\infty$ at fixed $x$ would follow a different part of the geometry and miss the trapping scale. We instead keep $z$ fixed and impose the flat-past data at each finite $n$ before taking the large-$D$ limit.

The large-$D$ equations simplify strongly in this region. The detailed reduction and the control of the large-$D$ limit are given in Appendix~\ref{app:css-response}. Here we summarize the radial profiles that enter the outgoing expansion. At leading order, the classical metric perturbations can be eliminated in favor of the scalar profile, leaving
\be
    z^2\frac{d^2f_c}{dz^2}
    +z\frac{df_c}{dz}+z^2f_c=0.
\label{eq:css_leading_scalar_bessel}
\ee
The solution compatible with the flat-past normalization is $J_0(z)$, where $J_\nu$ denotes the Bessel function of the first kind. The corresponding classical contribution to the outgoing expansion is proportional to $zJ_1(z)$, so it oscillates as one moves across the stretched radial region.

The quantum contribution has a simpler limiting profile. The flat-past state selects a regular solution of the sourced radial equations, and the combination relevant for the large-$D$ horizon balance approaches a positive constant. Explicitly,
\be
\begin{aligned}
    f_c&=J_0(z)+O(n^{-1}),
    \\
    \frac{nV_c}{h}&=zJ_1(z)+O(n^{-1}),
    \\
    \frac{V_q}{nR^{-n}h^2}
    &=\frac{1}{64\pi}\left(1-\frac{1}{n}\right)+O(n^{-2}).
\end{aligned}
\label{eq:css_large_D_outgoing_profiles}
\ee
The classical and quantum contributions therefore have qualitatively different radial behavior. The normalized classical profile oscillates, while the
normalized quantum profile approaches a positive constant.
Substituting them into the radial onset condition and using
$d[zJ_1(z)]/dz=zJ_0(z)$ gives
\be
    zJ_0(z)-J_1(z)=0.
\label{eq:css_large_D_bessel_root}
\ee
Positive classical focusing requires $\mathcal V_c<0$, which at large $D$ becomes $J_1(z)<0$. The smallest positive root satisfying this condition is
\be
    z_*=5.331\ldots .
\label{eq:css_large_D_root}
\ee
The root is nondegenerate, with the positive radial derivative required for a local minimum. The areal-radius factor and the ingoing null factor both approach one at large $D$, giving a positive-radius future MOTS. Moreover, the finite-$n$ profiles and their relevant derivatives converge uniformly near $z_*$. The implicit-function theorem therefore guarantees a unique nearby physical root of the exact finite-$n$ equations for every sufficiently large finite $D$. The corresponding estimates are given in Appendix~\ref{app:css-response}. This result is independent of the finite-$D$ numerical continuation discussed below.

Using \eqref{eq:css_marginal_radius}, the physical onset radius becomes
\be
    \frac{r_h}{\ell_q}
    =\frac{K_h}{n^2}
     \left[
       1+\frac{2\log n+2-\log(64\pi z_*^2)}{n}
       +O\!\left(\frac{(\log n)^2}{n^2}\right)
     \right],
    \quad
    K_h\equiv4z_*^2
    =113.697\ldots .
\label{eq:css_large_D_marginal_radius}
\ee
Here $K_h$ is the dimensionless coefficient that fixes the leading large-$D$ trapping radius. Thus
\be
    \frac{n^2r_h}{\ell_q}\longrightarrow K_h,
    \qquad
    \frac{r_h}{\ell_q}\longrightarrow0,
    \qquad n\to\infty .
\label{eq:css_large_D_radius_limit}
\ee
The numerical value of $K_h$ is fixed by the quantum prescription and the classical radial component chosen above. Different choices can change this coefficient, while the radius must be determined from their corresponding radial horizon problem.

The origin of the new scale can now be seen directly. The inverse-area factor $(\ell_q/r_*)^n$ changes by order one when the radius changes by a fractional amount of order $1/n$, so quantum vacuum polarization becomes sharply localized at large $D$. The scale $\ell_q$ characterizes this enhancement, but it does not by itself determine where the outgoing expansion vanishes. In fact, the calculation selects a smaller scale,
\be
    r_h\sim\frac{\ell_q}{(D-2)^2}.
\label{eq:css_quantum_sublayer_scale}
\ee
We refer to this parametrically narrow trapping region as the quantum wall. The extra $(D-2)^{-2}$ suppression is therefore a result of the horizon tracing calculation, not of the quantum growth exponent alone.

For this onset to remain semiclassical, the initial matter-loop coupling must be small and the trapping radius must remain well above the $D$-dimensional Planck length,
\be
    g_q\ll1,
    \qquad
    \ell_{P,D}\ll r_h\ll\ell,
    \qquad
    \ell_{P,D}^{\,n}=G_D\hbar.
\label{eq:css_semiclassical_hierarchy}
\ee
At fixed finite $D$, increasing the number of matter fields separates $\ell_q$ from $\ell_{P,D}$ and suppresses graviton loops relative to the matter loop.\footnote{An effectively large number of light matter modes can also arise without introducing many fundamental species. One example is provided by the lowest Landau levels of charged fermions around a magnetically charged black hole, as discussed in Appendix~D of \cite{Shi:2026wnk}.} Since
$r_h/\ell_q\sim K_h/n^2$, the condition $r_h\gg\ell_{P,D}$ requires
\be
    N\gg\frac{\Omega_n}{8\pi}
          \left(\frac{n^2}{K_h}\right)^n,
    \qquad
    \log N\gtrsim\frac32n\log n+O(n).
\label{eq:css_species_hierarchy}
\ee
The required number of fields therefore grows rapidly with $D$. A controlled weak-curvature window can be maintained at any fixed finite $D$ by taking $N$ sufficiently large, but it cannot survive the strict large-$D$ limit at fixed $N$.

These conditions control the loop expansion and the curvature scale, but they do not bound the nonlinear corrections to the critical-departure expansion. As in classical critical collapse, we use the growing first-order profiles to predict the scale at which the critical trajectory departs. The finite-radius MOTS is therefore a prediction of the first-order geometry constructed here. Establishing the same surface in the full semiclassical solution requires additional control of the nonlinear corrections near the onset. Appendix~\ref{app:css-response} discusses this limitation in more detail.

\begin{figure}[t!]
    \centering

    \begin{subfigure}[t]{0.495\linewidth}
        \centering
        \includegraphics[width=\linewidth]{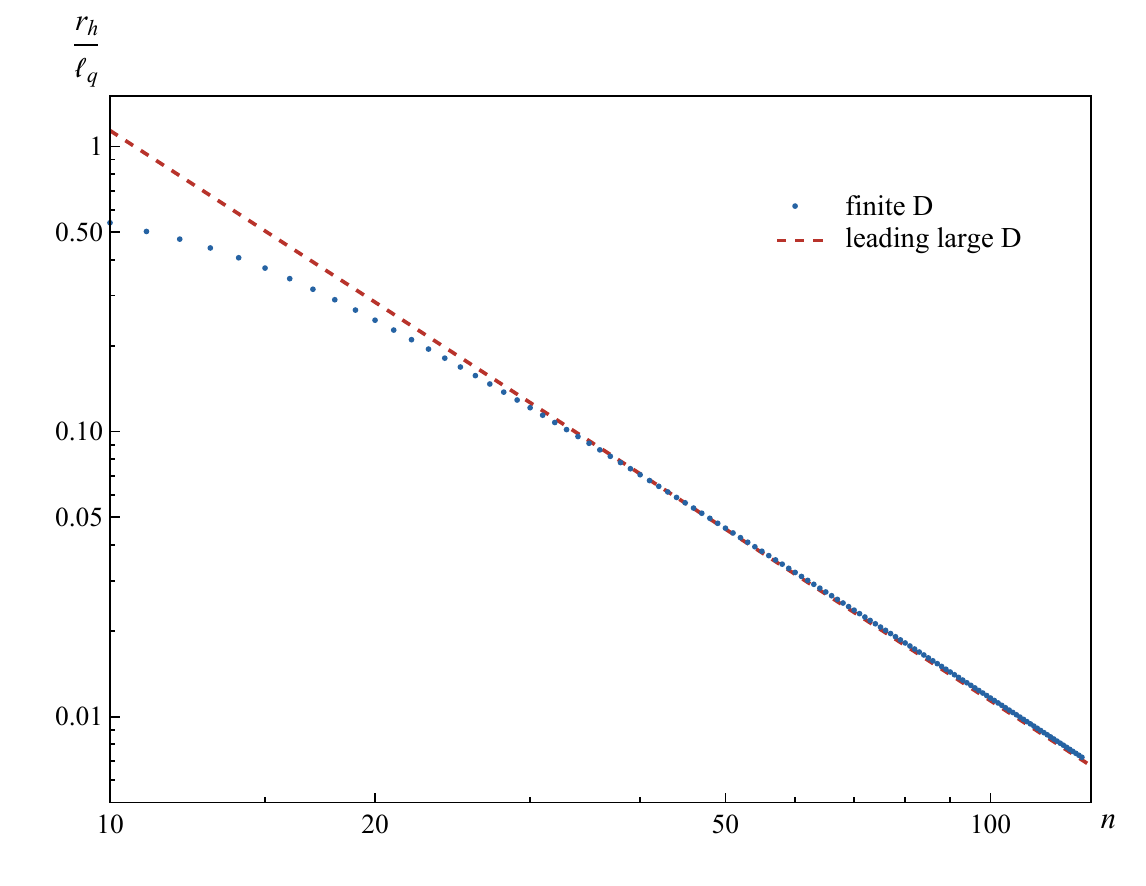}
        \caption{(a) Finite-$D$ trapping radius}
        \label{fig:finit_D}
    \end{subfigure}
    \hfill
    \begin{subfigure}[t]{0.495\linewidth}
        \centering
        \includegraphics[width=\linewidth]{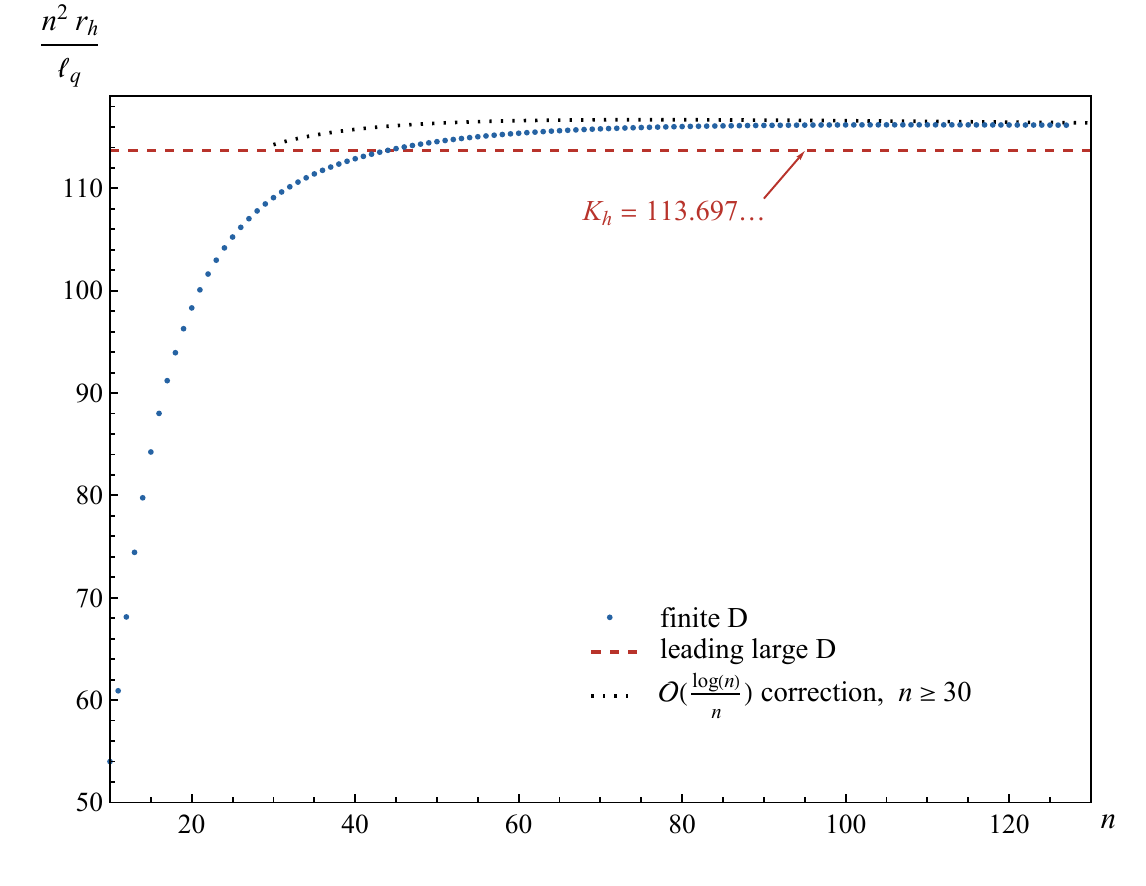}
        \caption{(b) Large-$D$ asymptotics}
        \label{fig:large_D}
    \end{subfigure}

    \caption{Finite-$D$ continuation of the trapping radius along the radial
component connected to the large-$D$ solution.
The blue points are the finite-$D$ numerical results, while the red
dashed curve is the leading large-$D$ prediction
$r_h/\ell_q=K_h/n^2$, with
$K_h=113.697\ldots$.
The numerical branch approaches this behavior as $D$ increases,
showing directly the inverse-square trapping scale
$r_h\sim\ell_q/(D-2)^2$.
The right panel removes this leading $n^{-2}$ scaling, so the
large-$D$ prediction becomes the constant value $K_h$.
The black dotted curve includes the first relative correction of
order $O(\log n/n)$ in
\eqref{eq:css_large_D_marginal_radius} and is shown from $n=30$
onward. Since the approach to $K_h$ is only $O(\log n/n)$,
finite-$n$ deviations remain visible even after the leading
$1/n^2$ scaling has been removed. The dotted curve displays this
predicted approach and agrees with the numerical branch at
sufficiently large $n$.}
    \label{fig:largeD_scaling}
\end{figure}

\paragraph{Finite-$D$ continuation.}

We now follow the radial component selected by the large-$D$ analysis toward finite $D$. Along this component, the ingoing expansion changes orientation at
$n\simeq9.73647$. The first integer dimension with the sign required for a future MOTS is therefore $n=10$, or $D=12$. At $D=11$, the corresponding stationary zero has the opposite ingoing orientation. Thus $D=12$ is the lower end of the continuation studied here, not a general lower bound on MOTS formation.

At $D=12$, the stationary root is located at
\be
    Q_h\simeq0.85812252,
    \qquad
    x_h\simeq-0.47360216.
\label{eq:css_D12_central_root}
\ee
With the normalization in \eqref{eq:css_classical_past_data}, this gives
\be
    p_*^{\,q}-p_*\simeq5.25158\,g_q^{1/2},
    \qquad
    r_h\simeq0.539925\,\ell_q.
\label{eq:css_D12_tuning_and_radius}
\ee
As we discussed, the coefficient in the threshold shift depends on the normalization of the classical perturbation, while the onset radius is a physical quantity and is unchanged by an overall rescaling of that mode. The numerical solution has positive areal radius and
negative ingoing expansion at the stationary zero.
The Hessian in $(T,x)$ has eigenvalues approximately
$1.98145$ and $18.5133$, so this zero is a local minimum.
The radial integration and consistency checks are
described in Appendix~\ref{app:css-response}.

We then apply the same prescription at every integer
\be
    10\leq n\leq127,
    \qquad
    12\leq D\leq129.
\label{eq:css_finite_dimension_continuation}
\ee
At all 118 dimensions, the selected root retains positive radius, future orientation, and a stationary zero with a positive-definite Hessian. The same flat-past normalization and radial selection are used throughout, with the selected roots approaching the Bessel component identified above as $D$ increases. Figure~\ref{fig:largeD_scaling} shows this approach to the large-$D$ regime, both in the $1/n^2$ decrease of the trapping radius and in the slower finite-$n$ convergence of the rescaled radius to $K_h$. This provides a numerical bridge between the $D=12$ solution and the analytic large-$D$ regime. This continuation does not classify the other radial roots. The large-$D$ argument independently guarantees a nearby physical root for all sufficiently large finite $D$.

The finite-$D$ calculation therefore reproduces the same local mechanism found analytically at large $D$. Fine-tuning toward the classical threshold delays the classical departure toward progressively smaller scales, while the quantum backreaction remains present and eventually competes with it. This competition produces a future MOTS at nonzero radius. This is also the point of contact with the earlier 4d Roberts analysis \cite{Tomasevic:2025clf,Tomasevic:2025kqy,Wu:2026xbr}. However, the relevant 4d classical exponents are complex, so the
real perturbation also depends on an oscillation phase. This belongs to a separate low-dimensional problem, while the component followed here develops the large-$D$ radius law \eqref{eq:css_quantum_sublayer_scale}.

\section{Quantum growing mode in the large-$D$ DSS spacetime}
\label{sec:dss}

This section applies the one-loop theory to the leading large-$D$ DSS geometry first constructed in \cite{Ecker:2026akf}. By following the quantum source through successive echoes, we show that self-similar contraction produces the same kinematical growth $\omega_q=D-2$ found in CSS. Echoing modulates this growth and can leave memory of earlier echoes in the stress tensor. The minimal anomaly-induced action gives a concrete realization of this memory, but its stress tensor is singular at the center. We therefore construct one-loop actions whose stress tensors remain smooth throughout the self-similar domain and have bounded periodic profiles. Finally, we use the radial constraint to relate the effective energy enclosed by each sphere to the change in the outgoing expansion, showing how the backreaction can favor or oppose trapping.

\subsection{Quantum growth and echo memory}
\label{subsec:dss-growing-mode}

At leading order in large $D$, the Einstein-scalar system admits an analytic family controlled by a smooth periodic function $\beta(T)$~\cite{Ecker:2026akf}. The construction uses an expansion in $1/(D-1)$ in which the leading radial equations can be integrated in sequence. At this order, their integration leaves both the function $\beta(T)$ and its echoing period $\Delta$ undetermined. Any smooth periodic $\beta(T)$ gives a leading-order DSS solution, while subleading orders impose additional consistency conditions that can constrain the $\beta(T)$ and $\Delta$. The leading scalar field itself depends only on $T$, while its radial dependence first appears at subleading order.

We hold $\beta(T)$ and its period $\Delta$ fixed as $D$ increases. This keeps the time derivatives of $\beta(T)$ of order one, preventing them from enhancing terms suppressed by inverse powers of $D$. This allows us to isolate the large-$D$ scaling of the one-loop source from any additional $D$ dependence of the classical DSS data. At finite $D$, the full DSS solution and its period are instead
determined by the classical equations together with regularity at
the center and the past self-similarity horizon (SSH) \cite{Ecker:2026joz}.\footnote{Throughout this section, $D$ is an integer greater than three. After spherical reduction, the classical equations admit an analytic 
continuation to real $D$, as used in \cite{Ecker:2026joz} to study their 
dimension dependence.} The function
$\beta(T)$ in the leading large-$D$ description must ultimately
match this finite-$D$ solution. Here we study the quantum source on a chosen member of the leading analytic family.

\paragraph{The DSS background.}

With $n=D-2$, the classical metric and areal radius are
\be
\begin{split}
    ds_D^2
    ={}&
    \ell^2 e^{-2T}
    \Bigg[
        \left(
            x^2-\frac{1+\beta^2(T)x^2}{1+\beta^2(T)}
        \right)dT^2
        -2x\,dT\,dx
        +dx^2
        +x^2d\Omega_n^2
    \Bigg],
    \\
    r_*={}&\ell e^{-T}x,
    \qquad
    0\leq x\leq1,
\end{split}
    \label{eq:dss_leading_metric}
\ee
where again $*$ denotes the classical critical background.
Here $x=r_*/(\ell e^{-T})$ measures the areal radius relative to the shrinking classical scale. It differs from the logarithmic null-coordinate ratio used in Section~\ref{sec:css}. The dimensionless function $\beta(T)$ describes how the scalar evolves within an echo through
\be
    \partial_T f_*
    =
    \sqrt n\,
    \frac{\beta(T)}{\sqrt{1+\beta^2(T)}},
    \qquad
    \partial_x f_*=0.
    \label{eq:dss_background_scalar}
\ee
Its sign determines whether the scalar increases or decreases with $T$, while its magnitude controls the rate of change. The metric depends only on $\beta^2$ and is therefore unchanged when the scalar evolution reverses sign.\footnote{For constant $\beta$, the metric inside the brackets in \eqref{eq:dss_leading_metric} is independent of $T$, giving a CSS geometry. The scalar is then linear in $T$, with slope fixed by \eqref{eq:dss_background_scalar}. These solutions have a regular center, which is absent from the CSS interior studied in Section~\ref{sec:css}. The local quantum calculations below also apply to this constant-$\beta$ case.}

The Einstein-scalar equations are invariant under $f\to-f$. The 4d Choptuik solution exhibits this sign reversal after half an echo \cite{Ecker:2026joz}. To retain this behavior in the analytic family, we follow \cite{Ecker:2026akf} and take
\be
    \beta(T+\Delta/2)=-\beta(T).
    \label{eq:dss_waveform_half_echo}
\ee
After fixing the additive constant of $f_*$, this implies
\be
    f_*(T+\Delta/2)=-f_*(T),
    \qquad
    g_{*\mu\nu}(T+\Delta_g,x)
    =
    e^{-2\Delta_g}g_{*\mu\nu}(T,x),
    \qquad
    \Delta_g=\frac{\Delta}{2}.
    \label{eq:dss_background_half_echo}
\ee
Because the metric depends only on $\beta^2$, the second half of each scalar oscillation reproduces the same geometry at a smaller scale. The metric therefore has period $\Delta_g$ modulo the overall scale. This shorter period follows from the scalar reflection symmetry and is not implied by DSS alone.

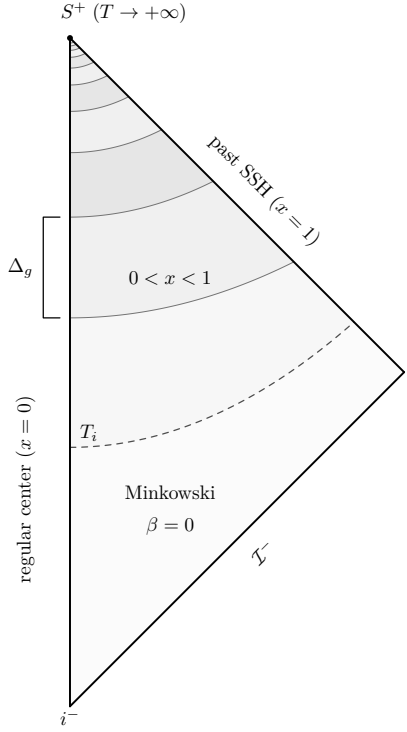
\begin{figure}[t]
\centering
\resizebox{0.35\linewidth}{!}{%
\begin{tikzpicture}[
  x=1.5cm, y=1.5cm,
  line cap=round,
  line join=round,
  font=\small,
  every node/.style={inner sep=1.5pt},
  boundary/.style={draw=black, line width=1.0pt},
  echoslice/.style={draw=black!55, line width=0.45pt},
  preparation/.style={
    draw=black!70, line width=0.65pt, dashed
  }
]

%------------------------------------------------------------
% Interior causal domain.
% Null boundaries have slope +/-1. The center is timelike.
%------------------------------------------------------------
\coordinate (S) at (0,4);
\coordinate (P) at (0,-4);
\coordinate (R) at (4,0);

\fill[gray!3] (P) -- (R) -- (S) -- cycle;

% A spacelike preparation slice T=T_i.
% Its slope ranges from 0 to 3/3.4 < 1.
\coordinate (IC) at (0,-0.90);
\coordinate (IH) at (3.40,0.60);

% First slice after the smooth switch to the periodic geometry.
% Echo slices are schematic spacelike curves, not numerical
% constant-T contours in a computed conformal coordinate system.
\def\hzero{3.35}
\def\qdraw{0.64}
\def\shape{0.80}

% Light shading above the initial preparation slice.
\fill[gray!6]
  (IC) -- (S) -- (IH)
  -- plot[domain=1:0, samples=50, variable=\s]
    ({3.40*\s},{-0.90+1.50*\s*\s})
  -- cycle;

%------------------------------------------------------------
% Successive geometric echoes.
% qdraw controls only their separation on the page.
%------------------------------------------------------------
\foreach \j in {0,...,9}{
  \pgfmathsetmacro{\ha}{\hzero*pow(\qdraw,\j)}
  \pgfmathsetmacro{\hb}{\hzero*pow(\qdraw,\j+1)}
  \pgfmathtruncatemacro{\shadelevel}{12+8*mod(\j,2)}

  \fill[gray!\shadelevel]
    (0,{4-\ha})
    -- plot[domain=0:1, samples=45, variable=\s]
      ({\shape*\ha*\s},
       {4-\ha+(1-\shape)*\ha*\s*\s})
    -- ({\shape*\hb},{4-\shape*\hb})
    -- plot[domain=1:0, samples=45, variable=\s]
      ({\shape*\hb*\s},
       {4-\hb+(1-\shape)*\hb*\s*\s})
    -- cycle;
}

% The unresolved accumulation of further echoes.
\pgfmathsetmacro{\hcap}{\hzero*pow(\qdraw,10)}
\fill[gray!16]
  (0,{4-\hcap})
  -- plot[domain=0:1, samples=30, variable=\s]
    ({\shape*\hcap*\s},
     {4-\hcap+(1-\shape)*\hcap*\s*\s})
  -- (S) -- cycle;

% Draw the echo slices.
% Their maximum slope is 2*(1-shape)/shape = 1/2.
\foreach \j in {0,...,9}{
  \pgfmathsetmacro{\hh}{\hzero*pow(\qdraw,\j)}

  \draw[echoslice]
    plot[domain=0:1, samples=45, variable=\s]
      ({\shape*\hh*\s},
       {4-\hh+(1-\shape)*\hh*\s*\s});
}

% Preparation slice, not a thin shell.
\draw[preparation]
  plot[domain=0:1, samples=50, variable=\s]
    ({3.40*\s},{-0.90+1.50*\s*\s});

%------------------------------------------------------------
% Causal boundaries.
%------------------------------------------------------------
\draw[boundary] (P) -- (S);  % x=0
\draw[boundary] (R) -- (S);  % x=1, past SSH
\draw[boundary] (P) -- (R);  % past null infinity in the flat region
\fill (S) circle[radius=1.5pt];

%------------------------------------------------------------
% Boundary and region labels.
%------------------------------------------------------------
\node[anchor=south west] at (-0.15,4.14)
  {$S^+\;(T\to+\infty)$};

\node[rotate=90] at (-0.53,-1.40)
  {regular center ($x=0$)};

\node[rotate=-45]
  at ($(S)!0.52!(R)+(0.25,0.25)$)
  {past SSH ($x=1$)};

\node[rotate=45]
  at ($(P)!0.52!(R)+(0.23,-0.23)$)
  {$\mathcal I^-$};

\node[anchor=north] at (P)
  {$i^-$};

\node[align=center] at (1.20,-1.65)
  {Minkowski\\[3pt]$\beta=0$};

\node[anchor=south west] at (0.08,-0.86)
  {$T_i$};

\node at (1.20,1.12)
  {$0<x<1$};

% One geometric echo, measured in logarithmic scale time.
\pgfmathsetmacro{\yfirst}{4-\hzero}
\pgfmathsetmacro{\ysecond}{4-\hzero*\qdraw}

\draw[line width=0.65pt]
  (-0.12,\yfirst) -- (-0.32,\yfirst)
  -- (-0.32,\ysecond) -- (-0.12,\ysecond);

\node[anchor=east]
  at (-0.40,{0.5*(\yfirst+\ysecond)})
  {$\Delta_g$};

\end{tikzpicture}%
}
\caption{
Penrose diagram of the leading large-$D$ DSS interior with a Minkowski
preparation. The regular timelike center ($x=0$) and the ingoing null
past SSH ($x=1$) meet at the classical accumulation point $S^+$.
The dashed spacelike slice marks $T=T_i$. We take $\beta(T)=0$ for
$T\leq T_i$ and switch it smoothly to a periodic function before the
first displayed echo slice. The solid interior curves represent
successive constant-$T$ slices separated by $\Delta_g=\Delta/2$ in
the periodic region, and the shading distinguishes adjacent echo
intervals. The preparation slice is drawn differently to emphasize
that it is not an echo image of the later periodic slices. The leading
geometry is untrapped, with $H_*=(\nabla r_*)^2=1$.
}
\label{fig:DSSInteriorEchoes}
\end{figure}

The center at $x=0$ is smooth at every finite $T$. At $x=1$, the coefficient of $dT^2$ vanishes and the self-similar generator $\partial_T$ becomes null, identifying the past SSH. To check whether this surface is also marginal, we evaluate $H=(\nabla r)^2$,
\be
    H_*
    =
    \ell^2 e^{-2T}
    \left(
        x^2 g_*^{TT}
        -2x g_*^{Tx}
        +g_*^{xx}
    \right)
    =1.
    \label{eq:dss_leading_H}
\ee
A marginal sphere requires $H=0$, so the leading geometry contains no marginal spheres. In particular, the past SSH is not a trapping horizon. The equality in \eqref{eq:dss_leading_H} holds at leading order in the classical large-$D$ expansion. The causal structure of this leading DSS interior is shown in Figure~\ref{fig:DSSInteriorEchoes}.

\paragraph{The quantum state.}

We now apply the one-loop theory of Section~\ref{subsec:one-loop}, beginning with the minimal action~\eqref{eq:minimal_anomaly_theory}. We retain its finite subtraction, chosen so that the Minkowski vacuum has vanishing stress tensor. On the present family, Minkowski space corresponds to $\beta=0$. This vacuum supplies the initial data for the quantum state evolved into the DSS region.

We retain the auxiliary fields of \eqref{eq:auxiliary_potentials} and the combination used in the CSS calculation,
\be
    \Box U=R^{(2)},
    \qquad
    \Box V=(\nabla\phi)^2,
    \qquad
    W=V-\frac{n}{2}\phi.
    \label{eq:dss_auxiliary_fields}
\ee
The background dilaton is
\be
    \phi_*=\frac{n}{2}(T-\log x).
    \label{eq:dss_background_dilaton}
\ee
Even in Minkowski space, the dilaton varies with the areal radius as
$\phi=-n\log(r/\ell)/2$. With the flat 2d metric, it satisfies
\be
    \Box\left(\frac{n}{2}\phi\right)
    =
    \frac{n^2}{4r^2}
    =
    (\nabla\phi)^2.
    \label{eq:dss_minkowski_auxiliary_identity}
\ee
The Minkowski vacuum can therefore be represented by $U=0$ and
$V=n\phi/2$. The variable $W=V-n\phi/2$ measures the departure from
this particular solution, so the vacuum data take the simple form
$U=W=0$. With the finite subtraction specified above, these data
give a vanishing stress tensor.

We prepare this vacuum by taking
$\beta(T)=0$ for $T\leq T_i$ and switching smoothly to the chosen
periodic function. We set $U=W=0$ with vanishing time derivatives in the initial Minkowski region and evolve both fields with the same retarded Green's function. We impose regular center data and add no independent incoming homogeneous solution at the past SSH. This prescription then follows the vacuum polarization generated by the
collapsing geometry without inserting an additional quantum flux. However, any flux produced by the evolution remains part of the final stress tensor. These conditions fix the auxiliary solutions without imposing echo symmetry on the quantum state, exactly the same philosophy as in Section~\ref{sec:css}.

All covariant derivatives below refer to the 2d part of \eqref{eq:dss_leading_metric}. The auxiliary equations become
\be
\begin{aligned}
    \ell^2 e^{-2T}\Box U
    &=
    -\frac{2\beta^2}{(1+\beta^2x^2)^2},
    \\
    \ell^2 e^{-2T}\Box W
    &=
    \frac{n^2\beta^2}{4(1+\beta^2x^2)}.
\end{aligned}
    \label{eq:dss_background_auxiliary_equations}
\ee
Once $\beta(T)$ is periodic, the coefficients and the RHS of these equations repeat with the geometry.

The trace is local and state-independent, so it can be evaluated before solving for $U$ and $W$. Substituting the background into \eqref{eq:minkowski_normalized_trace} gives
\be
    \langle T^{(2)a}{}_{a}\rangle
    =
    -\frac{N\hbar e^{2T}\beta^2}{24\pi\ell^2}
    \left[
        \frac{2}{(1+\beta^2x^2)^2}
        +
        \frac{3n^2}{2(1+\beta^2x^2)}
    \right].
    \label{eq:dss_minimal_trace}
\ee
The $x^{-2}$ terms from $(\nabla\phi)^2$ and $\Box\phi$ cancel, leaving a finite center limit. The trace also vanishes at phases where $\beta(T)=0$. Excluding the $e^{2T}$, its dependence on $\beta^2$ gives the same period $\Delta_g$ as the geometry.

It is still open whether vacuum
polarization accumulates between echoes. Even though the trace is local and periodic with the geometry, the complete stress tensor also depends on the solutions of \eqref{eq:dss_background_auxiliary_equations}, which integrate the source over its earlier history.   Returning to the same phase of $\beta(T)$ does not reset these fields
to their previous values. This distinction enters the stress tensor explicitly through the traceless term
\be
    \langle T^{(2)}_{ab}\rangle
    \supset
    -\frac{N\hbar}{4\pi}U
    \left(
        \nabla_a\phi\,\nabla_b\phi
        -\frac{1}{2}g_{ab}(\nabla\phi)^2
    \right).
    \label{eq:dss_traceless_auxiliary_term}
\ee
This term depends on $U$ itself and has zero 2d trace for any $U$. An increment accumulated by $U$ between echoes can therefore change the stress tensor while leaving \eqref{eq:dss_minimal_trace} unchanged. In fact, even at a phase where $\beta(T)=0$, the complete stress tensor can still retain contributions from earlier evolution. Below we calculate this increment and the resulting echo memory. The full variation of the minimal action and the dilaton source $E_\phi$ are collected in Appendix~\ref{app:dss}. The complete source satisfies the trace anomaly and the conservation law \eqref{eq:reduced_ward_identity} with $E_f=0$.

\paragraph{Echo memory and quantum growth.}

Once $\beta(T)$ has entered its periodic regime, the auxiliary equations repeat from one period to the next. Their solutions can still accumulate contributions from earlier echoes. A fixed increment per echo produces a term linear in the elapsed time. If an accumulated contribution is integrated again, higher powers can appear. A finite sequence of such accumulations gives the polynomial factors of a generalized Floquet solution.

For this periodic and polynomial part of the source, the inverse transverse area gives the form introduced in \eqref{eq:generalized_echo_growth} for the $D$-dimensional renormalized stress tensor itself
\be
    \langle T^{(D)}_{\mu\nu}\rangle
    =
    e^{nT}
    \sum_{j=0}^{J}
    (T-T_i)^j P_{\mu\nu}^{(j)}(T,x),
    \qquad
    P_{\mu\nu}^{(j)}(T+\Delta_s,x)
    =
    P_{\mu\nu}^{(j)}(T,x).
    \label{eq:dss_general_quantum_source}
\ee
The profiles $P_{\mu\nu}^{(j)}$ describe the variation within each echo, and $\Delta_s$ is their period. For a source preserving scalar reflection, $\Delta_s=\Delta_g=\Delta/2$. If only the full echo is preserved, $\Delta_s=\Delta$. The highest power $J$ is the Jordan degree, as we have described near~\eqref{eq:generalized_echo_growth}. The auxiliary equations and state conditions must determine whether the source belongs to this class and which degree is realized.

We now determine this structure in the minimal theory~\eqref{eq:minimal_anomaly_theory}. In the periodic regime, the derivatives of $U$ are periodic while the field itself acquires a constant increment. It can therefore be written as\footnote{The first few echoes depend on how $\beta(T)$ is switched on, so this form need not hold during the initial preparation.}
\be
    U(T,x)
    =
    U_{\rm per}(T,x)+m_U(T-T_i),
    \qquad
    U_{\rm per}(T+\Delta_s,x)
    =
    U_{\rm per}(T,x).
    \label{eq:dss_affine_auxiliary_field}
\ee
The coefficient $m_U$ is fixed by the field equation and the boundary conditions. Integrating the first of \eqref{eq:dss_background_auxiliary_equations} over the radial interval and one echo gives
\be
    m_U
    =
    \frac{U(T+\Delta_s,x)-U(T,x)}{\Delta_s}
    =
    \frac{2}{\Delta_s}
    \int_T^{T+\Delta_s}
    d\tau\,
    \frac{\beta^2(\tau)}{1+\beta^2(\tau)}.
    \label{eq:dss_memory_coefficient}
\ee
The derivation is given in Appendix~\ref{app:dss}. Every nontrivial smooth periodic $\beta(T)$ has $0<m_U<2$. Each echo therefore leaves a nonzero increment $m_U\Delta_s$ in $U$, which contributes directly to the accumulated vacuum polarization.

The term containing $U$ in \eqref{eq:dss_traceless_auxiliary_term} retains the factor $m_U(T-T_i)$ and gives a linear contribution to $\langle T^{(2)}_{ab} \rangle$. The corresponding dilaton source contains
$E_\phi\supset-N\hbar m_U(T-T_i)\Box\phi/(4\pi)$,
which determines the angular contribution. Using \eqref{eq:spherical_uplift}, these terms give
\be
    \langle T^{(D)}_{\mu\nu}\rangle
    =
    e^{nT}
    \left[
        (T-T_i)P_{\mu\nu}^{(1)}(T,x)
        +
        P_{\mu\nu}^{(0)}(T,x)
    \right].
    \label{eq:dss_minimal_memory_source}
\ee
The complete coefficient of the accumulated term is
\be
\begin{aligned}
    P_{xx}^{(1)}
    &=
    -\frac{N\hbar n^2m_U}
    {32\pi\Omega_n\ell^n x^{n+2}},
    &
    P_{Tx}^{(1)}
    &=-xP_{xx}^{(1)},
    \\
    P_{TT}^{(1)}
    &=
    \left[
        x^2+\frac{1+\beta^2x^2}{1+\beta^2}
    \right]P_{xx}^{(1)},
    \\
    P_{ij}^{(1)}
    &=
    -\frac{2x^2}{n(1+\beta^2x^2)}
    P_{xx}^{(1)}\gamma_{ij},
    &
    P_{ai}^{(1)}
    &=0.
\end{aligned}
    \label{eq:dss_memory_components}
\ee
Here $a\in\{T,x\}$ and $\gamma_{ij}$ is the metric on the unit $n$-sphere. The angular components follow from the linear term in $E_\phi$. The profile $P_{\mu\nu}^{(0)}$ contains the periodic auxiliary fields and their derivatives, including the constant derivatives of the linear terms. In particular, $W$ enters the stress tensor only through derivatives, so its linear part contributes to $P_{\mu\nu}^{(0)}$. This profile is determined by the same action and boundary conditions. 

The minimal theory therefore realizes $J=1$, and
the quantum Lyapunov exponent is
\be
    \omega_q=n=D-2.
    \label{eq:dss_quantum_growth_rate}
\ee
As in CSS, the inverse transverse area supplies this growth after the time dependence of the 2d source has been determined. Memory changes the amplitude through the number of elapsed echoes, yet any finite polynomial leaves the exponential growth rate unchanged.

The periodic profiles can also be resolved into Fourier harmonics,
\be
    P_{\mu\nu}^{(j)}(T,x)
    =
    \sum_{m\in\mathbb Z}
    P_{\mu\nu,m}^{(j)}(x)e^{2\pi imT/\Delta_s},
    \qquad
    \omega_{q,m}
    =
    D-2+\frac{2\pi im}{\Delta_s}.
    \label{eq:dss_general_harmonics}
\ee
This may look like an infinite tower of complex growing modes, but
the different values of $m$ do not represent independent
instabilities. After one period,
\be
    e^{\omega_{q,m}\Delta_s}
    =
    e^{(D-2)\Delta_s}e^{2\pi im}
    =
    e^{(D-2)\Delta_s},
\ee
so every harmonic has the same Floquet multiplier and the same real
growth rate. The integer $m$ only resolves the variation of a single
growing source within an echo. For the reflection-symmetric sources considered here,
$\Delta_s=\Delta_g=\Delta/2$, giving harmonic spacing
$4\pi i/\Delta$. Without the scalar half-echo symmetry, a generic DSS source has period $\Delta_s=\Delta$ and spacing $2\pi i/\Delta$.

However, the quantum source we have constructed from the minimal theory fails a crucial criterion. \emph{Smoothness} of the full $D$-dimensional stress tensor throughout the
self-similar domain is a necessary condition for a physical quantum
source. The classical geometry is regular at every finite $T$,
including the center and the past SSH. The minimal theory fails this requirement. At phases with
$\beta\neq0$, \eqref{eq:dss_minimal_trace} approaches a nonzero value
as $x\to0$. The trace over the 2d $(T,x)$ sector of the $D$-dimensional stress tensor is therefore divided by $\Omega_n r_*^n$ and diverges at the regular center. This divergence makes the minimal source unphysical for the present problem. 

In particular, smooth auxiliary fields
$U$ and $W$ could only ensure a regular 2d stress tensor, but physical smoothness concerns the full $D$-dimensional stress tensor. Furthermore, the trace is state-independent, hence changing the homogeneous auxiliary data cannot remove the
divergence. The accumulated memory term is traceless in 2d, so this trace obstruction is also independent of the memory.\footnote{This does not conflict with the CSS construction of
Section~\ref{sec:css}. The self-similar domain studied there contains
no regular timelike spherical center and is instead matched to the
flat past across a null boundary, so center smoothness is not a bottleneck. By contrast, the constant nonzero
$\beta$ members of the present family are CSS geometries with a
regular center, and the same obstruction applies to them. For these
solutions, $m_U=2\beta^2/(1+\beta^2)$. However, for CSS, $m_U$ should not be interpreted as an increment per echo. Since there are no echoes in
this limit, it simply measures the linear drift of $U$.}

This problem has a direct precedent in our earlier 3d
critical spacetime calculation~\cite{Tomasevic:2025clf,Tomasevic:2025kqy,Wu:2026xbr}, where the self-similar domain also contains a
regular center. There, regularity at the center and the past light cone fix the quantum state. Suitable finite terms
then give a smooth full stress tensor.

The present large-$D$ problem is considerably more restrictive.
Conversion from the 2d source to the $D$-dimensional stress tensor
introduces the factor $r_*^{-n}$. It is therefore not enough for the
2d source to remain finite at the center. It must vanish with powers
of $r_*$ that increase with $n=D-2$. We impose this stronger smoothness requirement in regular coordinates as a central criterion, and construct one-loop
actions that satisfy it. The resulting profiles turn out to be echo-periodic, realizing $J=0$.

\subsection{Smooth quantum sources and the trapping criterion}
\label{subsec:dss-memory-center}

The classical DSS geometry is smooth throughout the self-similar
interior at every finite $T$. We therefore promote smoothness as
a central criterion for constraining both the functional content of
the one-loop action and the choice of quantum state.

\paragraph{Regularity and the one-loop action.}

The relevant condition concerns the full $D$-dimensional stress
tensor. It is most transparent in a local orthonormal frame, where
regularity can be imposed directly on the energy density, flux, and
pressures measured by an observer. The same frame will also make the construction of the smooth source and its relation to trapping
particularly simple. With $L=\ell e^{-T}$ and
$r_*=Lx$, we choose
\be
    \partial_{\hat0}
    =
    \frac{1}{L}
    \sqrt{\frac{1+\beta^2}{1+\beta^2x^2}}
    \left(\partial_T+x\partial_x\right),
    \qquad
    \partial_{\hat1}
    =
    \frac{1}{L}\partial_x.
    \label{eq:dss_orthonormal_frame}
\ee
These vectors satisfy
$g(\partial_{\hat a},\partial_{\hat b})=\eta_{\hat a\hat b}$.
The timelike vector $\partial_{\hat0}$ follows worldlines of fixed
areal radius, while $\partial_{\hat1}$ is the outward radial unit
vector. The remaining orthonormal directions lie along the transverse
sphere. A hatted index denotes contraction with this frame, so
$\langle T_{\hat a\hat b}\rangle
=e_{\hat a}{}^a e_{\hat b}{}^b\langle T_{ab}\rangle$.

In this frame, the full stress tensor satisfies
\be
    \langle T^{(D)}_{\hat a\hat b}\rangle
    =
    \frac{\langle T^{(2)}_{\hat a\hat b}\rangle}
         {\Omega_n r_*^n}, \qquad
    \langle T^{(D)}_{\hat i\hat j}\rangle
    =
    -\frac{E_\phi}{2\Omega_n r_*^n}\delta_{ij},
    \qquad
    \langle T^{(D)}_{\hat a\hat i}\rangle=0.
    \label{eq:dss_physical_center_components}
\ee
Here $\hat a,\hat b\in\{\hat0,\hat1\}$ and
$\hat i,\hat j$ denote the angular directions.
The components
$\langle T^{(D)}_{\hat0\hat0}\rangle$,
$\langle T^{(D)}_{\hat0\hat1}\rangle$, and
$\langle T^{(D)}_{\hat1\hat1}\rangle$
are the locally measured energy density, radial energy flux, and
radial pressure, while
$-E_\phi/(2\Omega_n r_*^n)$ is the angular pressure.

At a smooth center, the 2d source must first compensate for the
factor $r_*^{-n}$ introduced in
\eqref{eq:dss_physical_center_components}. Since $r_*=Lx$ at fixed finite $T$,
\eqref{eq:dss_physical_center_components} gives
\[
    \langle T^{(D)}_{\hat0\hat0}\rangle
    =
    \frac{\langle T^{(2)}_{\hat0\hat0}\rangle}
         {\Omega_n r_*^n},
    \qquad
    \langle T^{(D)}_{\hat1\hat1}\rangle
    =
    \frac{\langle T^{(2)}_{\hat1\hat1}\rangle}
         {\Omega_n r_*^n},
    \qquad
    \langle T^{(D)}_{\hat i\hat i}\rangle
    =
    -\frac{E_\phi}{2\Omega_n r_*^n}.
\]
At fixed finite $T$, $r_*=Lx$, so the conversion in
\eqref{eq:dss_physical_center_components} introduces a factor
$x^{-n}$. Finiteness at the center therefore requires
$\langle T^{(2)}_{\hat0\hat0}\rangle$,
$\langle T^{(2)}_{\hat1\hat1}\rangle$, and $E_\phi$ to vanish at
least as $x^n$. The scalar source $E_f/(\Omega_n r_*^n)$ requires
the same falloff for $E_f$.

Smoothness requires more than finiteness. Across a regular center,
the outward radial direction reverses, so
$\langle T^{(D)}_{\hat0\hat1}\rangle$ changes sign and must vanish
at least linearly with $r_*$. By contrast,
$\langle T^{(D)}_{\hat1\hat1}\rangle
-\langle T^{(D)}_{\hat i\hat i}\rangle$
is unchanged by this reversal and vanishes at the center, where the
radial and angular directions become equivalent. Its smooth expansion
therefore begins at order $r_*^2$. Using
\eqref{eq:dss_physical_center_components}, these conditions become
\be
\begin{aligned}
    \langle T^{(2)}_{\hat0\hat0}\rangle,\quad
    \langle T^{(2)}_{\hat1\hat1}\rangle,\quad
    E_\phi,\quad E_f
    &=O(x^n),
    \\
    \langle T^{(2)}_{\hat0\hat1}\rangle
    &=O(x^{n+1}),
    \\
    \langle T^{(2)}_{\hat1\hat1}\rangle
    +\frac12 E_\phi
    &=O(x^{n+2}).
\end{aligned}
    \label{eq:dss_complete_center_conditions}
\ee
The first line compensates for
the inverse transverse-area factor. In the last line, the leading
terms in $\langle T^{(2)}_{\hat1\hat1}\rangle$ and $E_\phi/2$
must cancel so that the full $D$-dimensional stress tensor is smooth
at the center. These powers give the minimum radial falloff required at the
center. After factoring them out, the remaining radial dependence must
be smooth in $x^2$, as required for a smooth spherically symmetric
source at $r_*=0$. Thus we see a finite 2d source is not sufficient, as the
factor $r_*^{-n}$ requires increasingly high radial falloff as
$D$ increases.

We have seen that the minimal anomaly-induced action
does not satisfy these regularity conditions. Its 
$U$-dependent memory term and its nonvanishing center trace both
violate the required center behavior. The anomaly, however, does not
determine the complete one-loop effective action. As discussed in
Section~\ref{sec:framework}, it fixes $\Gamma_{\rm anom}$, while the
functional content of $\Gamma_W+\Gamma_{\rm ct}$ remains to be
specified together with the quantum state and renormalization
conditions. We therefore take smoothness of the full $D$-dimensional source,
the Minkowski normalization, and the collapse preparation as physical
criteria for this completion. These are imposed directly on covariant functionals in the
one-loop action. Their variation determines
$\langle T^{(2)}_{ab}\rangle$, $E_\phi$, and $E_f$ together, so the
stress tensor and field sources continue to satisfy the conservation law
\eqref{eq:reduced_ward_identity}.

The leading DSS family obeys two useful covariant identities,
\be
    \mathcal E
    \equiv
    (\nabla f)^2
    -2\Box\phi
    +\frac{4}{n}(\nabla\phi)^2,
    \qquad
    \mathcal C
    \equiv
    \nabla\phi\cdot\nabla f,
    \qquad
    \mathcal E_*=\mathcal C_*=0.
\label{eq:dss_background_combinations}
\ee
The first relates the scalar kinetic term to the dilaton geometry,
while the second expresses the orthogonality of the scalar and
dilaton gradients on the leading background. Because $\mathcal E_*=\mathcal C_*=0$, these combinations can be
included without changing the auxiliary equation on the DSS
background. Their first variations are nonzero, and we use this freedom
to make the same nonlocal auxiliary history enter the physical source
in a different way.

For compactness, define the two kinetic invariants
\be
    X\equiv (\nabla\phi)^2,
    \qquad
    Z\equiv (\nabla f)^2,
\label{eq:dss_finite_action_invariants}
\ee
and introduce the dimensionless combination
\be
    \sigma=-\frac{nZ}{4X+nZ}, \qquad \sigma_*=\beta^2x^2.
\label{eq:dss_sigma_definition}
\ee
Thus $\sigma$ is built entirely from covariant scalar invariants,
but on this background reproduces the combination $\beta^2x^2$.
This allows the action to depend on this combination without
introducing $\beta(T)$ or the coordinate $x$ by hand.
Consider the Weyl-invariant contribution
\be
    \Gamma_W
    =
    \frac{n^2N\hbar}{32\pi}
    \int d^2x\sqrt{-g}
    \left[
        (\nabla\Psi)^2
        +2\Psi\mathcal A
    \right],
    \qquad
    \Box\Psi=\mathcal A.
    \label{eq:dss_periodic_completion}
\ee
where
\be
    \mathcal A
    ={}
    -\left(\Box\phi-\frac{2X}{n}\right)
    \left(\frac{2}{n}+\frac{Z}{2X}\right)
    +\frac{\sigma}{n}\mathcal E
+
    \sigma(\sigma+2)
    \frac{\mathcal C\,\Box f}{X}.
\label{eq:dss_finite_weyl_source}
\ee
Thus $\mathcal A$ is the local covariant source for the auxiliary
field, while the solution
$\Psi=\Box^{-1}\mathcal A$ is a nonlocal functional that accumulates
the past history of this source.
Every term in $\mathcal A$ has Weyl weight $-2$. Thus
$\mathcal A\to e^{-2\lambda}\mathcal A$ under
$g_{ab}\to e^{2\lambda}g_{ab}$ with the scalar fields held fixed,
and \eqref{eq:dss_periodic_completion} is Weyl invariant in 2d. The structure of $\mathcal A$ is chosen so that, on the DSS
background,
\be
    \mathcal A_*=-\frac12 R_*^{(2)}.
\ee
The terms containing $\mathcal E$ and $\mathcal C$ do not change
this result because $\mathcal E_*=\mathcal C_*=0$. They do,
however, contribute when the action is varied. The auxiliary field
therefore obeys the same background equation while its contribution
to the physical source is changed.

Since $\Box U=R^{(2)}$, using the same retarded Green's function and
initial vacuum data gives
\be
    \Psi_*=-\frac{U}{2}.
    \label{eq:dss_periodic_auxiliary_relation}
\ee
Thus $\Psi$ carries the same nonlocal history as $U$. But the completion
changes how this history enters the physical source. Varying \eqref{eq:dss_periodic_completion} gives, among other terms,
\be
\begin{aligned}
    \langle T^{(2)}_{ab}\rangle_W
    &\supset
    \frac{N\hbar U}{4\pi}
    \left[
        \nabla_a\phi\,\nabla_b\phi
        -\frac12 g_{ab}(\nabla\phi)^2
    \right],
    \\
    E_{\phi,W}
    &\supset
    \frac{N\hbar U}{4\pi}\Box\phi.
\end{aligned}
    \label{eq:dss_cancellation_of_singular_memory}
\ee
These cancel the terms proportional to $U$ itself in
the minimal stress tensor and dilaton source. Since
$U=U_{\rm per}+m_U(T-T_i)$, the linear echo memory
drops out of the physical source. The auxiliary
derivatives remain and are periodic in the repeating regime.
 
The Weyl-invariant completion leaves the 2d trace unchanged.
Using the minimal trace relation together with
$\mathcal E_*=0$, the trace on the leading DSS family becomes
\be
    \langle T^{(2)a}{}_{a}\rangle_{{\rm min},*}
    =
    \frac{N\hbar}{24\pi}
    \left(
        R_*^{(2)}+\frac{3n}{2}Z_*
    \right).
\ee
The local part of the one-loop completion can modify this trace.
We choose
\be
    \Gamma_{\rm ct}^{\rm DSS}
    =
    \frac{N\hbar}{24\pi}
    \int d^2x\sqrt{-g}\,
    r^2X
    \left[
        \frac{2R^{(2)}}{n^2}
        +\frac{3Z}{n}
    \right].
    \label{eq:dss_finite_trace_action}
\ee
On the leading DSS family,
$r_*^2X_*=n^2/4$. Its Weyl variation therefore gives
\be
    \langle T^{(2)a}{}_{a}\rangle_{{\rm ct},*}
    =
    -\frac{N\hbar}{24\pi}
    \left(
        R_*^{(2)}+\frac{3n}{2}Z_*
    \right)
    =
    -\langle T^{(2)a}{}_{a}\rangle_{{\rm min},*}.
    \label{eq:dss_finite_trace_cancellation}
\ee
This cancellation is specific to the leading DSS family and does
not remove the anomaly as a functional identity on a general
geometry. The counterterm also preserves the Minkowski normalization. On the
Minkowski background, $R^{(2)}=Z=0$ and $r^2X=n^2/4$ is constant,
so its bulk first variation vanishes. The corresponding variations and their consistency with
\eqref{eq:reduced_ward_identity} are given in
Appendix~\ref{app:dss}.

The linear echo memory and the trace obstruction are therefore absent
in this completion. Full smoothness still
requires the componentwise radial falloff in
\eqref{eq:dss_complete_center_conditions}.

\paragraph{A smooth source throughout the interior.}

To control the remaining radial profiles without changing the
trace cancellation, we use the Weyl-invariant freedom in the
one-loop action. The combinations $\mathcal E$ and $\mathcal C$
in \eqref{eq:dss_background_combinations} both have Weyl weight
$-2$. We therefore consider
\be
    \Delta\Gamma_W
    =
    \frac{n^2N\hbar}{16\pi}
    \int d^2x\sqrt{-g}
    \left[
        e\,\mathcal E+c\,\mathcal C
    \right],
    \label{eq:dss_localized_profile_action}
\ee
where $e$ and $c$ are dimensionless scalar functionals with
zero Weyl weight. This contribution is Weyl invariant and
therefore leaves the 2d trace unchanged. 

We take $e$ and $c$ in
\eqref{eq:dss_localized_profile_action} as covariant scalar
functionals of the fields, and denote their values on the DSS
background by $e(T,x)$ and $c(T,x)$. The radial equations below
determine these background profiles. An admissible one-loop action
must realize them covariantly, without introducing explicit
dependence on the coordinates $T$ and $x$.

Since $\mathcal E_*=\mathcal C_*=0$, the contribution
\eqref{eq:dss_localized_profile_action} vanishes when evaluated
on the leading background. But the first variation does not,
\be
    \delta\Delta\Gamma_W
    \propto
    e\,\delta\mathcal E
    +
    c\,\delta\mathcal C,
\ee
since the terms
$\mathcal E\,\delta e+\mathcal C\,\delta c$ vanish.
The background profiles $e(T,x)$ and $c(T,x)$ can therefore change
the stress tensor and field sources while preserving the vanishing
2d trace. We need two profiles $e(T,x)$ and $c(T,x)$ since there are two independent source functions in this traceless sector.
The symmetric 2d stress tensor together with $E_\phi$ and $E_f$
contains five functions. The trace condition removes one, while the
two components of the conservation law
\eqref{eq:reduced_ward_identity} remove two more.

We take the two independent source profiles to be the energy density
$\Delta\langle T^{(2)}_{\hat0\hat0}\rangle$ and the scalar source
$\Delta E_f$. Direct variation gives
\be
\begin{aligned}
    x\partial_x e
    -
    \frac{1+2\beta^2x^2}{1+\beta^2x^2}\,e
    &=
    \frac{16\pi L^2x^2}{n^3N\hbar}
    \Delta\langle T^{(2)}_{\hat0\hat0}\rangle,
    \\
    x\partial_x c
    -
    \frac{c}{1+\beta^2x^2}
    &=
    \frac{32\pi L^2x^2}{n^3N\hbar}
    \left(
        \Delta E_f-\Delta E_f^{(e)}
    \right),
\end{aligned}
    \label{eq:dss_radial_profile_equations}
\ee
where
\be
    \Delta E_f^{(e)}
    =
    -\frac{n^2N\hbar}{8\pi}
    \nabla_a\left(e\,\nabla^a f_*\right)
    \label{eq:dss_scalar_source_from_e}
\ee
is the scalar source already generated by $e$. On the leading background,
$\partial_{\hat0}\phi_*=0$ and
$\partial_{\hat1}f_*=0$. As a result, $c$ does not contribute
to the energy density, and the first equation determines $e$
directly from
$\Delta\langle T^{(2)}_{\hat0\hat0}\rangle$.
Once $e$ is fixed, its contribution
\eqref{eq:dss_scalar_source_from_e} is known, and the second
equation determines $c$ from the remaining scalar source.
The system is therefore triangular.

At fixed $T$, the general solutions of
\eqref{eq:dss_radial_profile_equations} are
\be
\begin{aligned}
    e(T,x)
    ={}&
    x\sqrt{1+\beta^2x^2}
    \left[
        C_e(T)
        +
        \frac{16\pi L^2}{n^3N\hbar}
        \int_0^x
        du\,
        \frac{
            \Delta\langle T^{(2)}_{\hat0\hat0}\rangle(T,u)
        }{
            \sqrt{1+\beta^2u^2}
        }
    \right],
    \\
    c(T,x)
    ={}&
    \frac{x}{\sqrt{1+\beta^2x^2}}
    \left[
        C_c(T)
        +
        \frac{32\pi L^2}{n^3N\hbar}
        \int_0^x
        du\,
        \sqrt{1+\beta^2u^2}
        \left(
            \Delta E_f(T,u)
            -
            \Delta E_f^{(e)}(T,u)
        \right)
    \right].
\end{aligned}
\label{eq:dss_radial_profile_solution}
\ee
Near the center, the homogeneous parts behave as $C_e(T)x+O(x^3)$ and $C_c(T)x+O(x^3)$. These linear terms are incompatible with smooth scalar profiles at
a regular center. Equivalently, requiring $e/x^2$ and $c/x^2$ to
remain finite fixes
\be
    C_e(T)=C_c(T)=0.
\ee
Center regularity thus fixes the radial solutions uniquely.
Moreover, $1+\beta^2x^2$ remains strictly positive throughout
$0\leq x\leq1$, including phases where $\beta=0$. No independent boundary condition is needed for the past SSH at
$x=1$. Smooth input
profiles therefore determine regular $e$ and $c$ from the center
to the past SSH. 

The source generated by
$\Gamma_{\rm min}+\Gamma_W+\Gamma_{\rm ct}^{\rm DSS}$ is already
fixed by the action and state prescription, but its individual
components need not yet satisfy the required center falloff. Near
$x=0$, the lower radial powers are canceled by choosing
\be
\begin{aligned}
    \Delta\langle T^{(2)}_{\hat0\hat0}\rangle
    &=
    -\left.
    \langle T^{(2)}_{\hat0\hat0}\rangle
    \right|_{\Gamma_{\rm min}+\Gamma_W+\Gamma_{\rm ct}^{\rm DSS}}
    +O(x^n),
    \\
    \Delta E_f
    &=
    -\left.
    E_f
    \right|_{\Gamma_{\rm min}+\Gamma_W+\Gamma_{\rm ct}^{\rm DSS}}
    +O(x^n).
\end{aligned}
\ee
The radial equations then determine the corresponding center-regular
profiles $e(T,x)$ and $c(T,x)$.\footnote{More generally, for any prescribed
$\Delta\langle T^{(2)}_{\hat0\hat0}\rangle$ and $\Delta E_f$
that are smooth and have the required center behavior,
\eqref{eq:dss_radial_profile_solution} gives unique center-regular
profiles $e$ and $c$. In particular, if both prescribed changes begin at $O(x^n)$,
then $e,c=O(x^{n+2})$ and
$\Delta E_f^{(e)}=O(x^{n+2})$, so the radial solution does not
generate lower powers or spoil the center falloff.} 

The complete one-loop action used in this construction is therefore
\be
    \Gamma_{\rm min}
    +\Gamma_W
    +\Gamma_{\rm ct}^{\rm DSS}
    +\Delta\Gamma_W .
\ee
We now verify all the center conditions in
\eqref{eq:dss_complete_center_conditions}. The contribution of
$\Delta\Gamma_W$ to the $D$-dimensional stress tensor can be written
explicitly. For periodic $e$ and $c$, the 2d covariant stress tensor generated by
$\Delta\Gamma_W$ is periodic in $T$. The only remaining similarity
scale in its $D$-dimensional uplift is therefore
$r_*^{-n}=\ell^{-n}e^{nT}x^{-n}$. We write
\be
    \Delta\langle T^{(D)}_{\mu\nu}\rangle
    =
    e^{nT}\Delta P_{\mu\nu}(T,x).
\ee
On the leading background,
\be
\begin{aligned}
    \Delta P_{xx}
    &=
    \frac{n^3N\hbar}
         {16\pi\Omega_n\ell^n x^{n+2}}
    \left[
        x\partial_x e
        -
        \frac{1+2\beta^2x^2}{1+\beta^2x^2}\,e
    \right],
    \\
    \Delta P_{Tx}=\Delta P_{xT}
    &=
    \frac{n^3N\hbar}
         {16\pi\Omega_n\ell^n x^{n+1}}
    \left[
        \partial_T e
        +
        \frac{1+2\beta^2x^2}{1+\beta^2x^2}\,e
        +
        \frac{c}{2}\partial_T f_*
    \right],
    \\
    \Delta P_{TT}
    &=
    \frac{1-x^2}{1+\beta^2}\Delta P_{xx}
    -2x\Delta P_{Tx}.
\end{aligned}
    \label{eq:dss_regular_source_components}
\ee
The last relation follows from the vanishing 2d trace of
$\Delta\Gamma_W$. Since $\Gamma_{\rm ct}^{\rm DSS}$ cancels the
trace of $\Gamma_{\rm min}$ on the leading background, while
$\Gamma_W$ and $\Delta\Gamma_W$ are Weyl invariant, the complete
source also obeys
\be
    \langle T^{(D)}_{TT}\rangle
    +2x\langle T^{(D)}_{Tx}\rangle
    =
    \frac{1-x^2}{1+\beta^2(T)}
    \langle T^{(D)}_{xx}\rangle .
    \label{eq:dss_regular_source_trace_relation}
\ee 
The angular components are fixed by the dilaton variation of the
same action. The dilaton variation gives
\be
\begin{aligned}
    \Delta E_\phi
    &=
    -\frac{n^2N\hbar}{8\pi}
    \left[
        \Box e
        +\frac{4}{n}
        \nabla_a\left(e\,\nabla^a\phi_*\right)
        +\frac12
        \nabla_a\left(c\,\nabla^a f_*\right)
    \right],
    \\
    \Delta P_{ij}
    &=
    -\frac{L^2\Delta E_\phi}
          {2\Omega_n\ell^n x^{n-2}}\gamma_{ij},
    \qquad
    \Delta P_{ai}=0.
\end{aligned}
    \label{eq:dss_finite_angular_source}
\ee
All derivatives in these expressions use the classical 2d metric.
Together with \eqref{eq:dss_scalar_source_from_e} and
\eqref{eq:dss_radial_profile_solution}, these equations determine
every component generated by $\Delta\Gamma_W$ once
$\Delta\langle T^{(2)}_{\hat0\hat0}\rangle$ and $\Delta E_f$
have been fixed. There is no further freedom in the flux or
pressures.

To recap, the center conditions are checked on the sum of these contributions
and the source generated by
$\Gamma_{\rm min}+\Gamma_W+\Gamma_{\rm ct}^{\rm DSS}$.
Thus the same functions $e$ and $c$ that correct
$\langle T^{(2)}_{\hat0\hat0}\rangle$ and $E_f$ also fix the
accompanying radial flux, radial pressure, and angular pressure.
Substitution into
\eqref{eq:dss_regular_source_components} and
\eqref{eq:dss_finite_angular_source} gives the remaining conditions
in \eqref{eq:dss_complete_center_conditions}. The conservation law
\eqref{eq:reduced_ward_identity} is satisfied by construction because
all components follow from the same covariant action. The radial equations determine the background profiles $e(T,x)$ and
$c(T,x)$. In the action, these profiles must be realized as covariant,
Weyl-invariant scalar functionals of the fields. The covariant
variation of $\Delta\Gamma_W$, the center-regular radial
reconstruction, and the half-echo parity of $e$ and $c$ are derived
in Appendix~\ref{app:dss}.

Now, the full physical stress tensor takes the form
\be
    \langle T^{(D)}_{\mu\nu}\rangle
    =
    e^{nT}P_{\mu\nu}(T,x),
    \qquad
    P_{\mu\nu}(T+\Delta_s,x)
    =
    P_{\mu\nu}(T,x).
    \label{eq:dss_periodic_quantum_source}
\ee
With scalar reflection symmetry, $e$ is even and $c$ is odd
under a half echo, so the source has the geometric period
$\Delta_s=\Delta_g$. The construction applies separately to every integer $n=D-2\geq2$. Thus the result is an
ordinary Floquet source with
\be
    J=0,
    \qquad
    \omega_q=D-2.
\ee
This agrees with the bounded echo-periodic quantum behavior
conjectured in \cite{Tomasevic:2025clf,Tomasevic:2025kqy} for DSS spacetimes.

The construction shows that a smooth source can realize $J=0$,
but smoothness by itself does not prove that this behavior is
unique. This is because center regularity constrains the behavior as $x\to0$ at each
finite $T$, but not the dependence on $T$ from one echo to the
next. The result
nevertheless provides direct support for the same picture on the
genuine 4d Choptuik spacetime, where one expects the universal
growth to be modulated by the exact echoing symmetry. 

\paragraph{Implications for trapping.}

For any chosen $\beta(T)$, the source generated by
$\Gamma_{\rm min}+\Gamma_W+\Gamma_{\rm ct}^{\rm DSS}$ is completely
fixed. The role of $\Delta\Gamma_W$ is first to supply the
additional radial dependence required by the center conditions.
These conditions do not, however, fix all smooth finite terms in
$\Delta\Gamma_W$. Contributions that already satisfy the required
radial falloff can still be added without changing the regularity or
the Minkowski normalization. The backreaction on the geometry must
therefore be determined from the complete source after this remaining
finite part of the one-loop action has been specified.

A smooth quantum source allows us to ask whether the geometry
moves toward a marginal surface. The quantity
$H=(\nabla r)^2$ measures this directly. In the untrapped region, we use polar-areal coordinates with
areal radius $r$,
\be
    ds_D^2
    =
    -\mathcal N^2(t,r)\,dt^2
    +\frac{dr^2}{H(t,r)}
    +r^2d\Omega_n^2,
\ee
The effective energy density $\varepsilon_{\rm eff}$ entering the
radial Einstein equation is the total energy measured by observers
at fixed $r$. Its first-order change is
\be
    \delta\varepsilon_{\rm eff}
    =
    \langle T^{(D)}_{\hat0\hat0}\rangle
    +
    \delta\varepsilon_f.
    \label{eq:dss_total_effective_energy}
\ee
The first term is the complete one-loop energy density,
including the contributions from
$\Gamma_{\rm min}+\Gamma_W+\Gamma_{\rm ct}^{\rm DSS}
+\Delta\Gamma_W$.
The second term is the induced change in the classical scalar
energy. It must also be included because the quantum source
changes both the scalar field and the metric used to measure
its energy.

Linearizing the exact finite-$D$ radial constraint about the
classical background, and only then taking the large-$D$ limit,
gives
\be
    \delta H(t,r)
    =
    -\frac{16\pi G_D}{nr^{n-1}}
    \int_0^r
    ds\,s^n
    \delta\varepsilon_{\rm eff}(t,s).
    \label{eq:dss_cumulative_energy_response}
\ee
Regularity at the center fixes the integration function to
zero. The cumulative effective energy is the weighted integral
appearing in \eqref{eq:dss_cumulative_energy_response}.
If this integral is positive, $\delta H<0$ and the outgoing
null expansion moves toward zero. If it is negative,
$\delta H>0$ and the geometry moves away from marginality.

At large $D$, the factor $s^n$ strongly weights the region
nearest the sphere at radius $r$. For an energy profile that
varies on a radial scale of order $r$, the integral is dominated
by a thin layer immediately inside that sphere, with
$r-s=O(r/n)$. Thus the change in $H(t,r)$ is controlled mainly
by the effective energy in this local radial layer, while its
magnitude continues to vary with the echo phase. Large $D$ therefore localizes the geometric
effect of the source while retaining its evolution between
echoes.  The finite-$D$ radial
constraint, its relation to the outgoing null expansion, and
the large-$D$ localization are derived in
Appendix~\ref{app:dss}.

Center regularity still leaves a smooth finite freedom. Let $e_0(T,x)$ and $c_0(T,x)$
denote a center-regular profile for which the complete source
satisfies \eqref{eq:dss_complete_center_conditions}. These profiles
may still be changed by terms that already have the
required center behavior. For example, consider
\be
    \delta e_\pm(T,x)
    =
    \pm\frac{\beta^2(T)}{n^2}x^{n+2},
    \qquad
    \delta c_\pm(T,x)=0.
    \label{eq:dss_regular_finite_sign_choices}
\ee
The profiles $e_0+\delta e_\pm$ and $c_0+\delta c_\pm$ remain
center regular and preserve the Minkowski normalization. The
corresponding changes in the direct $D$-dimensional one-loop
energy density are
\be
    \delta\langle T^{(D)}_{\hat0\hat0}\rangle_\pm
    =
    \pm
    \frac{N\hbar n\beta^2}
         {16\pi\Omega_n L^{n+2}}
    \left(
        n+\frac{1}{1+\beta^2x^2}
    \right).
    \label{eq:dss_regular_finite_energy_sign}
\ee
These finite terms can raise or lower the quantum energy.
Their effect on trapping is determined by the complete
cumulative effective energy in
\eqref{eq:dss_cumulative_energy_response}, including
$\delta\varepsilon_f$. 

However, the leading DSS background has $H_*=1$. A controlled
deformation with $|\delta H|\ll1$ therefore remains away from
a marginal surface. Following the competition toward trapping
also requires the physical classical growing perturbation on
this same background. This mode has not been determined for the analytic large-$D$
family used here. The unstable modes in \cite{Ecker:2026joz}
are tied to the numerically determined finite-$D$ critical
backgrounds and cannot be inserted directly into the present
leading large-$D$ geometry. A DSS counterpart of the CSS quantum wall therefore requires the classical growing mode
and the full semiclassical backreaction.

There is nevertheless a suggestive indication from the
finite-$D$ critical branch. The exponent denoted $\gamma$ in
\cite{Ecker:2026joz} corresponds to the length exponent
$\gamma_L=1/\omega_c$ in our notation, and its large-$D$
behavior is compatible with $\gamma_L\to 1/2$. If this limit
persists on the branch that approaches the analytic large-$D$
geometry, then
\be
    \omega_c\to 2,
    \qquad
    \frac{\omega_q}{\omega_c}
    \sim \frac{D-2}{2}.
\ee
The quantum source would then grow parametrically faster than
the classical instability. This is qualitatively different from
the CSS sector in Section~\ref{sec:css}, where $\omega_c\sim(D-2)/2$ and
$\omega_q=2\omega_c$. The comparison remains suggestive because
the finite-$D$ DSS branch also selects a dimension-dependent
echoing period, while the present calculation holds $\beta(T)$
and $\Delta$ fixed. A direct test therefore requires the
classical growing mode on the same large-$D$ DSS background.

\section{Discussion and Outlook}
\label{sec:discussion}

The CSS and DSS calculations establish a common quantum growth rate $\omega_q=D-2$. In CSS, physical boundary conditions select a vacuum polarization state. Its backreaction overtakes the classical departure under sufficiently precise tuning, and the geometry develops a local future MOTS at finite radius. At large $D$, the competition becomes a Bessel problem, concentrating the onset of trapping within a thin quantum wall. In DSS, the same quantum growth can retain memory of previous echoes. The minimal one-loop theory accumulates memory but produces a stress tensor singular at the regular center. We construct effective actions whose stress tensors remain smooth throughout the self-similar interior and have bounded echo-periodic structure. Smoothness therefore provides a physical condition that can constrain the quantum behavior between echoes.

Together, the results show that vacuum polarization destabilizes classical self-similarity, while CSS and DSS channel the same large-$D$ quantum growth into fundamentally different semiclassical dynamics. Large $D$ exposes the common kinematical origin of this instability and separates the radial scales on which it acts. Physical boundary conditions and regularity guide the construction of the quantum source, suggesting a route toward building a smooth quantum state governed by the echoing symmetry on the genuine 4d Choptuik spacetime. The form of self-similarity therefore becomes a decisive ingredient in the quantum fate of critical collapse.

Finally, we discuss open problems suggested by this work. See also~\cite{Tomasevic:2025clf,Tomasevic:2025kqy,Wu:2026xbr} for a broader set of questions.

\begin{itemize}

    \item \textbf{Quantum effects in the 4d Choptuik spacetime.}
    The most important extension is to construct a smooth one-loop stress tensor on the Choptuik spacetime \cite{Choptuik:1992jv,Gundlach:1995kd,Gundlach:1996eg,Ecker:2026joz}. This would determine whether the quantum growth has exponent $\omega_q=2$ and whether regularity throughout the self-similar interior gives rise to echo-periodic structure. The numerical background and its classical growing perturbation would then allow the competition with quantum backreaction to be studied directly, testing whether vacuum polarization shifts the threshold and produces trapping at a finite radius.

    \item \textbf{The exterior spacetime and critical dynamics.}
    Our analysis has been restricted to the self-similar interior. Extending the semiclassical evolution into an asymptotically flat exterior following the approach in~\cite{Wu:2026xbr} would determine whether an event horizon forms and how the local onset radius $r_h$ is related to the final black hole mass. Following the full evolution would also establish whether the finite-radius onset produces a final mass gap and whether the local Type I-like behavior extends to a genuine Type I transition, with a metastable critical solution and an associated lifetime scaling law.

    \item \textbf{Holographic observables.}
Classical critical collapse and self-similar naked singularities in AdS can leave echoing and singular signals in boundary observables \cite{Chesler:2019ozd,Emparan:2021ewh}. An AdS exterior extension would test whether quantum backreaction makes these signals finite and how much of their echo structure survives at the boundary. Large-$D$ braneworld models of collapse and evaporation \cite{Emparan:2023dxm} provide a complementary setting for asking whether quantum effects interrupt self-similar collapse in a strongly coupled matter theory.

\item \textbf{Black strings and pure gravity.}
Black string dynamics provides another setting in which gravity generates thin neck regions. In 5d, unstable strings develop a cascade toward a pinching neck \cite{Lehner:2010pn}, while sufficiently large $D$ admits stable nonuniform final strings \cite{Emparan:2015gva}. The similarity between some reduced equations suggests that large-$D$ methods developed for black strings may also help organize nonlinear critical collapse \cite{Licht:2022rke,Clark:2025tqi}. It would be interesting to ask whether vacuum polarization changes the evolution of a shrinking neck and how this compares with string theory resolutions \cite{Emparan:2024mbp}. Extending the mechanism to vacuum critical collapse \cite{Bizon:2005cp} would instead require a one-loop treatment of the gravitational degrees of freedom themselves.

\end{itemize}

\paragraph{Acknowledgments} The author would like to thank Guanda Lin and Gustavo J. Turiaci for useful discussions on critical collapse. The author would also like to thank the anonymous referee of~\cite{Wu:2026xbr} for suggesting the large-$D$ limit as an interesting direction in which to extend the quantum mechanism. This work was supported by the DOE Early Career Award DE-SC0026287. This work also benefited from access to OpenAI models provided through the ChatGPT for Academic Researchers program. They contributed substantially to the analytic and numerical developments in the CSS analysis and to the construction developed in the DSS sector, especially the covariant one-loop action, which the author found particularly creative and insightful. The author has checked and interpreted the results and remains fully responsible for the content of this work.

\appendix

\section{CSS backreaction and the large-$D$ limit}
\label{app:css-response}

This appendix records the equations and numerical checks used in Section~\ref{subsec:css-horizon}. We also derive the large-$D$ limit and the first correction to the trapping radius. Primes denote derivatives with respect to $x$ unless stated otherwise.

\subsection*{Backreaction equations and flat-past data}

For a profile $Z_k=(r_k,\rho_k,f_k)^{\rm T}$ with
time dependence $e^{kT}$, the scalar equation is
\be
    \frac{4}{n}f_k''
    +2\left(\frac{2k}{n}-\chi\right)f_k'
    +k(1-\chi)f_k
    +4f_*'r_k'
    +2kf_*'r_k=0.
\label{eq:css_perturbation_scalar}
\ee
The three Einstein equations are
\be
\begin{split}
    2kr_k'+k(k-\chi)r_k+2\rho_k'
    +k(1+\chi)\rho_k+\frac{2k}{n}f_*'f_k
    =
    \begin{cases}
        0,&k=n/2,\\[2pt]
        -\dfrac{R^{-n}W_0'}{2\pi n},&k=n,
    \end{cases}
\end{split}
\label{eq:css_perturbation_mixed}
\ee
\be
    r_k''-\chi r_k'-(1-\chi)\rho_k'
    +\frac{2}{n}f_*'f_k'
    =
    \begin{cases}
        0,&k=n/2,\\[2pt]
        \dfrac{R^{-n}(W_0'-W_0'')}{4\pi n},&k=n,
    \end{cases}
\label{eq:css_perturbation_constraint}
\ee
and
\be
\begin{split}
    \frac{2}{n}r_k''
    +2\left(\frac{k}{n}-\chi\right)r_k'
    +\left[k(1-\chi)-Q\right]r_k+Q\rho_k
    =
    \begin{cases}
        0,&k=n/2,\\[2pt]
        -\dfrac{R^{-n}W_0''}{2\pi n^2},&k=n.
    \end{cases}
\end{split}
\label{eq:css_perturbation_radial}
\ee
The RHS vanishes for the classical perturbation. For the quantum backreaction it follows from \eqref{eq:css_2d_adapted_stress}, with the overall coupling included in $g_q$. The factor $R^{-n}$ carries the radial dependence associated with the transverse area.

We solve two instances of this system. The classical endpoint profile is the bounded homogeneous solution at $k=n/2$, while the quantum backreaction is the sourced solution at $k=n$. The hatted variables below have the known flat-past falloffs factored out, making the corresponding boundary data finite.

At fixed $T$, the limit $Q\to0$ approaches the null matching surface $v=0$ with $r_*=-u/\sqrt2>0$. For the classical perturbation, define
\be
    \widehat Y_c
    =(u_c,v_c,w_c,f_c,f_c')^{\rm T}
    =
    \left(
        \frac{n^2r_c}{h},
        \frac{nr_c'}{h},
        \frac{\rho_c}{h},
        f_c,
        f_c'
    \right)^{\rm T}.
\label{eq:app_css_classical_stripped_vector}
\ee
The normalization used in the main text is
\be
    \lim_{Q\to0}\widehat Y_c
    =
    \left(
        0,0,-\frac{\sqrt{\alpha}}{4},1,0
    \right)^{\rm T}.
\label{eq:css_classical_past_data}
\ee
At $k=n/2$, the equations become
\be
\begin{aligned}
    u_c'
    &=
    n\left(v_c-\frac{\chi}{2}u_c\right),
    \\
    v_c'
    &=
    \left(\chi-\frac{n}{2}\right)v_c
    -\frac{(1-\chi)(n/2-\chi)}{4}u_c
    -\frac{n^2(1-\chi^2)}{4}w_c
    \\
    &\quad
    -\frac{n^2\sqrt{\alpha}(1-\chi)}{4}f_c
    -n\sqrt{\alpha}\,f_c',
    \\
    w_c'
    &=
    -\frac{v_c}{2}
    -\frac{n/2-\chi}{4n}u_c
    -\frac{n(1+3\chi)}{4}w_c
    -\frac{n\sqrt{\alpha}}{4}f_c,
    \\
    f_c''
    &=
    -\frac{n\sqrt{\alpha}\,h^2}{8}u_c
    -\frac{n\sqrt{\alpha}\,h^2}{2}v_c
    -\frac{n^2(1-\chi)}{8}f_c
    -\frac{n(1-\chi)}{2}f_c'.
\end{aligned}
\label{eq:app_css_classical_stripped_system}
\ee
This form uses \eqref{eq:css_perturbation_constraint} to evolve the radial derivative. The remaining equation \eqref{eq:css_perturbation_radial} provides an independent consistency check.

The indicial polynomial at the flat-past endpoint is
\be
    \lambda^2(\lambda+n)
    \left(\lambda+\frac{n}{2}\right)
    \left(\lambda+\frac{n}{2}-1\right).
\label{eq:app_css_classical_indicial_polynomial}
\ee
The data in \eqref{eq:css_classical_past_data}, together with the exclusion of the logarithmic solution, select the bounded branch. For $n>2$, its first two Frobenius terms are
\be
    \widehat Y_c
    =
    \begin{pmatrix}
        0\\
        0\\
        -\sqrt{\alpha}/4\\
        1\\
        0
    \end{pmatrix}
    +e^x
    \begin{pmatrix}
        \dfrac{\sqrt{\alpha}\,n^2(n-1)}{n+2}\\[5pt]
        \dfrac{\sqrt{\alpha}\,n(n-1)}{2}\\[5pt]
        \dfrac{\sqrt{\alpha}\,n(n^3-4n^2-4n-8)}
              {16(n+1)(n+2)}\\[5pt]
        -n^2/4\\[2pt]
        -n^2/4
    \end{pmatrix}
    +O(e^{2x}).
\label{eq:app_css_classical_frobenius_start}
\ee
Higher coefficients follow recursively from \eqref{eq:app_css_classical_stripped_system}. These series supply the initial data at finite negative $x$.

For the quantum backreaction, define
\be
    s_n=R^{-n}h^2,
    \qquad
    \widehat Y_q
    =
    \left(
        \frac{r_q}{s_n},
        \frac{r_q'}{ns_n},
        \frac{\rho_q}{ns_n},
        \frac{f_q}{ns_nh},
        \frac{f_q'}{n^2s_nh}
    \right)^{\rm T}.
\label{eq:app_css_quantum_stripped_vector}
\ee
The source ratio has the bounded representation
\be
    \widehat m(Q)
    \equiv
    \frac{W_0'}{n^2h^2}
    =
    -\frac{n-1}{16}
    \int_0^1
    \frac{s^{n-1}}{\chi(Qs)}\,ds,
    \qquad
    \widehat m(0)=-\frac{\alpha}{16}.
\label{eq:app_css_bounded_source}
\ee
The exact rescaled system can be written as
\be
    \widehat Y_q'
    =
    n\left(M_n\widehat Y_q+J_n\right),
\label{eq:app_css_quantum_stripped_system}
\ee
where $M_n\widehat Y_q$ describes the homogeneous linear response, while $J_n$ is the forcing generated by the one-loop vacuum polarization. Here
\be
    M_n=
    \begin{pmatrix}
     -1+3\mathcal V_0 & 1 & 0 & 0 & 0\\
     -\mathcal V_0+\dfrac{Q}{2n}
       & -1+\mathcal V_0 & -\dfrac{Q}{2} & 0 & 0\\
     -\dfrac12+\dfrac{1}{2n}-\dfrac{\mathcal V_0}{n}
       & -1 & -2+4\mathcal V_0
       & -\dfrac{\sqrt{\alpha}\,h^2}{2} & 0\\
     0 & 0 & 0 & -\dfrac32+4\mathcal V_0 & 1\\
     -\dfrac{\sqrt{\alpha}}{4}
       & -\dfrac{\sqrt{\alpha}}{2}
       & 0 & -\dfrac{\mathcal V_0}{2}
       & -2+3\mathcal V_0
    \end{pmatrix},
\label{eq:app_css_quantum_matrix}
\ee
and
\be
    J_n
    =
    \left(
        0,
        \frac{\alpha}{64\pi},
        -\frac{\widehat m}{4\pi n},
        0,
        0
    \right)^{\rm T}.
\label{eq:app_css_quantum_forcing}
\ee
The finite equilibrium at $Q=0$ is
\be
    \lim_{Q\to0}\widehat Y_q
    =
    \frac{1}{64\pi}
    \left(
        \alpha,
        \alpha,
        -\frac{3\alpha^2}{4},
        -\frac{\alpha^{3/2}}{4},
        -\frac{3\alpha^{3/2}}{8}
    \right)^{\rm T}.
\label{eq:css_quantum_past_data}
\ee
The original fields still vanish in the flat past because $s_n$ and $h$ vanish there. The nonzero rescaled equilibrium is fixed by the inhomogeneous equations and does not add an incoming quantum pulse. Setting the entire rescaled vector to zero would instead change the boundary data.

The quantum integration uses \eqref{eq:css_perturbation_scalar}, \eqref{eq:css_perturbation_mixed}, and \eqref{eq:css_perturbation_radial}. The remaining Einstein equation is monitored through
\be
    \mathcal C_q
    =
    r_q''-\chi r_q'-(1-\chi)\rho_q'
    +\frac{2}{n}f_*'f_q'
    -\frac{R^{-n}(W_0'-W_0'')}{4\pi n}.
\label{eq:css_unused_quantum_constraint}
\ee
This unused equation provides an independent consistency check on both the quantum source and the integrated backreaction.

\paragraph{Endpoint packet estimate.}
The main calculation uses the normalized endpoint profile at $k=n/2$. A more general classical departure can approach this profile through modes strictly inside the allowed spectrum. Writing $\eta=n/2-k>0$, consider
\be
    \delta Z_c(T,x)
    =
    (p-p_*)\operatorname{Re}
    \int_0^{\eta_{\max}}
    d\eta\,
    \widehat A_n(\eta)
    e^{(n/2-\eta)T}
    Z_{n/2-\eta}(x).
\label{eq:css_endpoint_packet}
\ee
If the endpoint behavior is
\be
    \widehat A_n(\eta)Z_{n/2-\eta}(x)
    =
    \eta^{\beta-1}
    \left[Z_c(x)+O(\eta)\right],
    \qquad
    \beta>0,
\label{eq:css_endpoint_packet_weight}
\ee
Watson's lemma gives
\be
    \delta Z_c(T,x)
    =
    (p-p_*)\Gamma(\beta)
    e^{nT/2}T^{-\beta}
    \left[
        Z_c(x)+O(T^{-1})
    \right].
\label{eq:css_endpoint_packet_asymptotics}
\ee
The leading exponential rate therefore remains $n/2$. Since
$T_h=n^{-1}\log(1/g_q)+O(1)$ at fixed $n$ as $g_q\to0$, the corresponding threshold shift behaves as
\be
    p_*^{\,q}-p_*
    \propto
    g_q^{1/2}
    \left[
        \log\!\left(\frac{1}{g_q}\right)
    \right]^\beta.
\label{eq:css_packet_fold_tuning}
\ee
The square-root power is unchanged, while the packet can add logarithmic dressing. A complete finite-width calculation would require the full $k$ dependence of the radial profiles and is not carried out here.

\subsection*{Radial horizon onset condition}

The factor controlling the outgoing expansion was defined in \eqref{eq:css_outgoing_factor} as
\be
    \mathcal V(T,x)
    =
    \mathcal V_0(x)
    +(p-p_*)e^{nT/2}\mathcal V_c(x)
    +g_qe^{nT}\mathcal V_q(x).
\ee
The stationary onset conditions~\eqref{eq:css_stationary_conditions} can be reduced to a single radial equation. Define
\be
    G_n(x)
    \equiv
    n^2\mathcal V_0
    \left(
        \frac{\mathcal V_0'}{\mathcal V_0}
        -2\frac{\mathcal V_c'}{\mathcal V_c}
        +\frac{\mathcal V_q'}{\mathcal V_q}
    \right).
\label{eq:app_css_radial_function}
\ee
The radial position of the onset is determined by
$G_n(x_h)=0$. For numerical work and for the large-$D$ limit, it is useful to remove the known radial falloffs by defining
\be
    \widehat{\mathcal V}_c
    =
    \frac{n\mathcal V_c}{h}
    =
    v_c+\frac{\mathcal V_0}{n}u_c,
    \qquad
    \widehat{\mathcal V}_q
    =
    \frac{\mathcal V_q}{ns_n}.
\label{eq:app_css_stripped_outgoing_profiles}
\ee
Using
\be
    \frac{s_n'}{s_n}
    -2\frac{h'}{h}
    =
    -n\mathcal V_0,
\label{eq:app_css_stripping_identity}
\ee
the radial function becomes
\be
    G_n
    =
    n^2\mathcal V_0
    \left(
        \frac{\mathcal V_0'}{\mathcal V_0}
        -n\mathcal V_0
        -2\frac{\widehat{\mathcal V}_c'}
                {\widehat{\mathcal V}_c}
        +\frac{\widehat{\mathcal V}_q'}
               {\widehat{\mathcal V}_q}
    \right).
\label{eq:app_css_stripped_radial_function}
\ee

At a stationary zero, differentiating \eqref{eq:css_outgoing_factor} gives
\be
    (\partial_T^2\mathcal V)_h
    =
    \frac{n^2}{2}\mathcal V_{0,h}>0,
    \qquad
    \mathcal V_h''
    -
    \frac{
        [(\partial_T\mathcal V')_h]^2
    }{
        (\partial_T^2\mathcal V)_h
    }
    =
    \frac{G_n'(x_h)}{n^2}.
\label{eq:css_spatial_fold_curvature}
\ee
The second quantity is the remaining radial curvature after minimizing over the local time displacement. A simple root with $G_n'(x_h)>0$ therefore gives a positive-definite Hessian in the $(T,x)$ plane.

To display the local onset explicitly, keep $g_q$ fixed and write
\be
    p-p_*
    =
    (1+\delta)
    (p_*^{\,q}-p_*).
\ee
Near the stationary point,
\be
\begin{aligned}
    \mathcal V(T,x;p)
    ={}&
    -2\mathcal V_{0,h}\delta
    +\frac12
      (\partial_T^2\mathcal V)_h
      (T-T_h)^2
    \\
    &+
    (\partial_T\mathcal V')_h
    (T-T_h)(x-x_h)
    +\frac12
      \mathcal V_h''
      (x-x_h)^2
    +\cdots.
\end{aligned}
\label{eq:css_fold_control_term}
\ee
For a positive quadratic form, the local minimum remains above zero when $\delta<0$, reaches zero at $\delta=0$, and becomes negative for $\delta>0$. The nearby marginal branches then separate from $(T_h,x_h)$ with displacement proportional to $\sqrt{\delta}$.

This local expansion concerns the horizon function constructed from the first-order fields. It is distinct from the nonlinear terms omitted from the field equations. Extending the finite-radius MOTS to a solution of the full semiclassical equations requires independent control of those corrections.

\subsection*{Uniform large-$D$ limit}

The relevant radial region moves toward the flat-past boundary as $n$ grows. It is convenient to keep fixed
\be
    \zeta
    =
    x+2\log n
    =
    2\log z.
\label{eq:app_css_outer_log_coordinate}
\ee
At fixed $n$, $d/dx=d/d\zeta$, and primes below denote derivatives with respect to $\zeta$. All large-\(n\) remainder estimates below are uniform on fixed compact intervals in $\zeta$. The flat-past data are imposed at each finite $n$ before taking $n\to\infty$ at fixed $\zeta$.

The same fixed-$\zeta$ expansion used in \eqref{eq:css_outer_background_expansion} gives
\be
    R
    =
    \frac{1}{\sqrt2}
    \left(
        1+\frac{e^\zeta}{n^2}
    \right)
    +O(n^{-3}),
    \qquad
    \chi
    =
    1-\frac{2e^\zeta}{n^2}
    +O(n^{-4}).
\label{eq:css_outer_background_radius}
\ee
For the classical perturbation, the equations for $u_c$, $v_c$, and $w_c$ in \eqref{eq:app_css_classical_stripped_system} reduce at leading order to
\be
\begin{aligned}
    u_c'
    &=
    n\left(
        -\frac12u_c+v_c
    \right)+O(1),
    \\
    v_c'
    &=
    n\left(
        -\frac12v_c-f_c'
    \right)+O(1),
    \\
    w_c'
    &=
    n\left(
        -w_c-\frac14f_c
    \right)+O(1).
\end{aligned}
\label{eq:app_css_fast_equation}
\ee
On the bounded branch selected by the flat-past data, the leading $O(n)$ terms must cancel, leaving $u_c,v_c$, and $w_c$ fixed algebraically by the scalar profile. This gives
\be
    u_c=-4f_0'+O(n^{-1}),
    \qquad
    v_c=-2f_0'+O(n^{-1}),
    \qquad
    w_c=-\frac{f_0}{4}+O(n^{-1}),
\label{eq:css_outer_classical_metric}
\ee
where
\be
    f_c=f_0+O(n^{-1}).
\ee
The remaining scalar equation is
\be
    f_0''
    +\frac{e^\zeta}{4}f_0=0,
    \qquad
    f_0(-\infty)=1,
    \qquad
    f_0'(-\infty)=0.
\label{eq:css_outer_bessel_equation}
\ee
Since $z=e^{\zeta/2}$, the normalized solution is
\be
    f_0=J_0(z).
\ee
Collecting these, the finite-$n$ bounded solution therefore approaches
\be
    \widehat Y_{c}
    =
    \left(
        -4f_0',
        -2f_0',
        -\frac{f_0}{4},
        f_0,
        f_0'
    \right)^{\rm T}
    +O(n^{-1}).
\label{eq:app_css_classical_compact_convergence}
\ee
Writing
\be
    f_c
    =
    f_0+\frac{1}{n}f_1+O(n^{-2}),
\ee
the first correction obeys
\be
    f_1''
    +\frac{z^2}{4}f_1
    =
    -z^2f_0',
    \qquad
    f_1
    =
    zJ_1(z)
    -\frac{z^2}{2}J_0(z).
\label{eq:app_css_first_scalar_correction}
\ee
The classical outgoing profile is then
\be
    \widehat{\mathcal V}_c
    =
    zJ_1(z)
    +
    \frac{
        z(3-z^2)J_1(z)
        -3z^2J_0(z)
    }{2n}
    +O(n^{-2}).
\label{eq:app_css_classical_outgoing_correction}
\ee

The quantum sector simplifies further. In the fixed-$\zeta$ region,
$Q$ and $\mathcal V_0$ are $O(n^{-2})$, while $h^2$ is smaller than
any power of $1/n$. The matrix and source in
\eqref{eq:app_css_quantum_stripped_system} therefore approach
\be
    M_\infty=
    \begin{pmatrix}
        -1&1&0&0&0\\
        0&-1&0&0&0\\
        -1/2&-1&-2&0&0\\
        0&0&0&-3/2&1\\
        -1/4&-1/2&0&0&-2
    \end{pmatrix},
    \qquad
    J_\infty
    =
    \frac{1}{64\pi}
    \left(
        0,1,0,0,0
    \right)^{\rm T}.
\label{eq:app_css_quantum_limiting_matrix}
\ee
The eigenvalues of $M_\infty$ are
$-1,-1,-2,-3/2,-2$. The homogeneous equation admits terms
proportional to $e^{n\lambda\zeta}$. Since all eigenvalues satisfy
$\lambda<0$, these terms diverge as $\zeta\to-\infty$. The flat-past
boundary condition therefore sets their coefficients to zero. The
bounded quantum profile is consequently fixed by the one-loop source,
with no independent homogeneous contribution.

Since $M_\infty$ and $J_\infty$ are independent of $\zeta$, the
leading bounded solution is constant and satisfies
\be
    M_\infty\widehat Y_q+J_\infty=0.
\ee
Including the first $1/n$ correction gives
\be
    \widehat Y_q
    =
    \frac{1}{64\pi}\Bigg[
        \left(
            1,1,-\frac34,-\frac14,-\frac38
        \right)^{\rm T}
        +
        \frac1n
        \left(
            -1,-1,\frac32,\frac38,\frac9{16}
        \right)^{\rm T}
    \Bigg]
    +O(n^{-2}).
\label{eq:app_css_quantum_vector_expansion}
\ee
The combination entering the outgoing expansion is therefore
\be
    \widehat{\mathcal V}_q
    =
    \frac{1}{64\pi}
    \left(
        1-\frac1n
    \right)
    +O(n^{-2}).
\label{eq:app_css_quantum_outgoing_control}
\ee
To the order here, $\widehat{\mathcal V}_q$ is independent of
$\zeta$. The quantum term therefore contributes a nearly constant outgoing profile across this radial region, and its derivative
does not enter the leading large-$D$ condition from $G_n(x_h)=0$.

Substituting the classical and quantum limits into \eqref{eq:app_css_stripped_radial_function} gives
\be
    G_\infty(\zeta)
    =
    z^2
    \left(
        1-\frac{zJ_0(z)}{J_1(z)}
    \right).
\label{eq:app_css_limiting_radial_function}
\ee
The physical root satisfies
\be
    zJ_0(z)-J_1(z)=0,
    \qquad
    J_1(z)<0.
\ee
The smallest positive root satisfying this focusing condition is
\be
    z_*=5.3314427735\ldots.
\ee
At this root,
\be
    G_\infty'(2\log z_*)
    =
    \frac{z_*^2(z_*^2-1)}{2}>0.
\label{eq:app_css_limiting_root_derivative}
\ee
The root is therefore nondegenerate. Since the finite-$n$ profiles and their relevant derivatives converge uniformly near $z_*$, the implicit-function theorem gives a unique nearby root of the exact finite-$n$ onset equation for all sufficiently large $n$. 

The first correction to its position follows from
\be
\begin{aligned}
    \frac{G_n}{z^2}
    &=
    1-\frac{zJ_0(z)}{J_1(z)}
    -\frac{3}{n}
      \frac{d}{d\zeta}
      \left[
        1-\frac{zJ_0(z)}{J_1(z)}
      \right]
    +O(n^{-2}),
    \\
    \zeta_h
    &=
    2\log z_*+\frac3n+O(n^{-2}).
\end{aligned}
\label{eq:css_large_D_first_correction}
\ee
Finally, the coefficient in \eqref{eq:css_marginal_radius} can be written as
\be
    C_n
    =
    \mathcal F_hQ_h
    \left[
        \frac{
            n\widehat{\mathcal V}_{q,h}
        }{
            (n-1)\mathcal V_{0,h}
        }
    \right]^{1/n}.
\label{eq:css_marginal_radius_coefficient}
\ee
The required factors are
\be
\begin{aligned}
    Q_h
    &=
    \frac{4\alpha e^{\zeta_h}}{n^2}
    \left[
        1+O(n^{-2})
    \right],
    \\
    \mathcal V_{0,h}
    &=
    \frac{e^{\zeta_h}}{n^2}
    \left[
        1+O(n^{-2})
    \right],
    \\
    \mathcal F_h
    &=
    1+O(n^{-2}),
    \qquad
    \widehat{\mathcal V}_{q,h}
    =
    \frac{1}{64\pi}
    \left(
        1-\frac1n
    \right)
    +O(n^{-2}).
\end{aligned}
\label{eq:app_css_radius_factors}
\ee
Using $n(1-1/n)/(n-1)=1$ and
\eqref{eq:css_large_D_first_correction}, we obtain
\be
    \log C_n
    =
    \log\!\left(
        \frac{4z_*^2}{n^2}
    \right)
    +
    \frac{
        2\log n+2-\log(64\pi z_*^2)
    }{n}
    +
    O\!\left(
        \frac{(\log n)^2}{n^2}
    \right).
\label{eq:app_css_log_radius_expansion}
\ee
Exponentiating this expression gives \eqref{eq:css_large_D_marginal_radius}. This analytic result identifies the radial component that we continue numerically toward finite $D$ below.

\subsection*{Finite-$D$ continuation and numerical checks}

We now follow the radial component selected by the large-$D$ analysis toward finite $D$. The finite-$D$ radial equations are integrated from the flat-past series using the normalization above. The Einstein equation not used in the classical integration, together with the quantum residual \eqref{eq:css_unused_quantum_constraint}, provides an independent consistency check on the classical and quantum integrations, respectively.

Along this component, the ingoing orientation changes between $D=11$ and $D=12$. A continuous interpolation places the sign change near $n\simeq9.73647$. The first integer dimension on this component with the orientation required for a future MOTS is therefore $n=10$, or $D=12$. 

At $D=12$, the stationary root is
\be
    Q_h\simeq0.85812252,
    \qquad
    x_h\simeq-0.47360216.
\label{eq:app_css_D12_root}
\ee
The stationary amplitudes give
\be
    p_*^{\,q}-p_*
    \simeq
    5.25158\,g_q^{1/2},
    \qquad
    \frac{r_h}{\ell_q}
    \simeq
    0.539925.
\label{eq:app_css_D12_physical_values}
\ee
The first coefficient depends on the normalization of the classical perturbation, while the radius does not. At this solution,
\be
    \mathcal F_h>0,
    \qquad
    (\mathcal F-\partial_T\mathcal F)_h>0,
    \qquad
    \widehat{\mathcal V}_{c,h}<0
    <
    \widehat{\mathcal V}_{q,h},
\ee
so the marginal sphere has positive radius, positive amplitudes in the convention of the main text, and the required future orientation. The Hessian eigenvalues are approximately
\be
    \lambda_1\simeq1.98145,
    \qquad
    \lambda_2\simeq18.5133,
\label{eq:app_css_D12_hessian}
\ee
confirming that the stationary zero is a local minimum.

Following the same radial component through the integer dimensions
\be
    10\leq n\leq127
\ee
gives a future-oriented positive-radius local minimum at every sampled dimension. The same flat-past normalization is used throughout, and the selected roots approach the large-$D$ Bessel component identified above as $D$ increases. This provides a numerical continuation between the $D=12$ solution and the analytic large-$D$ regime.

The curvature invariants constructed from the first-order geometry remain finite at the onset. At fixed finite $D$, increasing the radius in Planck units can therefore place this geometry in a weak-curvature regime. This does not by itself control the nonlinear remainder of the critical-departure expansion. Quadratic scalar terms omitted from the first-order equations can become important near the marginal sphere, even when the curvature and loop expansion remain under control. Their size relative to the retained linear terms is not suppressed by increasing $N$ or shifting $T_h$. The continuation of the first-order MOTS into the full semiclassical solution therefore requires an additional nonlinear estimate.

\section{DSS auxiliary and variational identities}
\label{app:dss}

This appendix collects the derivations used in
Section~\ref{sec:dss}. All covariant derivatives refer to the 2d metric.
The action is varied before evaluating its sources on the
leading large-$D$ background, at fixed integer $n=D-2\geq2$.

\subsection*{Auxiliary fields and echo memory}

For the 2d part of \eqref{eq:dss_leading_metric}, define
\be
    L(T)=\ell e^{-T},
    \qquad
    a(T,x)
    =
    \sqrt{\frac{1+\beta^2(T)x^2}{1+\beta^2(T)}}.
    \label{eq:app_dss_metric_functions}
\ee
The 2d line element and its volume element are then
\be
    ds_2^2
    =
    L^2
    \left[
        (x^2-a^2)dT^2
        -2x\,dT\,dx
        +dx^2
    \right],
    \qquad
    \sqrt{-g}=L^2a.
\ee
For any scalar $\eta$, the wave operator takes the divergence
form
\be
    aL^2\Box\eta
    ={}
    \partial_T
    \left[
        -\frac{\partial_T\eta+x\partial_x\eta}{a}
    \right]
+
    \partial_x
    \left[
        -\frac{x\partial_T\eta}{a}
        +
        \left(
            a-\frac{x^2}{a}
        \right)\partial_x\eta
    \right].
    \label{eq:app_dss_flux_identity}
\ee
A smooth auxiliary field satisfies
$\partial_x\eta(T,0)=0$, so the $x$-boundary term vanishes at
the center. At $x=1$, $a=1$ and the coefficient of
$\partial_x\eta$ vanishes, leaving $-\partial_T\eta(T,1)$.

For $U$, the radial integral of the source in
\eqref{eq:dss_background_auxiliary_equations} is
\be
    \int_0^1dx\,
    \frac{\beta^2}
    {\sqrt{1+\beta^2}(1+\beta^2x^2)^{3/2}}
    =
    \frac{\beta^2}{1+\beta^2}.
    \label{eq:app_dss_memory_radial_integral}
\ee
Integrating \eqref{eq:app_dss_flux_identity} over the radial
interval therefore gives
\be
    \partial_T
    \left[
        U(T,1)
        +
        \int_0^1dx\,
        \frac{\partial_TU+x\partial_xU}{a}
    \right]
    =
    \frac{2\beta^2(T)}{1+\beta^2(T)}.
    \label{eq:app_dss_integrated_auxiliary_equation}
\ee
The initial Minkowski data set the integration constant to zero,
so
\be
    U(T,1)
    +
    \int_0^1dx\,
    \frac{\partial_TU+x\partial_xU}{a}
    =
    2\int_{T_i}^{T}d\tau\,
    \frac{\beta^2(\tau)}{1+\beta^2(\tau)}.
    \label{eq:app_dss_retarded_balance}
\ee
This exact relation holds through the preparation interval and
the subsequent periodic evolution.

Once $\partial_TU$ and $\partial_xU$ repeat with period
$\Delta_s$, the radial integral on the LHS of
\eqref{eq:app_dss_retarded_balance} also repeats. Taking the
difference after one period determines the change in
$U(T,1)$. Since $\partial_xU$ is periodic, this change is
independent of $x$, giving
\be
    U(T+\Delta_s,x)-U(T,x)
    =
    m_U\Delta_s,\qquad
    m_U
    =
    \frac{2}{\Delta_s}
    \int_T^{T+\Delta_s}d\tau\,
    \frac{\beta^2(\tau)}{1+\beta^2(\tau)}.
    \label{eq:app_dss_memory_increment_derivation}
\ee
This proves \eqref{eq:dss_memory_coefficient}. For every
nontrivial smooth periodic $\beta(T)$, the integrand lies
between zero and one and therefore
\be
    0<m_U<2.
\ee
The exact balance \eqref{eq:app_dss_retarded_balance} also
contains the preparation history. The constant increment above
arises once the auxiliary derivatives have become periodic.

For later use, define
\be
    \mathcal K_{ab}[F,G]
    =
    \nabla_aF\nabla_bG
    +
    \nabla_bF\nabla_aG
    -
    g_{ab}\nabla F\cdot\nabla G.
    \label{eq:app_dss_bilinear_tensor}
\ee
Before imposing the DSS background, the full variation of
$\Gamma_{\rm min}$ gives
\be
\begin{split}
    \frac{\langle T^{(2)}_{ab}\rangle_{\rm min}}{N\hbar}
    ={}&
    \frac{\mathcal K_{ab}[U,U]}{96\pi}
    -
    \frac{\mathcal K_{ab}[U,W]}{8\pi}
    -
    \frac{n\mathcal K_{ab}[U,\phi]}{16\pi}
    \\
    &-
    \frac{\nabla_a\nabla_bU-g_{ab}\Box U}{24\pi}
    +
    \frac{\nabla_a\nabla_bW-g_{ab}\Box W}{4\pi}
    \\
    &-
    \frac{U}{4\pi}
    \left[
        \nabla_a\phi\nabla_b\phi
        -
        \frac12g_{ab}(\nabla\phi)^2
    \right],
\end{split}
    \label{eq:dss_auxiliary_stress}
\ee
together with
\be
    \frac{E_{\phi,{\rm min}}}{N\hbar}
    =
    -
    \frac{
        \nabla U\cdot\nabla\phi+U\Box\phi
    }{4\pi}
    -
    \frac{nR^{(2)}}{16\pi},
    \qquad
    E_{f,{\rm min}}=0.
    \label{eq:dss_auxiliary_dilaton_source}
\ee
Using $\Box U=R^{(2)}$ and
$\Box W=(\nabla\phi)^2-n\Box\phi/2$, contraction of
\eqref{eq:dss_auxiliary_stress} reproduces
\eqref{eq:minkowski_normalized_trace}. Its divergence together
with \eqref{eq:dss_auxiliary_dilaton_source} satisfies
\eqref{eq:reduced_ward_identity}.

The only terms depending on undifferentiated $U$ are the last
term in \eqref{eq:dss_auxiliary_stress} and the $U\Box\phi$
term in \eqref{eq:dss_auxiliary_dilaton_source}. The linear
echo memory therefore enters the minimal source through these
terms. The initial Minkowski data fix the additive constant of
$U$.

\subsection*{Finite action variations and center regularity}

\paragraph{Memory and trace cancellation.}

For the variation of \eqref{eq:dss_periodic_completion}, the
background values needed below are
\be
\begin{aligned}
    X_*
    &=
    \frac{n^2}{4L^2x^2},
    &
    Z_*
    &=
    -\frac{n\beta^2}{L^2(1+\beta^2x^2)},
    \\
    \sigma_*
    &=
    \beta^2x^2,
    &
    \mathcal A_*
    &=
    \frac{\beta^2}{L^2(1+\beta^2x^2)^2}
    =
    -\frac{R_*^{(2)}}{2}.
\end{aligned}
    \label{eq:app_dss_carrier_background_value}
\ee
Here $X=(\nabla\phi)^2$ and $Z=(\nabla f)^2$, as in
Section~\ref{sec:dss}. Since
$\Box\Psi=\mathcal A$ and $\Box U=R^{(2)}$, using the same
retarded inverse and initial vacuum data gives
\be
    \Psi_*=-\frac{U}{2}.
\ee

Although $\mathcal E_*=\mathcal C_*=0$, their first variations
contribute to the source. To determine the coefficient of
$\mathcal E$ in $\mathcal A$, temporarily replace
$\sigma/n$ by $h(\sigma)$. Cancellation of the required metric
terms gives
\be
    2n\sigma(1+\sigma)h'
    -
    n(1+2\sigma)h
    -
    \sigma
    =
    0,
    \qquad
    h(\sigma)
    =
    \frac{\sigma}{n}
    +
    C_0\sqrt{\sigma(1+\sigma)},
    \label{eq:app_dss_carrier_coefficient}
\ee
where the prime denotes $d/d\sigma$. On the background,
$\sigma_*=\beta^2x^2$, so the homogeneous term is proportional
to
\[
    |\beta|x\sqrt{1+\beta^2x^2}.
\]
Smoothness at the center and through zeros of $\beta$ therefore
sets $C_0=0$. The term proportional to
$\mathcal C\Box f$ in \eqref{eq:dss_periodic_completion}
removes the remaining mixed metric contribution.

The terms containing undifferentiated $\Psi$ are then
\be
\begin{aligned}
    \langle T^{(2)}_{ab}\rangle_W
    &\supset
    -
    \frac{N\hbar}{2\pi}\Psi
    \left[
        \nabla_a\phi\nabla_b\phi
        -
        \frac12g_{ab}X
    \right],
    \\
    E_{\phi,W}
    &\supset
    -
    \frac{N\hbar}{2\pi}\Psi\Box\phi.
\end{aligned}
    \label{eq:app_dss_memory_cancellation}
\ee
Using $\Psi_*=-U/2$, these cancel the terms containing
undifferentiated $U$ in
\eqref{eq:dss_auxiliary_stress} and
\eqref{eq:dss_auxiliary_dilaton_source}. On the leading
background, $E_{f,W}$ contains no corresponding term
proportional to undifferentiated $\Psi$. The remaining
auxiliary dependence therefore involves derivatives, which
repeat once the auxiliary derivatives have entered the
periodic regime.

Before imposing the auxiliary equation, the variation of
$\Gamma_W$ satisfies
\be
\begin{split}
    \nabla^a\langle T^{(2)}_{ab}\rangle_W
    -
    E_{\phi,W}\nabla_b\phi
    -
    E_{f,W}\nabla_bf
    =
    \frac{n^2N\hbar}{16\pi}
    \left(
        \mathcal A-\Box\Psi
    \right)\nabla_b\Psi.
\end{split}
    \label{eq:app_dss_carrier_Ward_remainder}
\ee
Thus \eqref{eq:reduced_ward_identity} follows when
$\Box\Psi=\mathcal A$. Since $\Gamma_W$ is Weyl invariant, it
does not change the 2d trace. The remaining trace is canceled
by the finite term \eqref{eq:dss_finite_trace_action}, whose
Weyl variation gives
\be
\begin{split}
    g^{ab}\langle T^{(2)}_{ab}\rangle_{\rm ct}
    =
    -
    \frac{N\hbar}{24\pi}
    \left[
        \frac{4}{n^2}r^2XR^{(2)}
        +
        \frac{6}{n}r^2XZ
        +
        \frac{4}{n^2}\Box(r^2X)
    \right].
\end{split}
    \label{eq:app_dss_finite_trace}
\ee
On the leading background, $r_*^2X_*=n^2/4$ is constant.
Using the background identity $\mathcal E_*=0$ then gives
\be
    \left.
    g^{ab}\langle T^{(2)}_{ab}\rangle_{\rm ct}
    \right|_*
    =
    -
    \frac{N\hbar}{24\pi}
    \left[
        R_*^{(2)}
        +
        \frac{3n}{2}Z_*
    \right]
    =
    -
    \left.
    \langle T^{(2)a}{}_{a}\rangle_{\rm min}
    \right|_*.
    \label{eq:app_dss_trace_cancellation}
\ee
The counterterm also changes the scalar source. In particular,
\be
    E_{f,{\rm ct}}
    =
    -
    \frac{N\hbar}{4\pi n}
    \nabla_a
    \left(
        r^2X\nabla^af
    \right).
    \label{eq:app_dss_trace_scalar_source}
\ee
Its remaining variations must therefore be included when the
complete center conditions are imposed.

Both $\Gamma_W$ and $\Gamma_{\rm ct}^{\rm DSS}$ preserve the
Minkowski normalization. On the Minkowski background,
$\mathcal A=\Psi=0$. For $\Gamma_{\rm ct}^{\rm DSS}$,
$r^2X=n^2/4$ is constant while $R^{(2)}=Z=0$, so its bulk
first variation also vanishes.

\paragraph{Variation of $\Delta\Gamma_W$.}

We now vary \eqref{eq:dss_localized_profile_action}. The
coefficients $e$ and $c$ are covariant scalar functionals with
zero Weyl weight. Since
$\mathcal E_*=\mathcal C_*=0$, their variations do not enter the
first-order source on the leading background,
\be
    \delta(e\mathcal E+c\mathcal C)\big|_*
    =
    e\,\delta\mathcal E+c\,\delta\mathcal C.
\ee
Only the background profiles $e(T,x)$ and $c(T,x)$ are therefore
needed in the variations below.

The metric variation gives
\be
\begin{split}
    \Delta\langle T^{(2)}_{ab}\rangle
    =
    \frac{n^2N\hbar}{8\pi}
    \Bigg[
        &e(\Box\phi)g_{ab}
        -
        \frac{4e}{n}\nabla_a\phi\nabla_b\phi
        -
        e\nabla_af\nabla_bf
        \\
        &-
        \mathcal K_{ab}[e,\phi]
        -
        \frac{c}{2}
        \left(
            \nabla_a\phi\nabla_bf
            +
            \nabla_b\phi\nabla_af
        \right)
    \Bigg],
\end{split}
    \label{eq:app_dss_finite_metric_variation}
\ee
while variation with respect to $\phi$ and $f$ gives
\be
\begin{aligned}
    \Delta E_\phi
    &=
    -
    \frac{n^2N\hbar}{8\pi}
    \left[
        \Box e
        +
        \frac{4}{n}\nabla_a(e\nabla^a\phi)
        +
        \frac12\nabla_a(c\nabla^af)
    \right],
    \\
    \Delta E_f
    &=
    -
    \frac{n^2N\hbar}{8\pi}
    \nabla_a
    \left(
        e\nabla^af
        +
        \frac{c}{2}\nabla^a\phi
    \right).
\end{aligned}
    \label{eq:app_dss_finite_field_variations}
\ee
The fields on the RHS are evaluated on the leading DSS
background after the variation has been taken.

Contracting
\eqref{eq:app_dss_finite_metric_variation} gives
\be
    g^{ab}\Delta\langle T^{(2)}_{ab}\rangle
    =
    -
    \frac{n^2N\hbar}{8\pi}
    \left(
        e\mathcal E+c\mathcal C
    \right),
\ee
which vanishes on
$\mathcal E_*=\mathcal C_*=0$.
Together with
\eqref{eq:app_dss_finite_field_variations}, the same variation
satisfies \eqref{eq:reduced_ward_identity}.

Projecting
\eqref{eq:app_dss_finite_metric_variation} and
\eqref{eq:app_dss_finite_field_variations} onto the orthonormal
frame of Section~\ref{sec:dss} gives
\eqref{eq:dss_radial_profile_equations}. On the leading
background,
\be
    \partial_{\hat0}\phi_*=0,
    \qquad
    \partial_{\hat1}f_*=0,
\ee
so $c$ does not contribute directly to
$\Delta\langle T^{(2)}_{\hat0\hat0}\rangle$. The energy-density
equation therefore determines $e$ first. Its contribution to the
scalar source is
\be
    \Delta E_f^{(e)}
    =
    \frac{
        n^{5/2}N\hbar\sqrt{1+\beta^2}
    }{8\pi L^2}
    \Bigg[
        \frac{
            \beta\left(
                \partial_Te+x\partial_xe
            \right)
        }{1+\beta^2x^2}
+
        \frac{
            (\partial_T\beta+\beta)e
        }{(1+\beta^2x^2)^2}
    \Bigg].
    \label{eq:app_dss_explicit_scalar_source}
\ee
This is
$-n^2N\hbar\nabla_a(e\nabla^af_*)/(8\pi)$ evaluated on the
background. At a zero of $\beta$,
\be
    \Delta E_f^{(e)}
    =
    \frac{n^{5/2}N\hbar}{8\pi L^2}
    (\partial_T\beta)e,
\ee
so the contribution generated by $e$ remains finite through the scalar turning points. Once $e$ is known, the second equation in
\eqref{eq:dss_radial_profile_equations} determines $c$ from the
remaining scalar source $\Delta E_f-\Delta E_f^{(e)}$.

At first order, these bulk expressions depend only on the background
profiles of $e$ and $c$. A local covariant realization of the radial
reconstruction can be made explicit. On the region $X=(\nabla \phi)^2>0$, define
\be
    \bar g_{ab}
    =
    \frac{4r^2X}{n^2}g_{ab}.
\ee
This metric is Weyl invariant and satisfies $\bar g_{ab}=g_{ab}$ on
the leading DSS family. The covariant operator
\be
    \mathcal D
    =
    r\bar g^{ab}\nabla_a r\,\nabla_b
\ee
then reduces to $x\partial_x$ on this background. The radial equations
for $e$ and $c$ can therefore be formulated covariantly without
introducing the coordinates $T$ and $x$ explicitly. Their extension
away from the leading background remains part of the definition of
the finite action. We take the variations to have compact support in
the interior, with boundary values obtained from the smooth limits of
the resulting source.

\paragraph{Regularity throughout the interior.}

The radial equations
\eqref{eq:dss_radial_profile_equations} can be written as
\be
\begin{aligned}
    \partial_x
    \left(
        \frac{e}{x\sqrt{1+\beta^2x^2}}
    \right)
    &=
    \frac{16\pi L^2}{n^3N\hbar}
    \frac{
        \Delta\langle T^{(2)}_{\hat0\hat0}\rangle
    }{\sqrt{1+\beta^2x^2}},
    \\
    \partial_x
    \left(
        \frac{\sqrt{1+\beta^2x^2}}{x}c
    \right)
    &=
    \frac{32\pi L^2}{n^3N\hbar}
    \sqrt{1+\beta^2x^2}
    \left(
        \Delta E_f-\Delta E_f^{(e)}
    \right).
\end{aligned}
    \label{eq:app_dss_integrating_factors}
\ee
At fixed $T$, the corresponding homogeneous solutions are
proportional to
$x\sqrt{1+\beta^2x^2}$ and
$x/\sqrt{1+\beta^2x^2}$. Both begin linearly in $x$ and are
incompatible with a smooth scalar profile at the center.
Equivalently, requiring $e/x^2$ and $c/x^2$ to remain finite
removes both integration functions. Thus, for fixed
prescribed
$\Delta\langle T^{(2)}_{\hat0\hat0}\rangle$ and $\Delta E_f$,
the center-regular profiles $e$ and $c$ are unique.

If both prescribed changes begin at $O(x^n)$ with coefficients
smooth in $x^2$, the first radial integral gives
$e=O(x^{n+2})$. This follows directly by setting $u=xz$ in
the integral, which extracts an overall factor $x^{n+2}$.
\eqref{eq:app_dss_explicit_scalar_source} then gives
$\Delta E_f^{(e)}=O(x^{n+2})$, and the second radial integral
gives
\be
    e,c=O(x^{n+2}),
    \qquad
    \Delta E_f^{(e)}=O(x^{n+2}).
    \label{eq:app_dss_profile_center_falloff}
\ee
The radial reconstruction therefore does not generate lower
powers or spoil the prescribed center falloff.

Substitution into
\eqref{eq:app_dss_finite_metric_variation} and
\eqref{eq:app_dss_finite_field_variations} gives
\be
    \Delta\langle T^{(2)}_{\hat0\hat0}\rangle,
    \quad
    \Delta\langle T^{(2)}_{\hat1\hat1}\rangle,
    \quad
    \Delta E_\phi,
    \quad
    \Delta E_f
    =
    O(x^n),
\ee
together with
\be
    \Delta\langle T^{(2)}_{\hat0\hat1}\rangle
    =
    O(x^{n+1}),
    \qquad
    \Delta\langle T^{(2)}_{\hat1\hat1}\rangle
    +
    \frac12\Delta E_\phi
    =
    O(x^{n+2}).
    \label{eq:app_dss_regular_finite_additions}
\ee
After the inverse transverse-area factor is included, the first
relation makes the physical radial flux vanish linearly at the
center, while the second makes the radial and angular pressures
agree up to $O(x^2)$. After the displayed powers are removed,
the remaining coefficients are smooth functions of $x^2$.

The radial solutions remain regular throughout
$0\leq x\leq1$. The kernels in
\eqref{eq:app_dss_integrating_factors} are smooth through zeros
of $\beta$ and remain finite at $x=1$, since
$1+\beta^2x^2>0$. No second radial integration function is needed at the past SSH. If the prescribed changes
vanish in the initial Minkowski region, the same branch gives
$e=c=0$ there.

The falloff estimate above applies when the prescribed changes already
have the required center behavior. The changes used to cancel
lower-order radial terms in the source generated by
$\Gamma_{\rm min}+\Gamma_W+\Gamma_{\rm ct}^{\rm DSS}$ need not satisfy
these conditions separately. The center conditions
\eqref{eq:dss_complete_center_conditions} are imposed on the complete
source after the cancellation.

The radial reconstruction also preserves the scalar
half-echo symmetry. The scaled quantity
$L^2\Delta\langle T^{(2)}_{\hat0\hat0}\rangle$ is even under
$T\to T+\Delta_g$, while $L^2\Delta E_f$ is odd. Since the
first kernel in \eqref{eq:app_dss_integrating_factors} depends
only on $\beta^2$, it follows that
\be
    e(T+\Delta_g,x)=e(T,x).
\ee
Under the same shift,
$\beta$ and $\partial_T\beta$ change sign, so
\eqref{eq:app_dss_explicit_scalar_source} implies that
$L^2\Delta E_f^{(e)}$ is odd. The second radial integral then
gives
\be
    c(T+\Delta_g,x)=-c(T,x).
\ee
These are the half-echo parities used in
\eqref{eq:dss_periodic_quantum_source}.

\subsection*{Radial constraint and null expansion}

In the untrapped region, use polar-areal coordinates
\be
    ds_D^2
    =
    -\mathcal N^2(t,r)\,dt^2
    +
    \frac{dr^2}{H(t,r)}
    +
    r^2d\Omega_n^2.
\ee
The unit timelike vector at fixed areal radius is
$u^\mu=\mathcal N^{-1}(\partial_t)^\mu$. If
\be
    \varepsilon
    =
    T_{\mu\nu}^{\rm tot}u^\mu u^\nu
\ee
is the total orthonormal energy density, the $tt$ Einstein
equation gives
\be
    r\partial_rH
    +
    (n-1)(H-1)
    =
    -
    \frac{16\pi G_D}{n}r^2\varepsilon.
    \label{eq:app_dss_radial_Einstein_constraint}
\ee
Subtracting the classical equation at the same $(t,r)$ and
retaining the first-order change gives
\be
    \partial_r
    \left(
        r^{n-1}\delta H
    \right)
    =
    -
    \frac{16\pi G_D}{n}
    r^n\delta\varepsilon_{\rm eff}.
    \label{eq:app_dss_linear_radial_constraint}
\ee
Here $\delta\varepsilon_{\rm eff}$ is the first-order change in
the total energy density, including the complete one-loop
energy and the induced change in the classical scalar energy.
The homogeneous radial solution is proportional to
$r^{1-n}$. Center regularity removes it for $n\geq2$, and
integration gives
\eqref{eq:dss_cumulative_energy_response}. This subtraction is
performed at finite $D$ before the leading large-$D$ background
is imposed.

For a smooth source,
\be
    \delta\varepsilon_{\rm eff}(t,r)
    =
    \delta\varepsilon_{\rm eff}(t,0)
    +
    O(r^2)
\ee
near the center.
\eqref{eq:dss_cumulative_energy_response} then gives
\be
    \delta H(t,r)
    =
    -
    \frac{16\pi G_D}{n(n+1)}
    r^2\delta\varepsilon_{\rm eff}(t,0)
    +
    O(r^4).
    \label{eq:app_dss_center_backreaction}
\ee
The radial backreaction is therefore regular at the center.

To relate $H$ to the null expansion, let
$s^\mu=\sqrt H(\partial_r)^\mu$ be the outward radial unit
vector and define
\be
    k_\pm^\mu
    =
    \frac{u^\mu\pm s^\mu}{\sqrt2}.
\ee
The null expansions of the symmetry spheres are
\be
    \theta_\pm
    =
    \frac{n}{r}k_\pm^\mu\nabla_\mu r
    =
    \pm\frac{n\sqrt H}{\sqrt2\,r}.
\ee
Linearizing the outgoing expansion about $H_*=1$ gives
\be
    \delta\theta_+
    =
    \frac{n}{2\sqrt2\,r}\delta H.
    \label{eq:app_dss_null_expansion_variation}
\ee
Hence positive cumulative effective energy lowers both $H$
and the outgoing expansion at fixed areal radius, while
negative cumulative energy raises them.

The large-$D$ radial localization follows from
\be
    \int_0^rds\,s^n\delta\varepsilon_{\rm eff}(t,s)
    =
    r^{n+1}
    \int_0^1dz\,z^n
    \delta\varepsilon_{\rm eff}(t,rz).
    \label{eq:app_dss_energy_localization}
\ee
Writing
\be
    z=1-\frac{y}{n}
\ee
gives $z^n\to e^{-y}$ at fixed $y$ as $n\to\infty$.
The integral is therefore concentrated in the layer
\be
    r-s=O(r/n)
\ee
immediately inside the sphere at radius $r$. If the energy
profile varies slowly across this layer,
\be
    \int_0^rds\,s^n\delta\varepsilon_{\rm eff}(t,s)
    =
    \frac{r^{n+1}}{n+1}
    \left[
        \delta\varepsilon_{\rm eff}(t,r)
        +
        O(n^{-1})
    \right],
\ee
and therefore
\be
    \delta H(t,r)
    =
    -
    \frac{16\pi G_D}{n(n+1)}
    r^2
    \left[
        \delta\varepsilon_{\rm eff}(t,r)
        +
        O(n^{-1})
    \right].
    \label{eq:app_dss_local_radial_constraint}
\ee
If the source varies appreciably across a distance $r/n$, the
full integral in
\eqref{eq:dss_cumulative_energy_response} must instead be
retained.

Finally, consider the residual finite profiles used in
\eqref{eq:dss_regular_finite_sign_choices},
\be
    \delta e_\pm
    =
    \pm\frac{\beta^2}{n^2}x^{n+2},
    \qquad
    \delta c_\pm=0.
\ee
These represent different smooth finite profile choices around
a center-regular completion, not homogeneous solutions of
\eqref{eq:dss_radial_profile_equations} for fixed prescribed
sources. Their energy projection obeys
\be
    x\partial_x\delta e_\pm
    -
    \frac{1+2\beta^2x^2}{1+\beta^2x^2}\delta e_\pm
    =
    \delta e_\pm
    \left(
        n+\frac{1}{1+\beta^2x^2}
    \right).
    \label{eq:app_dss_sign_pair_projection}
\ee
The bracket is strictly positive, giving the opposite direct
one-loop energy shifts in
\eqref{eq:dss_regular_finite_energy_sign}.

%%%%%%%%%%%%%%%%%%%%%%%%%%%%%%%%%%%%%%%%%%%%%%%%%%
\addcontentsline{toc}{section}{References}
\bibliographystyle{JHEP}
\bibliography{bibliography}

\end{document}